\documentclass[12pt,a4paper]{article}

\usepackage{ifthen} 
\newboolean{pdflatex}
\setboolean{pdflatex}{true} 

\newboolean{articletitles}
\setboolean{articletitles}{true} 

\newboolean{uprightparticles}
\setboolean{uprightparticles}{false} 

\def\paperauthors{LHCb collaboration} 
\def\paperasciititle{Simultaneous determination of CKM angle gamma and charm mixing parameters} 
\def\papertitle{Simultaneous determination\\ of CKM angle $\gamma$ and \\charm mixing parameters} 
\def\paperkeywords{{High Energy Physics}, {LHCb}} 
\def\papercopyright{\the\year\ CERN for the benefit of the LHCb collaboration} 
\def\paperlicence{CC BY 4.0 licence}
\def\paperlicenceurl{https://creativecommons.org/licenses/by/4.0/}

\usepackage[top=1in, bottom=1.25in, left=1in, right=1in]{geometry}

\usepackage{microtype}
\usepackage{lineno}  
\usepackage{xspace} 
\usepackage{caption} 

\usepackage{graphicx}  
\usepackage{color}
\usepackage{colortbl}
\graphicspath{{./figs/}} 

\usepackage{amsmath} 
\usepackage{amssymb}
\usepackage{amsfonts}
\usepackage{upgreek} 

\newcommand*\patchAmsMathEnvironmentForLineno[1]{%
\expandafter\let\csname old#1\expandafter\endcsname\csname #1\endcsname
\expandafter\let\csname oldend#1\expandafter\endcsname\csname
end#1\endcsname
 \renewenvironment{#1}%
   {\linenomath\csname old#1\endcsname}%
   {\csname oldend#1\endcsname\endlinenomath}%
}
\newcommand*\patchBothAmsMathEnvironmentsForLineno[1]{%
  \patchAmsMathEnvironmentForLineno{#1}%
  \patchAmsMathEnvironmentForLineno{#1*}%
}
\AtBeginDocument{%
\patchBothAmsMathEnvironmentsForLineno{equation}%
\patchBothAmsMathEnvironmentsForLineno{align}%
\patchBothAmsMathEnvironmentsForLineno{flalign}%
\patchBothAmsMathEnvironmentsForLineno{alignat}%
\patchBothAmsMathEnvironmentsForLineno{gather}%
\patchBothAmsMathEnvironmentsForLineno{multline}%
\patchBothAmsMathEnvironmentsForLineno{eqnarray}%
}

\usepackage{hyperxmp}

\usepackage[pdftex,
            pdfauthor={\paperauthors},
            pdftitle={\paperasciititle},
            pdfkeywords={\paperkeywords},
            pdfcopyright={Copyright (C) \papercopyright},
            pdflicenseurl={\paperlicenceurl}]{hyperref}

\usepackage[colorinlistoftodos,textsize=scriptsize]{todonotes}

\usepackage[bottom,flushmargin,hang,multiple]{footmisc}

\usepackage[all]{hypcap} 

\usepackage{xspace} 
\usepackage{upgreek}

\def\MagUp {\mbox{\em Mag\kern -0.05em Up}\xspace}

\ifthenelse{\boolean{uprightparticles}}%
{
 
 \def\Pgamma      {\ensuremath{\upgamma}\xspace}

 \def\Ppi         {\ensuremath{\uppi}\xspace}

 \def\PDelta      {\ensuremath{\Delta}\xspace}                 
 \def\PXi         {\ensuremath{\Xi}\xspace}                 
 \def\PLambda     {\ensuremath{\Lambda}\xspace}                 
 \def\PSigma      {\ensuremath{\Sigma}\xspace}                 
 \def\POmega      {\ensuremath{\Omega}\xspace}                 
 \def\PUpsilon    {\ensuremath{\Upsilon}\xspace}

 \def\PB      {\ensuremath{\mathrm{B}}\xspace}                 
                  
 \def\PD      {\ensuremath{\mathrm{D}}\xspace}

 \def\PK      {\ensuremath{\mathrm{K}}\xspace}

 \def\Pb      {\ensuremath{\mathrm{b}}\xspace}                 
 \def\Pc      {\ensuremath{\mathrm{c}}\xspace}

 \def\Pi      {\ensuremath{\mathrm{i}}\xspace}

 \def\Ps      {\ensuremath{\mathrm{s}}\xspace}                 
                  
 \def\Pu      {\ensuremath{\mathrm{u}}\xspace}

 \def\thebaroffset{0.0em}
}
{
 
 \def\Pgamma      {\ensuremath{\gamma}\xspace}

 \def\Ppi         {\ensuremath{\pi}\xspace}

 \mathchardef\PDelta="7101
 \mathchardef\PXi="7104
 \mathchardef\PLambda="7103
 \mathchardef\PSigma="7106
 \mathchardef\POmega="710A
 \mathchardef\PUpsilon="7107
                  
 \def\PB      {\ensuremath{B}\xspace}                 
                  
 \def\PD      {\ensuremath{D}\xspace}

 \def\PK      {\ensuremath{K}\xspace}

 \def\Pb      {\ensuremath{b}\xspace}                 
 \def\Pc      {\ensuremath{c}\xspace}

 \def\Pi      {\ensuremath{i}\xspace}

 \def\Ps      {\ensuremath{s}\xspace}                 
                  
 \def\Pu      {\ensuremath{u}\xspace}

 \def\thebaroffset{0.18em}
}
\newcommand{\offsetoverline}[2][\thebaroffset]{\kern #1\overline{\kern -#1 #2}}%

\makeatletter
\ifcase \@ptsize \relax
  \newcommand{\miniscule}{\@setfontsize\miniscule{4}{5}}
\or
  \newcommand{\miniscule}{\@setfontsize\miniscule{5}{6}}
\or
  \newcommand{\miniscule}{\@setfontsize\miniscule{5}{6}}
\fi
\makeatother

\DeclareRobustCommand{\optbar}[1]{\shortstack{{\miniscule (\rule[.5ex]{1.25em}{.18mm})}
  \\ [-.7ex] $#1$}}

\def\g      {{\ensuremath{\Pgamma}}\xspace}

\def\uquark    {{\ensuremath{\Pu}}\xspace}

\def\squark    {{\ensuremath{\Ps}}\xspace}

\def\cquark    {{\ensuremath{\Pc}}\xspace}

\def\bquark    {{\ensuremath{\Pb}}\xspace}
\def\bquarkbar {{\ensuremath{\overline \bquark}}\xspace}

\def\pion   {{\ensuremath{\Ppi}}\xspace}
\def\piz    {{\ensuremath{\pion^0}}\xspace}
\def\pip    {{\ensuremath{\pion^+}}\xspace}
\def\pim    {{\ensuremath{\pion^-}}\xspace}
\def\pipm   {{\ensuremath{\pion^\pm}}\xspace}
\def\pimp   {{\ensuremath{\pion^\mp}}\xspace}

\def\kaon    {{\ensuremath{\PK}}\xspace}
\def\Kbar    {{\ensuremath{\offsetoverline{\PK}}}\xspace}

\def\KorKbar {\kern \thebaroffset\optbar{\kern -\thebaroffset \PK}{}\xspace}
\def\Kz      {{\ensuremath{\kaon^0}}\xspace}
\def\Kzb     {{\ensuremath{\Kbar{}^0}}\xspace}
\def\Kp      {{\ensuremath{\kaon^+}}\xspace}
\def\Km      {{\ensuremath{\kaon^-}}\xspace}
\def\Kpm     {{\ensuremath{\kaon^\pm}}\xspace}

\def\KS      {{\ensuremath{\kaon^0_{\mathrm{S}}}}\xspace}

\def\Kstarz  {{\ensuremath{\kaon^{*0}}}\xspace}

\def\Kstarpm {{\ensuremath{\kaon^{*\pm}}}\xspace}

\def\Dbar    {{\ensuremath{\offsetoverline{\PD}}}\xspace}
\def\D       {{\ensuremath{\PD}}\xspace}

\def\DorDbar {\kern \thebaroffset\optbar{\kern -\thebaroffset \PD}\xspace}
\def\Dz      {{\ensuremath{\D^0}}\xspace}
\def\Dzb     {{\ensuremath{\Dbar{}^0}}\xspace}
\def\Dp      {{\ensuremath{\D^+}}\xspace}
\def\Dm      {{\ensuremath{\D^-}}\xspace}

\def\Dmp     {{\ensuremath{\D^\mp}}\xspace}
\def\DpDm    {\ensuremath{\Dp {\kern -0.16em \Dm}}\xspace}
\def\Dstar   {{\ensuremath{\D^*}}\xspace}

\def\Dstarz  {{\ensuremath{\D^{*0}}}\xspace}
\def\Dstarzb {{\ensuremath{\Dbar{}^{*0}}}\xspace}

\def\Ds      {{\ensuremath{\D^+_\squark}}\xspace}

\def\Dsmp    {{\ensuremath{\D^{\mp}_\squark}}\xspace}

\def\B       {{\ensuremath{\PB}}\xspace}

\def\BorBbar {\kern \thebaroffset\optbar{\kern -\thebaroffset \PB}\xspace}
\def\Bz      {{\ensuremath{\B^0}}\xspace}

\def\Bd      {{\ensuremath{\B^0}}\xspace}

\def\BdorBdbar {\kern \thebaroffset\optbar{\kern -\thebaroffset \Bd}\xspace}
\def\Bu      {{\ensuremath{\B^+}}\xspace}
\def\Bub     {{\ensuremath{\B^-}}\xspace}
\def\Bp      {{\ensuremath{\Bu}}\xspace}
\def\Bm      {{\ensuremath{\Bub}}\xspace}
\def\Bpm     {{\ensuremath{\B^\pm}}\xspace}

\def\Bs      {{\ensuremath{\B^0_\squark}}\xspace}

\def\BsorBsbar {\kern \thebaroffset\optbar{\kern -\thebaroffset \Bs}\xspace}

\def\Y#1S{\ensuremath{\PUpsilon{(#1S)}}\xspace}

\def\LorLbar     {\kern \thebaroffset\optbar{\kern -\thebaroffset \PLambda}\xspace}

\def\to                 {\ensuremath{\rightarrow}\xspace}

\def\epsK  {{\ensuremath{\varepsilon_K}}\xspace}

\def\CP                {{\ensuremath{C\!P}}\xspace}

\def\Vub  {{\ensuremath{V_{\uquark\bquark}^{\phantom{\ast}}}}\xspace}
\def\Vcb  {{\ensuremath{V_{\cquark\bquark}^{\phantom{\ast}}}}\xspace}

\newcommand{\phis}{{\ensuremath{\phi_{\squark}}}\xspace}

\def\AT#1     {\ensuremath{A_{\mathrm{T}}^{#1}}\xspace}           

\def\C#1      {\ensuremath{\mathcal{C}_{#1}}\xspace}                       
\def\Cp#1     {\ensuremath{\mathcal{C}_{#1}^{'}}\xspace}                    
\def\Ceff#1   {\ensuremath{\mathcal{C}_{#1}^{\mathrm{(eff)}}}\xspace}        
\def\Cpeff#1  {\ensuremath{\mathcal{C}_{#1}^{'\mathrm{(eff)}}}\xspace}       
\def\Ope#1    {\ensuremath{\mathcal{O}_{#1}}\xspace}                       
\def\Opep#1   {\ensuremath{\mathcal{O}_{#1}^{'}}\xspace}                    

\newcommand{\ket}[1]{\ensuremath{|#1\rangle}}              

\newcommand{\aunit}[1]{\ensuremath{\text{\,#1}}}       

\newcommand{\tev}{\aunit{Te\kern -0.1em V}\xspace}
\newcommand{\gev}{\aunit{Ge\kern -0.1em V}\xspace}
\newcommand{\mev}{\aunit{Me\kern -0.1em V}\xspace}
\newcommand{\kev}{\aunit{ke\kern -0.1em V}\xspace}
\newcommand{\ev}{\aunit{e\kern -0.1em V}\xspace}
 
\newcommand{\mevc}{\ensuremath{\aunit{Me\kern -0.1em V\!/}c}\xspace}
\newcommand{\gevc}{\ensuremath{\aunit{Ge\kern -0.1em V\!/}c}\xspace}
\newcommand{\mevcc}{\ensuremath{\aunit{Me\kern -0.1em V\!/}c^2}\xspace}
\newcommand{\gevcc}{\ensuremath{\aunit{Ge\kern -0.1em V\!/}c^2}\xspace}

\newcommand{\chisq}{\ensuremath{\chi^2}\xspace}

\def\gsim{{~\raise.15em\hbox{$>$}\kern-.85em
          \lower.35em\hbox{$\sim$}~}\xspace}
\def\lsim{{~\raise.15em\hbox{$<$}\kern-.85em
          \lower.35em\hbox{$\sim$}~}\xspace}

\def\degrees{\ensuremath{^{\circ}}\xspace}

\def\tell1  {TELL1\xspace}
\def\ukl1   {UKL1\xspace}

\newcommand{\eg}{\mbox{\itshape e.g.}\xspace}
\newcommand{\ie}{\mbox{\itshape i.e.}\xspace}

\usepackage{cite} 
\usepackage{mciteplus}

\usepackage{multirow}
\usepackage{placeins}

\newcommand{\omcl}{\ensuremath{1-{\rm CL}}\xspace}

\newcommand{\rb}{\ensuremath{r_{B}}\xspace}
\newcommand{\db}{\ensuremath{\delta_{B}}\xspace}
\newcommand{\rd}{\ensuremath{r_{D}}\xspace}
\newcommand{\dd}{\ensuremath{\delta_{D}}\xspace}
\newcommand{\xD}{\ensuremath{x}\xspace}
\newcommand{\yD}{\ensuremath{y}\xspace}
\newcommand{\qopD}{\ensuremath{|q/p|}\xspace}
\newcommand{\phiD}{\ensuremath{\phi}\xspace}

\newcommand{\rdkpi}{\ensuremath{r_{D}^{K\pi}}\xspace}
\newcommand{\ddkpi}{\ensuremath{\delta_{D}^{K\pi}}\xspace}
\newcommand{\rbdk} {\ensuremath{r_{\Bpm}^{D\Kpm}}\xspace}
\newcommand{\dbdk} {\ensuremath{\delta_{\Bpm}^{D\Kpm}}\xspace}
\newcommand{\rbdpi} {\ensuremath{r_{\Bpm}^{D\pipm}}\xspace}
\newcommand{\dbdpi} {\ensuremath{\delta_{\Bpm}^{D\pipm}}\xspace}
\newcommand{\rbdstk} {\ensuremath{r_{\Bpm}^{\Dstar\Kpm}}\xspace}
\newcommand{\dbdstk} {\ensuremath{\delta_{\Bpm}^{\Dstar\Kpm}}\xspace}

\newcommand{\rbdstpi} {\ensuremath{r_{\Bpm}^{\Dstar\pipm}}\xspace}
\newcommand{\dbdstpi} {\ensuremath{\delta_{\Bpm}^{\Dstar\pipm}}\xspace}

\newcommand{\rbdkst} {\ensuremath{r_{\Bpm}^{D\Kstarpm}}\xspace}
\newcommand{\dbdkst} {\ensuremath{\delta_{\Bpm}^{D\Kstarpm}}\xspace}
\newcommand{\kbdkst} {\ensuremath{\kappa_{\Bpm}^{D\Kstarpm}}\xspace}
\newcommand{\rbdkstz} {\ensuremath{r_{\Bd}^{D\Kstarz}}\xspace}
\newcommand{\dbdkstz} {\ensuremath{\delta_{\Bd}^{D\Kstarz}}\xspace}
\newcommand{\kbdkstz} {\ensuremath{\kappa_{\Bd}^{D\Kstarz}}\xspace}
\newcommand{\rbdkpipi} {\ensuremath{r_{\Bpm}^{D\Kpm\pip\pim}}\xspace}

\newcommand{\rbdpipipi} {\ensuremath{r_{\Bpm}^{D\pipm\pip\pim}}\xspace}

\newcommand{\rbdsk} {\ensuremath{r_{\Bs}^{\Dsmp\Kpm}}\xspace}
\newcommand{\dbdsk} {\ensuremath{\delta_{\Bs}^{\Dsmp\Kpm}}\xspace}
\newcommand{\rbdskpipi} {\ensuremath{r_{\Bs}^{\Dsmp\Kpm\pip\pim}}\xspace}
\newcommand{\dbdskpipi} {\ensuremath{\delta_{\Bs}^{\Dsmp\Kpm\pip\pim}}\xspace}
\newcommand{\rbdmpi} {\ensuremath{r_{\Bd}^{\Dmp\pipm}}\xspace}
\newcommand{\dbdmpi} {\ensuremath{\delta_{\Bd}^{\Dmp\pipm}}\xspace}

\newcommand{\Fppp}    {\ensuremath{F^{+}_{\pi\pi\piz}}\xspace}
\newcommand{\Fkkp}    {\ensuremath{F^{+}_{K\pi\piz}}\xspace}
\newcommand{\rdkpp}   {\ensuremath{r_{\D}^{K\pi\piz}}\xspace}
\newcommand{\ddkpp}   {\ensuremath{\delta_{\D}^{K\pi\piz}}\xspace}
\newcommand{\kdkpp}   {\ensuremath{\kappa_{\D}^{K\pi\piz}}\xspace}
\newcommand{\Fpppp}   {\ensuremath{F^{+}_{4\pi}}\xspace}
\newcommand{\rdkppp}  {\ensuremath{r_{\D}^{K3\pi}}\xspace}
\newcommand{\ddkppp}  {\ensuremath{\delta_{\D}^{K3\pi}}\xspace}
\newcommand{\kdkppp}  {\ensuremath{\kappa_{\D}^{K3\pi}}\xspace}
\newcommand{\rdkskpi}  {\ensuremath{r_{\D}^{\KS K\pi}}\xspace}
\newcommand{\ddkskpi}  {\ensuremath{\delta_{\D}^{\KS K\pi}}\xspace}
\newcommand{\kdkskpi}  {\ensuremath{\kappa_{\D}^{\KS K\pi}}\xspace}

\newcommand{\Afav}{\ensuremath{\mathcal{A}_{\text{fav}}}\xspace}
\newcommand{\Asup}{\ensuremath{\mathcal{A}_{\text{sup}}}\xspace}

\newcommand{\hp}{\ensuremath{h^+}\xspace}
\newcommand{\hm}{\ensuremath{h^-}\xspace}
\newcommand{\hpm}{\ensuremath{h^\pm}\xspace}

\newcommand{\BuDh}{\ensuremath{\Bpm\to \D\hpm}\xspace}
\newcommand{\BuDK}{\ensuremath{\Bpm\to \D\Kpm}\xspace}
\newcommand{\BuDpi}{\ensuremath{\Bpm\to \D\pipm}\xspace}
\newcommand{\BuDsth}{\ensuremath{\Bpm\to\Dstar\hpm}\xspace}
\newcommand{\BuDKst}{\ensuremath{\Bpm\to\D\Kstarpm}\xspace}
\newcommand{\BuDhpipi}{\ensuremath{\Bpm\to\D\hpm\pip\pim}\xspace}
\newcommand{\BdDKst}{\ensuremath{\Bd\to\D\Kstarz}\xspace}
\newcommand{\BdDKpi}{\ensuremath{\Bd\to\D\Kp\pim}\xspace}
\newcommand{\BdDpi}{\ensuremath{\Bd\to\Dmp\pipm}\xspace}
\newcommand{\BsDsK}{\ensuremath{\Bs\to\Dsmp\Kpm}\xspace}
\newcommand{\BsDsKpipi}{\ensuremath{\Bs\to\Dsmp\Kpm\pip\pim}\xspace}

\newcommand{\Dhh}      {\ensuremath{D\to \hp\hm}\xspace}
\newcommand{\Dzhh}      {\ensuremath{\Dz\to \hp\hm}\xspace}

\newcommand{\DzKK}      {\ensuremath{\Dz\to \Kp\Km}\xspace}

\newcommand{\Dzpipi}    {\ensuremath{\Dz\to \pip\pim}\xspace}
\newcommand{\Dhhpiz}   {\ensuremath{D\to \hp\hm\piz}\xspace}
\newcommand{\Dhpipipi} {\ensuremath{D\to \hp\pim\pip\pim}\xspace}
\newcommand{\Dpipipipi}{\ensuremath{D\to \pip\pim\pip\pim}\xspace}

\newcommand{\DzKpi}     {\ensuremath{\Dz\to \Kp\pim}\xspace}
\newcommand{\DKpipiz}  {\ensuremath{D\to \Kp\pim\piz}\xspace}
\newcommand{\DKpipipi} {\ensuremath{D\to \Kpm\pimp\pip\pim}\xspace}
\newcommand{\DzKpipipi} {\ensuremath{\Dz\to \Kpm\pimp\pip\pim}\xspace}
\newcommand{\DKSpipi}  {\ensuremath{D\to\KS\pip\pim}\xspace}
\newcommand{\DzKSpipi}  {\ensuremath{\Dz\to\KS\pip\pim}\xspace}
\newcommand{\DKShh}    {\ensuremath{D\to\KS\hp\hm}\xspace}
\newcommand{\DzKShh}    {\ensuremath{\Dz\to\KS\hp\hm}\xspace}
\newcommand{\DKSKpi}   {\ensuremath{D\to \KS\Kpm\pimp}\xspace}
\newcommand{\DKpipi}   {\ensuremath{\Dp\to\Km\pip\pip}\xspace}
\newcommand{\Dshhh}    {\ensuremath{\Ds\to\hp\hm\pip}\xspace}

\newcommand{\rdk}{\rbdk}
\newcommand{\ddk}{\dbdk}
\newcommand{\rdpi}{\rbdpi}
\newcommand{\ddpi}{\dbdpi}
\newcommand{\rdstk}{\rbdstk}
\newcommand{\ddstk}{\dbdstk}
\newcommand{\rdstpi}{\rbdstpi}
\newcommand{\ddstpi}{\dbdstpi}
\newcommand{\rdkst}{\rbdkst}
\newcommand{\ddkst}{\dbdkst}

\newcommand{\rdkstz}{\rbdkstz}
\newcommand{\ddkstz}{\dbdkstz}

\newcommand{\rdkpipi}{\rbdkpipi}

\newcommand{\rdpipipi}{\rbdpipipi}

\newcommand{\ldsk}{\rbdsk}
\newcommand{\ddsk}{\dbdsk}
\newcommand{\ldskpipi}{\rbdskpipi}
\newcommand{\ddskpipi}{\dbdskpipi}
\newcommand{\ldmpi}{\rbdmpi}
\newcommand{\ddmpi}{\dbdmpi}
\newcommand{\rDkpi}{\rdkpi}
\newcommand{\dDkpi}{\ddkpi}
\newcommand{\rDkpipipi}{\rdkppp}
\newcommand{\dDkpipipi}{\ddkppp}
\newcommand{\kDkpipipi}{\kdkppp}
\newcommand{\rDkskpi}{\rdkskpi}
\newcommand{\dDkskpi}{\ddkskpi}
\newcommand{\kDkskpi}{\kdkskpi}

\usepackage{longtable} 
\usepackage{rotating}
\begin{document}

\renewcommand{\thefootnote}{\fnsymbol{footnote}}
\setcounter{footnote}{1}


\begin{titlepage}
\pagenumbering{roman}

\vspace*{-1.5cm}
\centerline{\large EUROPEAN ORGANIZATION FOR NUCLEAR RESEARCH (CERN)}
\vspace*{1.5cm}
\noindent
\begin{tabular*}{\linewidth}{lc@{\extracolsep{\fill}}r@{\extracolsep{0pt}}}
\ifthenelse{\boolean{pdflatex}}
{\vspace*{-1.5cm}\mbox{\!\!\!\includegraphics[width=.14\textwidth]{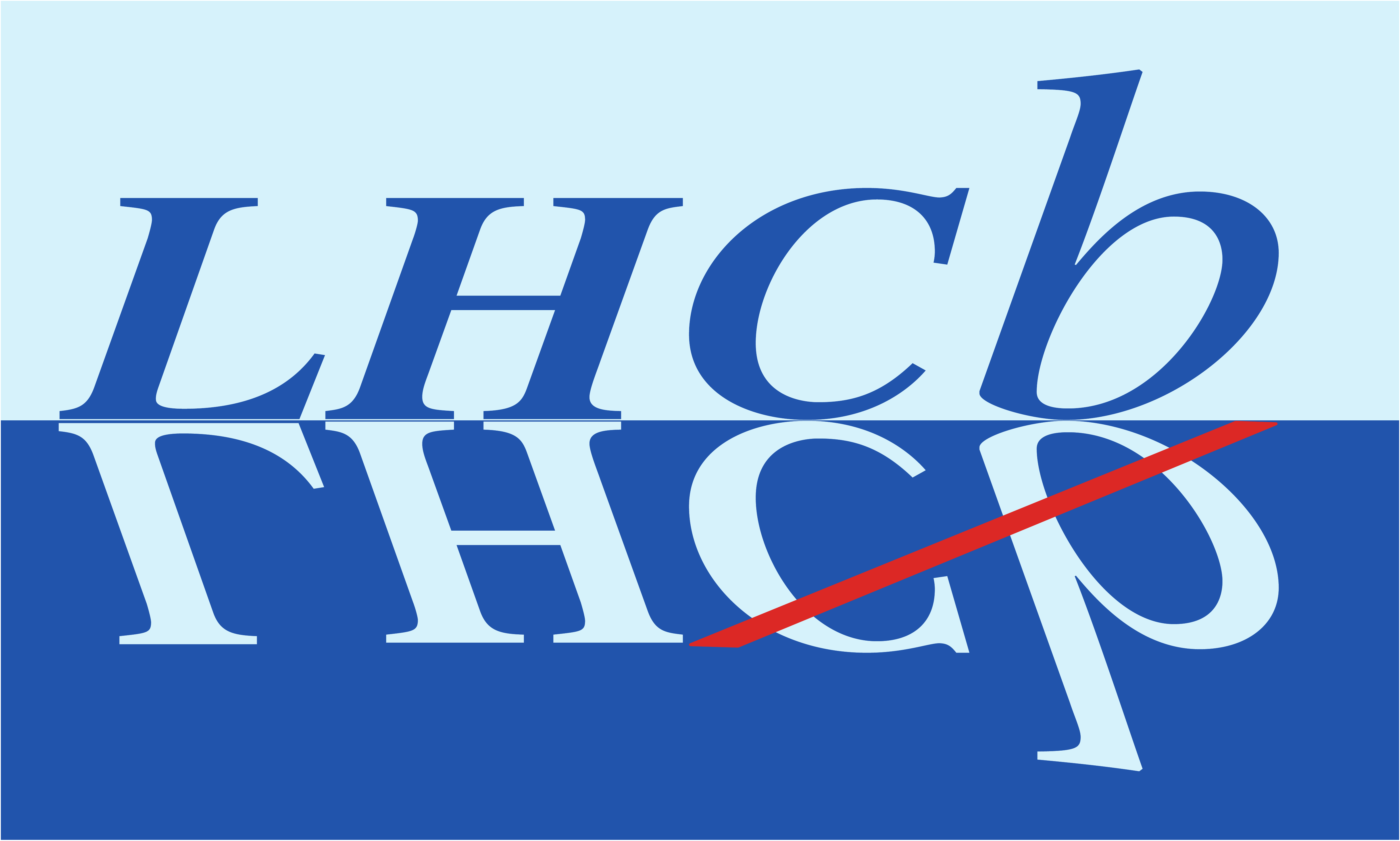}} & &}%
{\vspace*{-1.2cm}\mbox{\!\!\!\includegraphics[width=.12\textwidth]{figs/lhcb-logo.eps}} & &}%
\\
 & & CERN-EP-2021-183 \\  
 & & LHCb-PAPER-2021-033 \\  
 & & 21 Dec 2021 \\ 
 & & \\
\end{tabular*}

\vspace*{4.0cm}

{\normalfont\bfseries\boldmath\huge
\begin{center}
  \papertitle
\end{center}
}

\vspace*{2.0cm}

\begin{center}
\paperauthors\footnote{Authors are listed at the end of this paper.}
\end{center}

\vspace{\fill}

\begin{abstract}
\noindent
A combination of measurements sensitive to the \CP violation angle \g of the Cabibbo--Kobayashi--Maskawa unitarity triangle and to the charm mixing parameters that describe oscillations between \Dz and \Dzb mesons is performed.
Results from the charm and beauty sectors, based on data collected with the LHCb detector at CERN's Large Hadron Collider, are combined for the first time.
This method provides an improvement on the precision of the charm mixing parameter \yD by a factor of two with respect to the current world average.
The charm mixing parameters are determined to be \mbox{$x = (0.400^{\,+0.052}_{\,-0.053})\%$} and \mbox{$y = (0.630^{\,+0.033}_{\,-0.030})\%$}.
The angle \g is found to be \mbox{$\g = (65.4^{\,+3.8}_{\,-4.2})^\circ$} and is the most precise determination from a single experiment.
\end{abstract}

\vspace*{2.0cm}

\begin{center}
	Published in JHEP 12 (2021) 141
\end{center}

\vspace{\fill}

{\footnotesize
\centerline{\copyright~\papercopyright. \href{\paperlicenceurl}{\paperlicence}.}}
\vspace*{2mm}

\end{titlepage}


\newpage
\setcounter{page}{2}
\mbox{~}


\renewcommand{\thefootnote}{\arabic{footnote}}
\setcounter{footnote}{0}

\cleardoublepage


\pagestyle{plain} 
\setcounter{page}{1}
\pagenumbering{arabic}


\section{Introduction}


Precise measurements of the Cabibbo--Kobayashi--Maskawa (CKM) unitarity triangle provide a strict test of the Standard Model (SM) and allow for indirect new physics searches in the quark sector up to very high mass scales. 
The \CP violating phase  $\gamma\equiv\arg[-V^{}_{ud}V^*_{ub}/V^{}_{cd}V^*_{cb}]$, where $V_{qq^\prime}$ is the relevant CKM matrix element, is the only angle of the unitarity triangle that can be determined using solely measurements of tree-level \B-meson decays~\cite{Gronau:1991dp,Gronau:1990ra,Atwood:1996ci,Atwood:2000ck,Bondar,Giri:2003ty,Belle:2004bbr,Grossman:2002aq} with negligible theoretical uncertainty~\cite{Brod:2013sga}, assuming no sizeable new physics effects are present at tree level~\cite{Brod:2014bfa}.
Deviations between direct measurements of \g and the value derived from global CKM fits, which assume validity of the SM and hence unitarity of the CKM matrix, would be a clear indication of physics beyond the SM.
Furthermore, comparisons between the value of \g measured using decays of different $B$-meson species provide sensitivity to possible new physics effects at tree level given the different decay topologies involved.
The world average for direct measurements of $\g = (66.2^{\,+3.4}_{\,-3.6})^\circ$~\cite{HFLAV18} is dominated by LHCb results.
The experimental uncertainty on \g is larger than that obtained from global CKM fits, $\g = (65.6^{+0.9}_{-2.7})^\circ$~\cite{CKMfitter2015} using a frequentist framework, and $\g = (65.8 \pm 2.2)^\circ$\cite{UTfit-UT} with a Bayesian approach.
Closing this sensitivity gap is a key physics goal of the LHCb experiment and the comparison between the direct and indirect determinations of \g is an important test of the SM.

\begin{figure}[!b]
    \centering
    \includegraphics[width=0.48\textwidth]{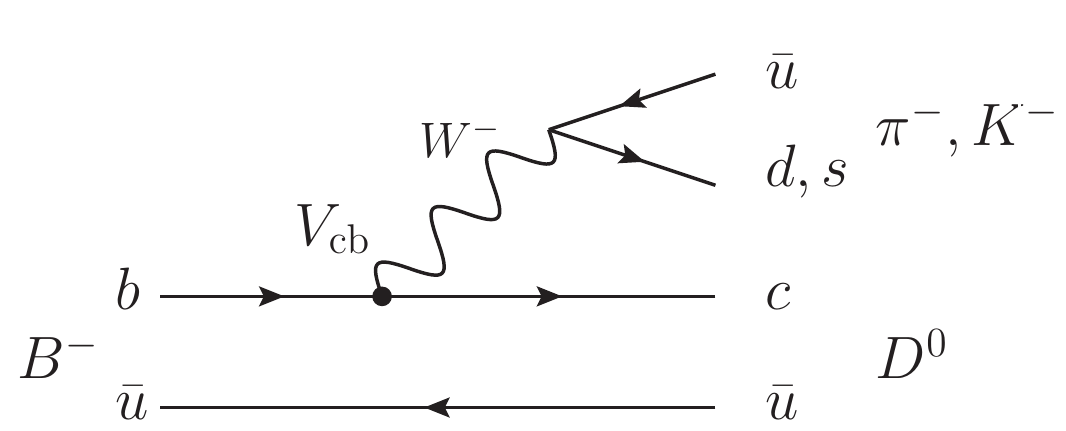}
    \includegraphics[width=0.48\textwidth]{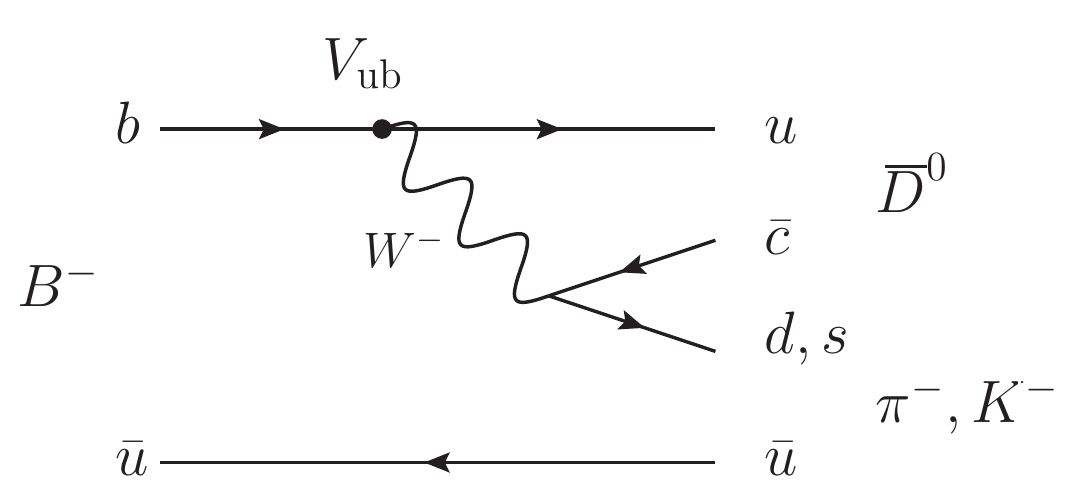}
    \caption{Leading-order Feynman diagrams for $\Bm\to Dh^-$ decays with a (left) favoured $\bquark\to \cquark$ and (right) suppressed $\bquark\to \uquark$ quark transition.}
    \label{fig:feyn}
\end{figure}

The CKM angle \g is measured in decays which are sensitive to interference between favoured $\bquark\to \cquark$ and suppressed $\bquark\to \uquark$ quark transition amplitudes that are proportional to \Vcb and \Vub, respectively.\footnote{Charge conjugation is implied throughout unless stated otherwise.}
The ratio of these two amplitudes is given by \mbox{$\Asup/\Afav = \rb e ^{i\db\pm\g}$}, where the $+$ or $-$ sign indicates whether the initial state contains a \bquarkbar- or \bquark-quark, \rb is the ratio of the amplitude magnitudes, and \db their \CP-conserving strong-phase difference.
This interference effect is typically measured in $B$-meson decays such as \BuDh, where $D$ is an admixture of the \Dz and \Dzb flavour states, and \hpm is either a charged kaon or pion.
Figure~\ref{fig:feyn} shows the leading-order Feynman diagrams for the favoured and suppressed processes.
Interference effects, providing sensitivity to \g, only occur when the $D$ meson decays to a final state, $f$, accessible to both \Dz and \Dzb mesons.
Neglecting mixing in the neutral charm system, the decay rate for a \BuDh decay is given by
\begin{equation}
	\Gamma(\BuDh) \propto | \rd e^{-i\dd} + \rb e ^{i(\db\pm\g)} | ^2 \Rightarrow \rd^2 + \rb^2 + 2\kappa_D\kappa_B \rd \rb \cos(\db+\dd\pm\g),
	\label{eq:no_mix_rate}
\end{equation}
where \rd and \dd are the magnitude ratio and strong-phase difference between the $\Dz\to f$ and $\Dzb\to f$ amplitudes.
For \D decays to \CP eigenstates, \eg $D\to\Kp\Km$, these values are $\rd=1$ and $\dd=0$.
The coherence factors of \B and \D decays, $\kappa_B$ and $\kappa_D$, are equal to unity for two-body decays, and account for a dilution of the interference term due to incoherence (strong phase variation) between contributing intermediate resonances in multibody decays.
The hadronic parameters, \rb, \db, \rd, \dd, are specific to each \B decay and subsequent \D decay, respectively.
However, the \CP-violating weak phase difference between \Bp and \Bm amplitudes, \g, is shared by all such decays.

Equation~\eqref{eq:no_mix_rate} has at least five unknown parameters, even more if the coherence factors are not set to unity, hence they cannot be determined using a single pair of \Bpm decay rates.
This is overcome by combining the results from many different \D-decay modes to overconstrain the parameters of the \B-meson decay, provided that the corresponding \rd, \dd and $\kappa_D$ parameters are constrained by other measurements.
In past combinations these parameters have been taken as external inputs using dedicated charm-meson measurements.
The large $B$-meson samples now constrain \g and \db so precisely that \ddkpi, the strong phase difference between \Dz\to\Km\pip and \Dzb\to\Km\pip decays, can be measured with similar precision as \g and \db, a factor of about two better than the previous world average~\cite{HFLAV18}.
This improved precision on \ddkpi can then be used to improve knowledge of charm mixing as described below.

The mass eigenstates of the neutral charm mesons can be written as \mbox{$\ket{D_{1,2}} \equiv p\ket{\Dz}\pm q\ket{\Dzb}$}, where $p$ and $q$ are complex parameters such that $|p|^2 + |q|^2=1$.
The $D_1$ ($D_2$) state corresponds to the $+$ ($-$) sign and is approximately \CP even (odd) in the chosen convention.
The mixing of charm flavour states can be described by two dimensionless parameters, $x\equiv (m_1 - m_2) / \Gamma$ and $y\equiv (\Gamma_1 - \Gamma_2)/2\Gamma$, where $m_i$ ($\Gamma_i$) is the mass (width)  of the appropriate $D$ mass state, and $\Gamma$ their average decay width.\footnote{Natural units, with $c=\hbar=1$, are used throughout.}
Effects of \CP violation in \Dz and \Dzb decays to a common final state, $f$, can be seen in mixing if $\qopD \neq 1$, or in the interference between mixing and decay if $\phiD \equiv \arg(q/p) \neq 0,\,\pi$.\footnote{The Wolfenstein parametrisation and the convention that $\CP\ket{\Dz} = \ket{\Dzb}$ is used.}
Study of the charm mixing parameters is of high interest in its own right, because the flavour-changing neutral currents responsible for the mixing transition do not occur at tree-level in the SM, and thus can be significantly affected by contributions from new heavy particles. The world averages for \mbox{$\xD = (4.09^{\,+0.48}_{\,-0.49})\times 10^{-3}$} and $\yD = (6.15^{\,+0.56}_{\,-0.55})\times 10^{-3}$~\cite{HFLAV18} are dominated by LHCb results.

The mixing parameters, \xD and \yD, can be determined using the ratio of wrong-sign (WS), \Dz\to\Kp\pim, and right-sign (RS), \Dz\to\Km\pip, time-dependent decay rates. This ratio is
\begin{align}
	R^\pm(t) \approx R^\pm + \sqrt{R^\pm} y'^\pm \left( \frac{t}{\tau} \right) + \frac{(x'^\pm)^2+(y'^\pm)^2}{4} \left( \frac{t}{\tau} \right) ^2 ,
	\label{eq:ws_rs_rate}
\end{align}
up to second order in the mixing parameters,
where $t$ is the decay time, $\tau$ is the \Dz meson lifetime, and the $+$ ($-$) signs correspond to the decay-rate ratio for a flavour-tagged \Dz (\Dzb) initial state.\footnote{
It should be noted that there are multiple conventions in the literature for the strong phase \ddkpi, depending on whether the discussion involves the CKM angle $\gamma$ or charm mixing.
The convention in which $\dd\to\pi$ in the SU(3) limit is used, which is shifted by $\pi$ with respect to the convention employed by the HFLAV Charm group.} The parameter \mbox{$R^\pm = r_D^2 (1\pm A_D)$} is the ratio of suppressed-to-favoured decay rates, modulated by the direct \CP asymmetry, $A_D$, between \Dz and \Dzb WS decays.
The parameters \mbox{$x'^\pm \equiv -\left|q/p\right|^{\pm 1}\left[x\cos(\ddkpi\pm \phiD) + y\sin(\ddkpi\pm\phiD)\right]$} and \mbox{$y'^\pm \equiv -\left|q/p\right|^{\pm 1}\left[y\cos(\ddkpi\pm\phiD) - x\sin(\ddkpi\pm\phiD)\right]$} encode the mixing.
Since \ddkpi is close to $\pi$ and \phiD is almost zero~\cite{HFLAV18}, it follows that $R^\pm(t)$ 
is mostly sensitive to the parameter \yD through the term linear in decay time and mixing parameters, and currently the precision on \yD is limited by the precision with which \ddkpi is known.
Consequently, a simultaneous combination using both beauty and charm observables from LHCb is performed for the first time, improving the precision on \yD (\xD) by about 50\% (2\%).

A further motivation for the simultaneous combination of both beauty and charm measurements is that non-negligible effects due to charm-meson mixing give rise to additional terms in Eq.~\eqref{eq:no_mix_rate}.
Incorporating the effect of \D-meson mixing, up to first order in $x$ and $y$, means the decay rate of Eq.~\eqref{eq:no_mix_rate} becomes~\cite{Rama:2013voa}
\begin{align}
	\Gamma(\BuDh) \propto & \phantom{+} \rd^2 + \rb^2 + 2\kappa_D\kappa_B \rd \rb \cos(\db+\dd\pm\g) \nonumber \\
									& - \alpha\left[ (1+\rb^2)\kappa_D\rd\cos(\dd) + (1+\rd^2)\kappa_B\rb\cos(\db\pm\g)\right] \yD \nonumber \\
									& + \alpha\left[ (1-\rb^2)\kappa_D\rd\,\sin(\dd) - (1-\rd^2)\kappa_B\rb\sin(\db\pm\g)\right] \xD,
	\label{eq:with_mix_rate}
\end{align}
where the $\alpha$ coefficient accounts for the non-uniform decay-time acceptance of the LHCb detector.
For cases where $\rb \gg \xD,\, \yD$, such as the \BuDK decay, the effect of \D mixing is small. However, for decays like \BuDpi, where $\rb \sim \xD,\, \yD$, the effect is significant~\cite{Rama:2013voa}. Studies of this combination, which includes both \BuDK and \BuDpi modes, suggest that not accounting for the effect of \D-meson mixing results in a bias on \g of approximately $1.8\degrees$, and an ever larger bias for the hadronic parameters, \rbdpi and \dbdpi, of the \BuDpi system.
Thus an unbiased determination of \g, \xD and \yD requires the simultaneous combination produced in this article.

This article presents results for the weak phase \g and charm mixing and \CP violation parameters \xD, \yD, \qopD and \phiD, as well as for several additional amplitude ratios and strong phases, using data collected at the LHCb experiment during the first two runs of the LHC.
The statistical procedure is identical to that described in Ref.~\cite{LHCb-PAPER-2016-032} and follows a frequentist treatment which is described in detail in Ref.~\cite{LHCb-PAPER-2013-020} and briefly recapped in Sec.~\ref{sec:stat}.
The results have additionally been cross-checked using Bayesian inference, which finds very similar values.
The results presented here supersede previous LHCb \mbox{combinations~\cite{LHCb-PAPER-2013-020,LHCb-PAPER-2016-032,LHCb-CONF-2018-002, LHCb-CONF-2020-003}.}

The full list of LHCb measurements that are used as inputs to the combination is provided in Table~\ref{tab:inputs}.
In the beauty sector this includes decay-rate ratios and charge asymmetries of \BuDh, \BuDsth, \BuDKst, \BuDhpipi, \mbox{\BdDKst}, \BdDpi, \BsDsK and \BsDsKpipi decays, where $\Dstar$ is an admixture of $\Dstarz$ and $\Dstarzb$ flavour states.
In the charm sector this includes time-dependent measurements of \Dzhh, \DzKpi, \DzKpipipi and \DzKSpipi decays.
There are seven new or updated measurements from beauty-meson decays since the last combination, including LHCb Run 2 updates from the highly sensitive \BuDh with \DKShh~\cite{LHCb-PAPER-2020-019} and \Dhh~\cite{LHCb-PAPER-2020-036} decays.
The eight inputs from LHCb charm analyses are included in the combination for the first time.

Additional external constraints are summarised in Table~\ref{tab:inputs_aux}; these are used predominantly to provide auxiliary information on the hadronic parameters and coherence factors in multibody $B$ and $D$ decays.
In the case of quasi-\CP-eigenstate decays, such as \mbox{\Dpipipipi}, the coherence factor is determined by the fraction of \CP-even content in the final-state amplitude, $F^+=(\kappa_D+1)/2$.
In the case of the \BdDpi, \BsDsK and \BsDsKpipi modes, the weak phases measured through the time-dependent \CP asymmetry are $(\g+2\beta)$ and $(\g-2\beta_{s})$, induced via interference between $B^0_{(s)}$ mixing and decay.
Therefore, in order to obtain sensitivity to \g, external constraints from the world averages of $\beta \equiv \arg[-V_{cd}^{}V_{cb}^{*}/V_{td}^{}V_{tb}^{*}]$ and $\phi_s \approx -2\beta_s \equiv -2 \arg[-V_{ts}^{} V_{tb}^{*}/V_{cs}^{} V_{cb}^{*}]$~\cite{HFLAV18} are included.

\begin{table}
  \caption{Measurements used in the combination. Inputs from the charm system appear in the lower part of the table. Those that are new, or that have changed, since the previous combination~\cite{LHCb-CONF-2018-002} are highlighted in bold. Measurements denoted by (*) include only a fraction of the Run 2 sample, corresponding to data taken in 2015 and 2016. Where multiple references are cited, measured values are taken from the most recent results, which include information from the others.
  }
  \centering
      \renewcommand{\arraystretch}{1.1}
      \resizebox{\textwidth}{!}{
      \begin{tabular}{l l l l l}
        \hline
        $\B$ decay  & $\D$ decay & Ref. & Dataset & Status since\\
        & & & & Ref.~\cite{LHCb-CONF-2018-002}  \\
        \hline
        \BuDh     & \Dhh          & \cite{LHCb-PAPER-2020-036} & Run 1\&2 & {\bf Updated}       \\
        \BuDh     & \Dhpipipi     & \cite{LHCb-PAPER-2016-003} & Run 1 & As before          \\
        \BuDh     & \Dhhpiz       & \cite{LHCb-PAPER-2015-014} & Run 1 & As before          \\
				\BuDh     & \DKShh        & \cite{LHCb-PAPER-2020-019} & Run 1\&2   & {\bf Updated}                \\
        \BuDh     & \DKSKpi       & \cite{LHCb-PAPER-2019-044} & Run 1\&2 & {\bf Updated}          \\
        \BuDsth   & \Dhh          & \cite{LHCb-PAPER-2020-036} & Run 1\&2 & {\bf Updated}       \\
        \BuDKst   & \Dhh          & \cite{LHCb-PAPER-2017-030} & Run 1\&2(*) & As before    \\
        \BuDKst   & \Dhpipipi     & \cite{LHCb-PAPER-2017-030} & Run 1\&2(*) & As before                \\
        \BuDhpipi & \Dhh          & \cite{LHCb-PAPER-2015-020} & Run 1    & As before          \\
        \BdDKst   & \Dhh         & \cite{LHCb-PAPER-2019-021} & Run 1\&2(*) & {\bf Updated}         \\
        \BdDKst   & \Dhpipipi     & \cite{LHCb-PAPER-2019-021} & Run 1\&2(*) & {\bf New}     \\
        \BdDKst   & \DKSpipi      & \cite{LHCb-PAPER-2016-007} & Run 1    & As before  \\
        \BdDpi    & \DKpipi       & \cite{LHCb-PAPER-2018-009} & Run 1   & As before   \\
        \BsDsK    & \Dshhh        & \cite{LHCb-PAPER-2017-047} & Run 1    & As before    \\
        \BsDsKpipi& \Dshhh        & \cite{LHCb-PAPER-2020-030}   & Run 1\&2 & {\bf New} \\
        \hline
        $\D$ decay  & Observable(s) & Ref. & Dataset & Status since\\
         & & & & Ref.~\cite{LHCb-CONF-2018-002}  \\
        \hline
        \Dzhh &  $\Delta A_{\CP}$ & \cite{LHCb-PAPER-2019-006,LHCb-PAPER-2014-013,LHCb-PAPER-2015-055} & Run 1\&2 & {\bf New} \\
        \Dzhh & $y_{\CP}$ & \cite{LHCb-PAPER-2018-038} & Run 1    & {\bf New} \\
        \Dzhh & $\Delta Y$ & \cite{LHCb-PAPER-2014-069,LHCb-PAPER-2016-063,LHCb-PAPER-2019-032,LHCb-PAPER-2020-045} & Run 1\&2 & {\bf New}  \\
        \DzKpi (Single Tag) & $R^{\pm}$, $(x'^\pm)^2$, $y'^\pm$ & \cite{LHCb-PAPER-2016-033} & Run 1 & {\bf New} \\
        \DzKpi (Double Tag) & $R^{\pm}$, $(x'^\pm)^2$, $y'^\pm$ & \cite{LHCb-PAPER-2017-046} & Run 1\&2(*) & {\bf New} \\
        \DzKpipipi & $(x^2+y^2)/4$ & \cite{LHCb-PAPER-2015-057} & Run 1    & {\bf New} \\
        \DzKSpipi & $x$, $y$ & \cite{LHCb-PAPER-2015-042} & Run 1     & {\bf New} \\
		\DzKSpipi  & $x_{\CP}$, $y_{\CP}$, $\Delta x$, $\Delta y$ & \cite{LHCb-PAPER-2019-001} & Run 1    & {\bf New} \\
		\DzKSpipi  & $x_{\CP}$, $y_{\CP}$, $\Delta x$, $\Delta y$ & \cite{LHCb-PAPER-2021-009} & Run 2    & {\bf New} \\
        \hline
      \end{tabular}
      }
  \label{tab:inputs}
\end{table}

\begin{table}
  \caption{Auxiliary inputs used in the combination. Those highlighted in bold have changed since the previous combination~\cite{LHCb-CONF-2018-002}.}
  \label{tab:inputs_aux}
  \renewcommand{\arraystretch}{1.3}
  \centering
  \resizebox{\textwidth}{!}{
      \begin{tabular}{l l l l l}
        \hline
        Decay      & Parameters                  & Source & Ref. & Status since \\
        & & & &Ref.~\cite{LHCb-CONF-2018-002} \\
        \hline
				 \BuDKst              & \kbdkst                              & LHCb              &  \cite{LHCb-PAPER-2017-030} & As before \\
         \BdDKst              & \kbdkstz                             & LHCb              &  \cite{LHCb-PAPER-2015-059} & As before \\
         \BdDpi               & $\beta$                              & HFLAV             &  \cite{HFLAV18} & {\bf Updated} \\
         \BsDsK\!\!$(\pi\pi)$ & \phis                                & HFLAV             &  \cite{HFLAV18} & {\bf Updated} \\
         \Dhhpiz              & \Fppp, \Fkkp                         & CLEO-c              &  \cite{Malde:2015mha}       & As before\\
         \Dpipipipi           & \Fpppp                               & CLEO-c              &  \cite{Malde:2015mha}       & As before\\
         \DKpipiz             & \rdkpp, \ddkpp, \kdkpp              & CLEO-c+LHCb+BESIII  &  \cite{Libby:2014rea,Evans:2016tlp,Ablikim:2021cqw}     & {\bf Updated} \\
         \DKpipipi            & \rdkppp, \ddkppp, \kdkppp           & CLEO-c+LHCb+BESIII  &  \cite{Libby:2014rea,Evans:2016tlp,LHCb-PAPER-2015-057,Ablikim:2021cqw}     & {\bf Updated} \\
         \DKSKpi              & \rdkskpi, \ddkskpi, \kdkskpi         & CLEO              &  \cite{Insler:2012pm}       & As before\\
         \DKSKpi              & \rdkskpi                             & LHCb              &  \cite{LHCb-PAPER-2015-026} & As before\\
        \hline
      \end{tabular}
      }
\end{table}

\section{Assumptions}

The mathematical formulae relating the input observables to the parameters of interest, via Eq.~\eqref{eq:with_mix_rate}, contain a few assumptions.
These are detailed below and their impact on the results has been checked to be negligible at the current precision.
In the future, as the precision on \g approaches one degree, many of them will need to be reassessed.

\subsubsection*{Neutral kaon mixing}
The extraction of \g from decays where the final state of the \D meson decay contains a neutral kaon is affected by \CP violation in \Kz--\Kzb mixing and decay and by regeneration~\cite{Bjorn:2019kov}.
For the \DKShh final state, a relative shift of approximately $\Delta\gamma / \gamma \approx \mathcal{O}(10^{-3})$ is expected; this has been studied in detail in Ref.~\cite{Bjorn:2019kov}.
Furthermore, the result of the relevant input analysis includes a small systematic uncertainty to account for this~\cite{LHCb-PAPER-2020-019}, so these effects are not considered further in this combination.
The size of the effect in \DKSKpi final states is larger, $\Delta\gamma / \gamma \approx \mathcal{O}(\epsK/\rb)$, where $\epsK=(2.10^{+0.27}_{-0.20})\times 10^{-3}$ quantifies \CP violation in neutral kaon mixing~\cite{CKMfitter2015}. However, the impact on \g is negligible at present because the sensitivity of the input measurement is relatively low~\cite{LHCb-PAPER-2019-044}.

\subsubsection*{{\boldmath \CP} violation in {\boldmath \D}-meson decays}
The effect of \CP violation in the direct decay of \DzKShh is not considered because it is negligible in the SM for the Cabibbo-favoured (CF) and doubly Cabibbo-suppressed (DCS) amplitudes contributing to that process.
However, the effect of \CP violation in the direct decay of \DzKpi is allowed for in the charm part of the combination, and is denoted by $A_D$.
A non-zero value of $A_D$ would cause a small shift in the charge asymmetries measured for \DzKpi final states in the beauty system, which is not accounted for in this combination.
The impact on the determination of \g is found to be smaller than $0.2\degrees$.
The difference between the size of direct \CP violation in \DzKK and \Dzpipi decays is included as an input in the charm part of the fit~\cite{LHCb-PAPER-2019-006}. In the beauty system, this value is used to account for direct \CP violation in \Dzhh decays for the most sensitive analyses, where the \D meson is produced in \BuDh or \BuDsth decays, under the hypothesis of $U$-spin symmetry, $A_{\CP}(KK) = -A_{\CP}(\pi\pi) = \Delta A_{\CP}/2$, where $A_{\CP}(f)$ is the \CP asymmetry of the \D meson decay to the final state $f$.
Any $U$-spin breaking effects are negligible given that ignoring any direct \CP violation in \Dhh decays only has a small impact, below $0.3\degrees$, on the determination of \g.
Time-dependent \CP violation in charm mixing, which would add additional terms to Eq.~\eqref{eq:with_mix_rate}, is also neglected in the beauty system, since its impact on the determination of \g is smaller than $0.1\degrees$~\cite{Rama:2013voa}.

\subsubsection*{Strong phases in {\boldmath\DKShh} decays}
The input measurements containing \DKShh final states~\cite{LHCb-PAPER-2020-019,LHCb-PAPER-2016-007,LHCb-PAPER-2019-001,LHCb-PAPER-2021-009}, both in the charm and beauty systems, require external knowledge of the strong-phase difference between the $\Dz\to\KS\hp\hm$ and $\Dzb\to\KS\hp\hm$ amplitudes across the phase space of the \D decay.
These values are taken from a combination of CLEO-c and BES-III measurements~\cite{CLEO:2010iul,BESIII:2020khq,BESIII:2020hlg,BESIII:2020hpo} and their uncertainties propagated to the uncertainties of the input measurements listed in Table~\ref{tab:inputs}.
In this combination, each set of input measurements is treated as statistically independent and thus a small part of the already sub-dominant systematic uncertainties of these measurements is being counted twice in the \DKSpipi system \ie the appropriate correlation is not accounted for.
This correlation is non-trivial to compute owing to the different binning schemes employed by the different input analyses.
In any case, the effect on this combination is small since the uncertainty on the strong phases accounts for approximately $0.5\degrees$ of the uncertainty on \g, 40\% of the uncertainty on \xD and 1\% of the uncertainty on \yD, and these parameters are nearly uncorrelated. Studies suggest that incorporation of this correlation will become important only with a three times larger data sample.

\subsubsection*{Correlations of systematic uncertainties between input measurements}
In addition to the effect of strong phases in \DKShh decays, there are various other potential systematic correlations that are not accounted for.
Whilst the individual input analyses provide both statistical and systematic covariance matrices between the sets of observables they measure, there are in principle sub-leading systematic correlations between input analyses which are not accounted for.
For example, systematic uncertainties originating from production and detection asymmetries will be correlated for most time-integrated measurements and those originating from knowledge of decay-time acceptance and resolution will be correlated for time-dependent measurements.
The impact of ignoring these small correlations is a marginal underestimation of the uncertainties (assuming the correlation is positive), but given that the combination is still statistically dominated ($3.3\degrees$ out of $3.6\degrees$) the effect is expected to be negligible.

\section{Statistical treatment}
\label{sec:stat}
The results are obtained using a frequentist treatment, with a likelihood function built from the product of the probability density functions, $f_i$, of experimental observables $\vec{A}_i$, defined as
\begin{equation}
	\label{eq:comblh}
	\mathcal{L}(\vec{\alpha}) = \prod_i f_i(\vec{A}_i^{\rm obs} | \vec{\alpha}).
\end{equation}
Here, $\vec{A}_i^{\rm obs}$ denotes the measured observables from analysis $i$, and $\vec{\alpha}$ is the set of underlying physics parameters on which they depend.
The observables of each input are assumed to follow a multi-dimensional Gaussian distribution
\begin{equation}
  f_i(\vec{A}_{i}^{\rm obs} | \vec{\alpha}) \propto \exp\left( -\frac{1}{2} (\vec{A}_i(\vec{\alpha})-\vec{A}_{i}^{\rm obs})^T \,
	V_i^{-1} \, (\vec{A}_i(\vec{\alpha})-\vec{A}_{i}^{\rm obs}) \right)\,,
  \label{eq:gpdf}
\end{equation}
where $V_i$ is the experimental covariance matrix, including both statistical and systematic uncertainties and their correlations.

A $\chi^2$-function is defined as $\chi^2(\vec{\alpha}) = -2 \ln \mathcal{L}(\vec{\alpha})$, with the best-fit point given by the global minimum of the $\chi^2$ function, $\chi^2(\vec{\alpha}_{\min})$.
The confidence level (CL) for a parameter at a given value, denoted $\alpha_0$, is determined in the following way.
		First, for every fixed $\alpha_0$, a new minimum of $\alpha^{\prime}$ is found, $\chi^2(\vec{\alpha}^{\,\prime}_{\min})$, and the deviation from the global minimum, \mbox{$\Delta\chi^2 = \chi^2(\vec{\alpha}^{\,\prime}_{\min}) - \chi^2(\vec{\alpha}_{\min})$}, is computed.
		Second, an ensemble of pseudoexperiments, $\vec{A}^{\,\rm MC}_{j}$, is generated according to the probability distribution of Eq.~\eqref{eq:gpdf}, with parameters $\vec{\alpha} = \vec{\alpha}^{\,\prime}_{\min}$.
		Finally, for each pseudoexperiment the \chisq-function is minimised once with the parameter of interest free to vary and once with it at a fixed value $\alpha_0$, to obtain the difference, $(\Delta\chi^{2})^{\rm MC}$, from $\vec{A}^{\rm \,MC}_{j}$, in the same way as $\Delta\chi^2$ was computed from $\vec{A}^{\rm obs}_{i}$.
The $p$-value, or \omcl, is then defined as the fraction of pseudoexperiments with $(\Delta\chi^2)^{\rm MC} > \Delta\chi^2$.
This method is often referred to as the $\hat\mu$ or {\em Plugin}
method; see Ref.~\cite{woodroofe} for details. Its coverage is not guaranteed~\cite{woodroofe}
for the full parameter space, but can be evaluated at various points across the phase space.
The coverage of the intervals quoted in this combination has been computed at several points across the phase space, including at the global minimum, by generating large samples of pseudoexperiments and computing the fraction which contains the generated value within a given confidence level.
The coverage of the quoted 68.3\% interval for \g is $(67.3\pm1.5)\%$, for \xD is $(68.2\pm 1.5)\%$, for \yD is $(67.6\pm 1.5)\%$, for \qopD is $(66.6\pm 1.5)\%$, and for \phiD is $(67.7\pm 1.5)\%$. Similar coverage is seen for the $95.4\%$ intervals and no correction to the quoted intervals is applied.

\section{Results}

\begin{figure}[!tb]
    \centering
    \includegraphics[scale=0.45]{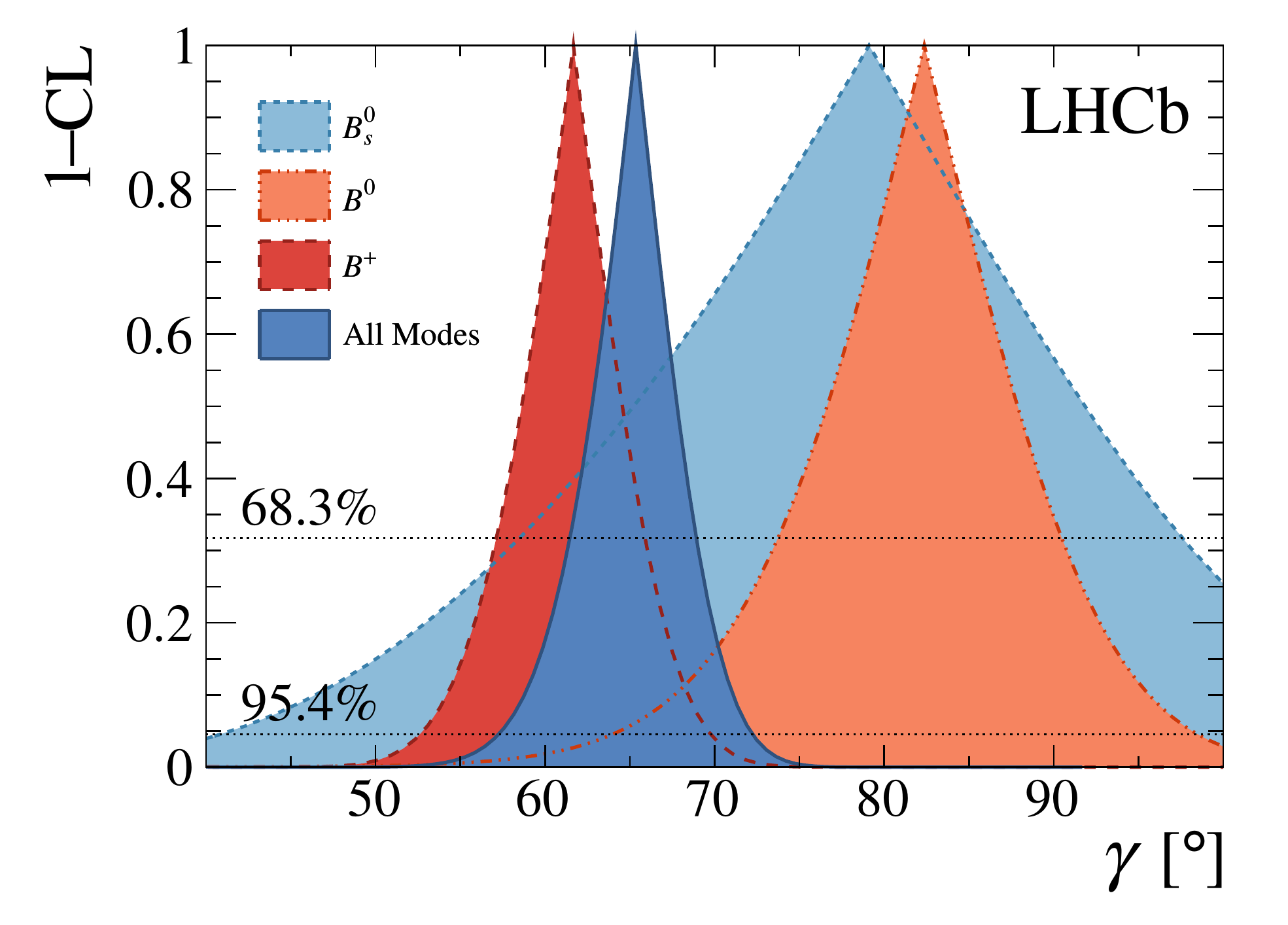}
    \vspace{-0.3cm}
    \caption{One dimensional \omcl profiles for \g from the combination using inputs from \Bs (light blue), \Bz (orange), \Bp  mesons (red) and all species together (dark blue).}
    \label{fig:scan}
\end{figure}

\begin{figure}[!tb]
    \centering
    \includegraphics[width=0.48\textwidth]{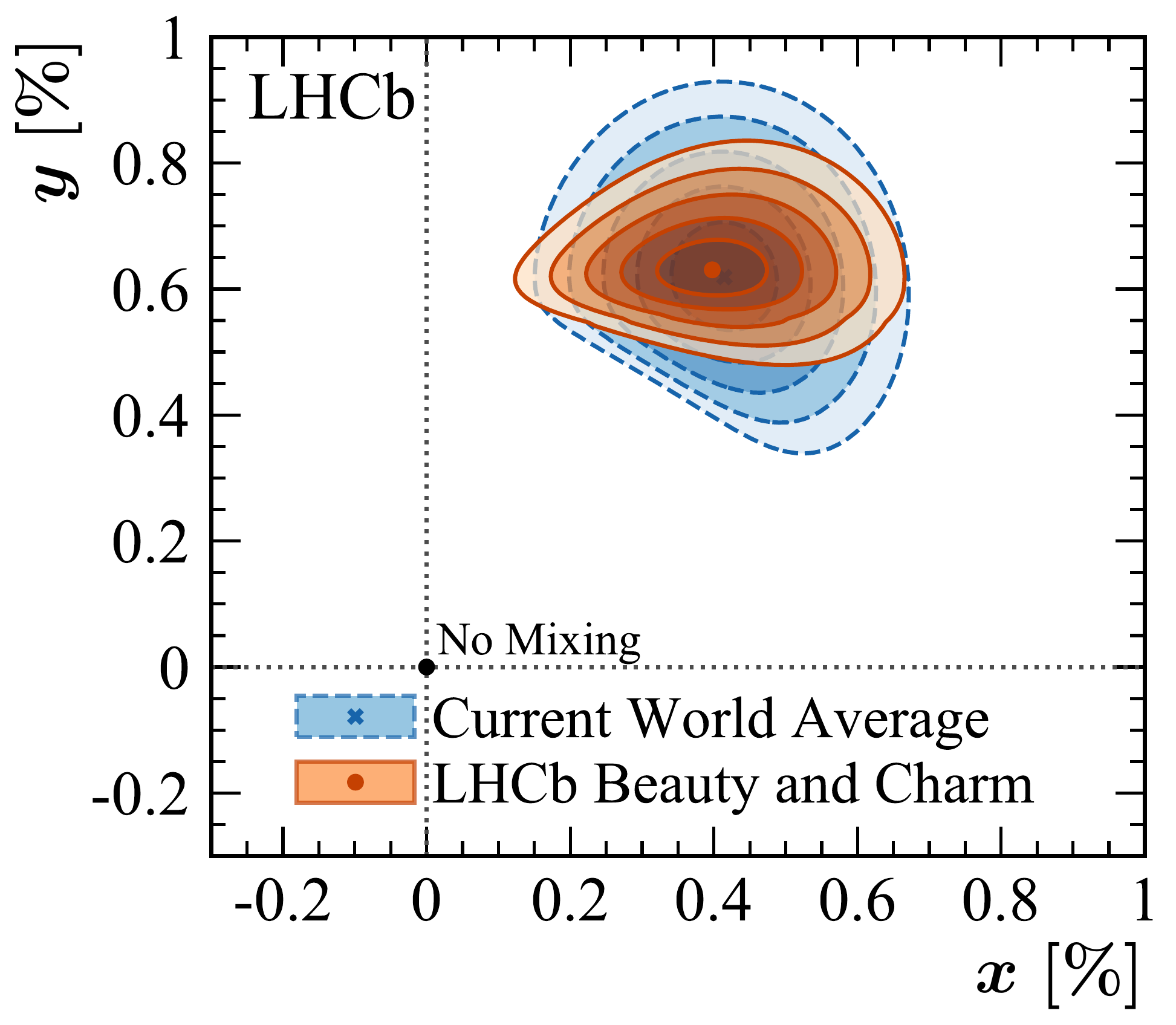}
    \includegraphics[width=0.48\textwidth]{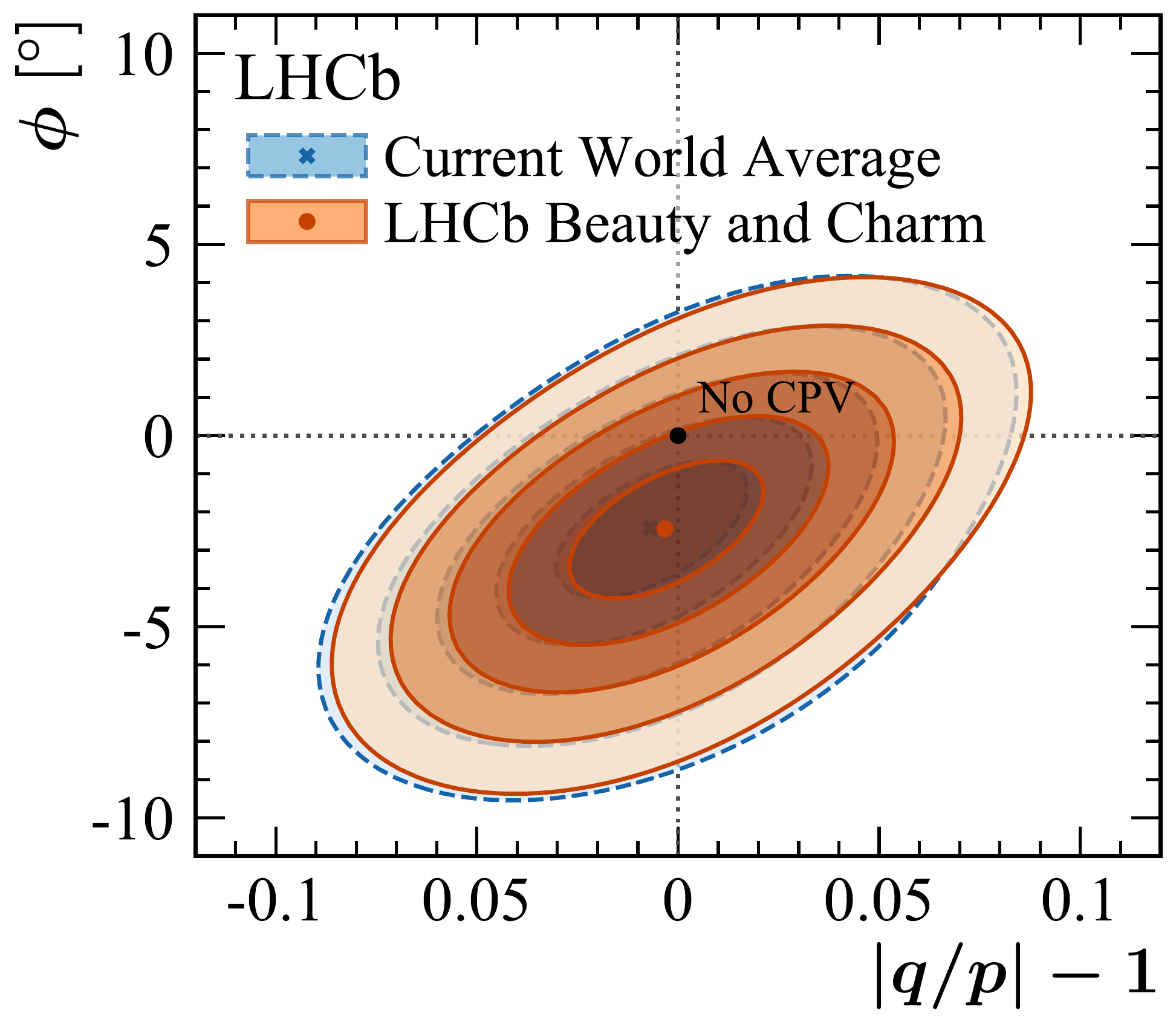}
		\caption{Two-dimensional profile likelihood contours for (left) the charm mixing parameters \xD and \yD, and (right) the \phiD and \qopD parameters. The blue contours show the current charm world average from Ref.~\cite{HFLAV18}; the brown contours show the result of this combination. Contours are drawn out from 1 ($68.3\%$) to 5 standard deviations.}
    \label{fig:2Dscan}
\end{figure}

The combination uses a total of 151 input observables to determine 52 free parameters, and the goodness of fit is found to be $84\%$, evaluated using the best-fit \chisq and cross-checked with pseudoexperiments.
The resulting confidence intervals for each parameter of interest, except for externally constrained nuisance parameters, are provided in Table~\ref{tab:results}. The correlation matrix of the parameters in Table~\ref{tab:results} is given in Appendix~\ref{sec:app:corr}, Tables~\ref{tab:corrs1},~\ref{tab:corrs2} and~~\ref{tab:corrs3}.
The $p$-value (or \omcl) distribution as a function of \g is shown in Fig.~\ref{fig:scan} for the total combination and for subsets in which the input observables are split by the species of the initial $B$ meson.
The corresponding confidence intervals are provided in Table~\ref{tab:bmesons}.
Significant differences between initial state $B$ mesons could be an indication of new physics entering at tree-level, as the decay topologies for charged and neutral initial states are different.
Figure~\ref{fig:scan} shows a moderate tension, 2.2 standard deviations ($\sigma$), between the charged and neutral $B$ states. The uncertainties in the \Bd and \Bs modes are considerably larger than in the dominant \Bp modes.
The sensitivity of the \Bd and \Bs modes is expected to improve by approximately a factor of $2$ with the analysis of \BdDKpi with \DKShh and \BsDsK decays using the full LHCb data sample.
Table~\ref{tab:time} presents the confidence intervals for \g as determined from inputs of time-dependent methods and time-integrated methods only.
Two-dimensional profile likelihood contours in the $(\xD,\yD)$ and $(\qopD,\phiD)$ planes are shown in Fig.~\ref{fig:2Dscan}.
The significant improvement, of a factor of two, in the precision to \yD demonstrates the advantage of this combination over the current world average in the charm system.

\begin{table}
  \begin{center}
		\caption{Confidence intervals and central values for each of the parameters of interest. Entries marked with an asterisk show where the scan has hit a physical boundary at the lower limit.}
		\label{tab:results}
		\setlength{\tabcolsep}{5pt}
		\renewcommand{\arraystretch}{1.2}
		\resizebox{\textwidth}{!}{
		\begin{tabular}{lccccc}
		\hline
		\multirow{2}{*}{Quantity}& \multirow{2}{*}{Value}     & \multicolumn{2}{c}{68.3\% CL}                    & \multicolumn{2}{c}{95.4\% CL}                    \\
													  &					  & Uncertainty 							& Interval             & Uncertainty 							& Interval              \\
		\hline
		 $\g \,[^{\circ}]$        & $65.4$    & $^{+3.8} _{-4.2}$         & $[61.2,69.2]$        & $^{+7.5} _{-8.7}$         & $[56.7,72.9]$        \\
		 $\rdk$                 & $0.0984$  & $^{+0.0027} _{-0.0026}$   & $[0.0958,0.1011]$    & $^{+0.0056} _{-0.0052}$   & $[0.0932,0.1040]$    \\
		 $\ddk \,[^{\circ}]$      & $127.6$   & $^{+4.0} _{-4.2}$         & $[123.4,131.6]$      & $^{+7.8} _{-9.2}$         & $[118.4,135.4]$      \\
		 $\rdpi$                & $0.00480$ & $^{+0.00070} _{-0.00056}$ & $[0.00424,0.00550]$  & $^{+0.0017} _{-0.0011}$   & $[0.0037,0.0065]$    \\
		 $\ddpi \,[^{\circ}]$     & $288$     & $^{+14} _{-15}$           & $[273,302]$          & $^{+26} _{-31}$           & $[257,314]$          \\
		 $\rdstk$               & $0.099$   & $^{+0.016} _{-0.019}$     & $[0.080,0.115]$      & $^{+0.030} _{-0.038}$     & $[0.061,0.129]$      \\
		 $\ddstk \,[^{\circ}]$    & $310$     & $^{+12} _{-23}$           & $[287,322]$          & $^{+20} _{-71}$           & $[239,330]$          \\
		 $\rdstpi$              & $0.0095$  & $^{+0.0085} _{-0.0061}$   & $[0.0034,0.0180]$    & $^{+0.017} _{-0.0089}$    & $[0.0006,0.026]$     \\
		 $\ddstpi \,[^{\circ}]$   & $139$     & $^{+22} _{-86}$           & $[53,161]$           & $^{+32} _{-129}$      & $[10,171]$      \\
		 $\rdkst$               & $0.106$   & $^{+0.017} _{-0.019}$     & $[0.087,0.123]$      & $^{+0.031} _{-0.040}$     & $[0.066,0.137]$      \\
		 $\ddkst \,[^{\circ}]$    & $35$      & $^{+20} _{-15}$           & $[20,55]$            & $^{+57} _{-28}$           & $[7,92]$             \\
		 $\rdkstz$              & $0.250$   & $^{+0.023} _{-0.024}$     & $[0.226,0.273]$      & $^{+0.044} _{-0.052}$     & $[0.198,0.294]$      \\
		 $\ddkstz \,[^{\circ}]$   & $197$     & $^{+10} _{-9.3}$          & $[187.7,207]$        & $^{+24} _{-18}$           & $[179,221]$          \\
		 $\ldsk$                & $0.310$   & $^{+0.098} _{-0.092}$     & $[0.218,0.408]$      & $^{+0.20} _{-0.21}$       & $[0.10,0.51]$        \\
		 $\ddsk \,[^{\circ}]$     & $356$     & $^{+19} _{-18}$           & $[338,375]$          & $^{+39} _{-39}$           & $[317,395]$          \\
		 $\ldskpipi$            & $0.460$   & $^{+0.081} _{-0.084}$     & $[0.376,0.541]$      & $^{+0.16} _{-0.17}$       & $[0.29,0.62]$        \\
		 $\ddskpipi \,[^{\circ}]$ & $345$     & $^{+13} _{-12}$           & $[333,358]$          & $^{+26} _{-25}$           & $[320,371]$          \\
		 $\ldmpi$               & $0.030$   & $^{+0.014} _{-0.012}$     & $[0.018,0.044]$      & $^{+0.036} _{-0.028}$     & $[0.002,0.066]$      \\
		 $\ddmpi \,[^{\circ}]$    & $30$      & $^{+26} _{-37}$           & $[-7,56]$            & $^{+45} _{-81}$           & $[-51,75]$           \\
		 $\rdkpipi$             & $0.079$   & $^{+0.028} _{-0.034}$     & $[0.045,0.107]$      & $^{+0.050} _{-0.079}$     & $[0.000,0.129]^*$      \\
		 $\rdpipipi$            & $0.067$   & $^{+0.025} _{-0.029}$     & $[0.038,0.092]$      & $^{+0.040} _{-0.067}$     & $[0.000,0.107]^*$      \\
		 $\xD \,[\%]$             & $0.400$   & $^{+0.052} _{-0.053}$     & $[0.347,0.452]$      & $^{+0.10} _{-0.11}$       & $[0.29,0.50]$        \\
		 $\yD \,[\%]$             & $0.630$   & $^{+0.033} _{-0.030}$     & $[0.600,0.663]$      & $^{+0.069} _{-0.058}$     & $[0.572,0.699]$      \\
		 $\rDkpi$               & $0.05867$ & $^{+0.00015} _{-0.00015}$ & $[0.05852,0.05882]$  & $^{+0.00031} _{-0.00030}$ & $[0.05837,0.05898]$  \\
		 $\dDkpi \,[^{\circ}]$    & $190.0$   & $^{+4.2} _{-4.1}$         & $[185.9,194.2]$      & $^{+8.6} _{-8.3}$         & $[181.7,198.6]$      \\
		 $\rDkskpi$            & $0.6150$  & $^{+0.0056} _{-0.0054}$   & $[0.6096,0.6206]$    & $^{+0.011} _{-0.011}$   & $[0.604,0.626]$      \\
         $\dDkskpi [^{\circ}]$ & $17$      & $^{+13} _{-14}$           & $[3,30]$             & $^{+26} _{-30}$         & $[-13,43]$           \\
         $\kDkskpi$            & $0.82$    & $\pm0.10$       & $[0.72,0.92]$        & $\pm0.20$     & $[0.62,1.02]$        \\
         $\rDkpipipi$             & $0.05560$ & $^{+0.00059} _{-0.00058}$ & $[0.05502,0.05619]$  & $^{+0.0012} _{-0.0012}$ & $[0.0544,0.0568]$    \\
         $\dDkpipipi [^{\circ}]$  & $154$     & $^{+15} _{-16}$           & $[138,169]$          & $^{+29} _{-33}$         & $[121,183]$          \\
         $\kDkpipipi$             & $0.480$   & $\pm0.064$     & $[0.416,0.544]$      & $\pm0.13$     & $[0.35,0.60]$        \\
		 $\qopD$                & $0.997$   & $\pm0.016$     & $[0.981,1.013]$      & $\pm0.033$     & $[0.964,1.030]$      \\
		 $\phiD \,[^{\circ}]$                & $-2.4$  & $\pm1.2$     & $[-3.6,-1.2]$    & $\pm2.5$     & $[-4.9,0.1]$     \\
		 $\Delta a_{\CP}^{\text{dir}}$ & $-0.00152$ & $\pm 0.00029$ & $[-0.00181, -0.00123]$ & $\pm 0.00058$ & $[-0.00210, -0.00094]$ \\
		\hline
		\end{tabular}
		}
	\end{center}
\end{table}
\begin{table}[!tb]
  \centering
  \caption{Confidence intervals and best-fit values for \g when splitting the combination inputs by initial $B$ meson species.}
  \label{tab:bmesons}
		\renewcommand{\arraystretch}{1.2}
		\begin{tabular}{lccccc}
		\hline
		 \multirow{2}{*}{Species}& \multirow{2}{*}{Value $[\degrees]$}     & \multicolumn{2}{c}{68.3\% CL}                    & \multicolumn{2}{c}{95.4\% CL}                    \\
													  &					  & Uncertainty 							& Interval             & Uncertainty 							& Interval              \\
		\hline
		 \Bp        & $61.7$  & $^{+4.4} _{-4.8}$ & $[56.9,66.1]$        & $^{+8.6} _{-9.5}$ & $[52.2,70.3]$        \\
		 \Bd        & $82.0$  & $^{+8.1} _{-8.8}$ & $[73.2,90.1]$        & $^{+17} _{-18}$   & $[64,99]$            \\
		 \Bs        & $79$    & $^{+21} _{-24}$   & $[55,100]$           & $^{+51} _{-47}$   & $[32,130]$           \\
		\hline
		\end{tabular}
\end{table}

\begin{table}[!tb]
  \centering
  \caption{Confidence intervals and best-fit values for \g when splitting the combination inputs by time-dependent and time-integrated methods.}
  \label{tab:time}
		\renewcommand{\arraystretch}{1.2}
		\begin{tabular}{lccccc}
		\hline
		 \multirow{2}{*}{Method}& \multirow{2}{*}{Value $[\degrees]$}     & \multicolumn{2}{c}{68.3\% CL}                    & \multicolumn{2}{c}{95.4\% CL}                    \\
		 &					  & Uncertainty 							& Interval             & Uncertainty 							& Interval              \\
		\hline
		 Time-dependent        & $79$  & $^{+21} _{-24}$ & $[55,100]$        & $^{+51} _{-47}$ & $[32,130]$        \\
		 Time-integrated        & $64.9$  & $^{+3.9} _{-4.5}$ & $[60.4,68.8]$        & $^{+7.8} _{-9.6}$   & $[55.3,72.7]$            \\
		\hline
		\end{tabular}
\end{table}

\begin{figure}[!tb]
  \centering
  \includegraphics[width=0.48\textwidth]{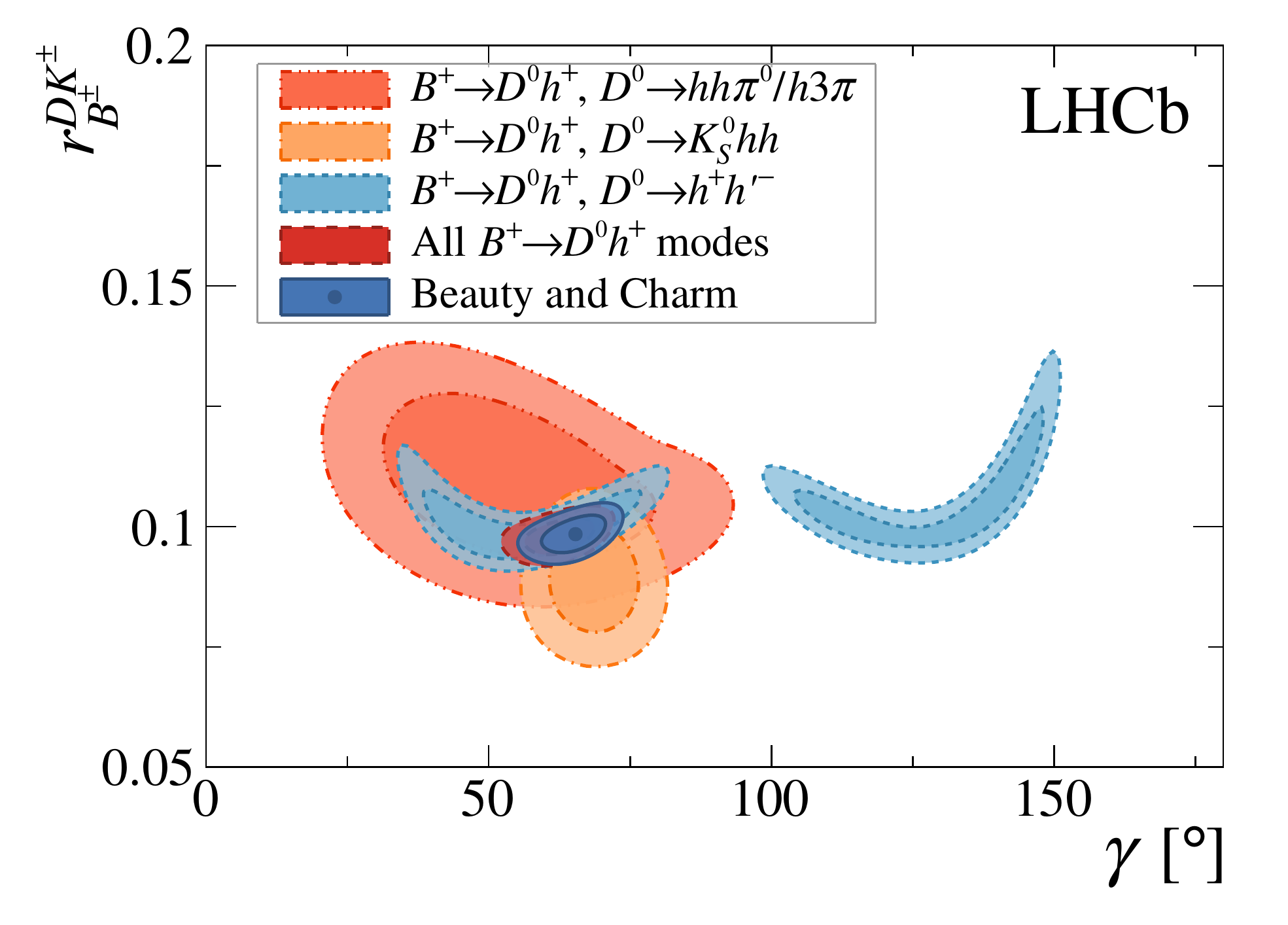}
  \includegraphics[width=0.48\textwidth]{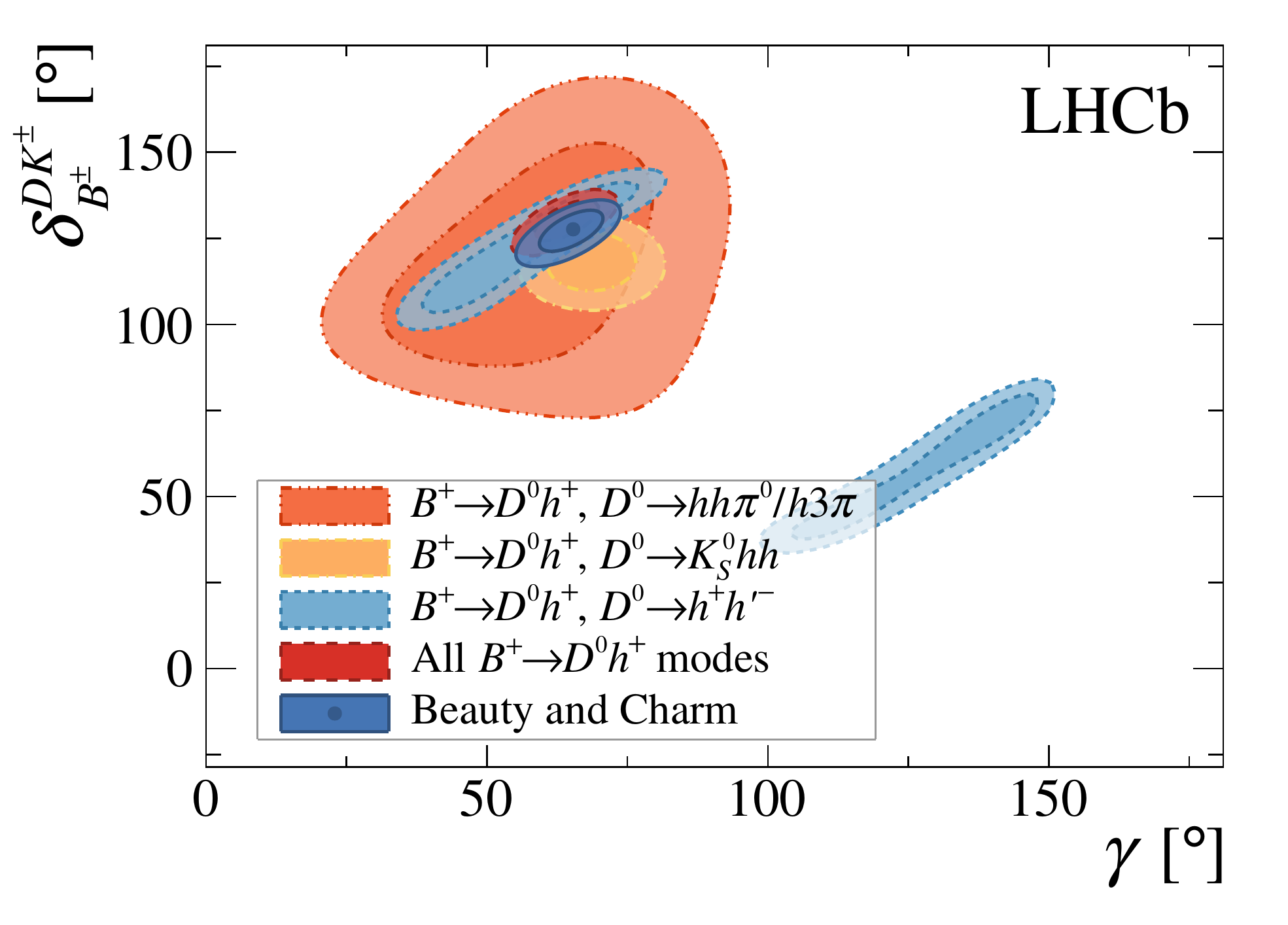}
  \includegraphics[width=0.48\textwidth]{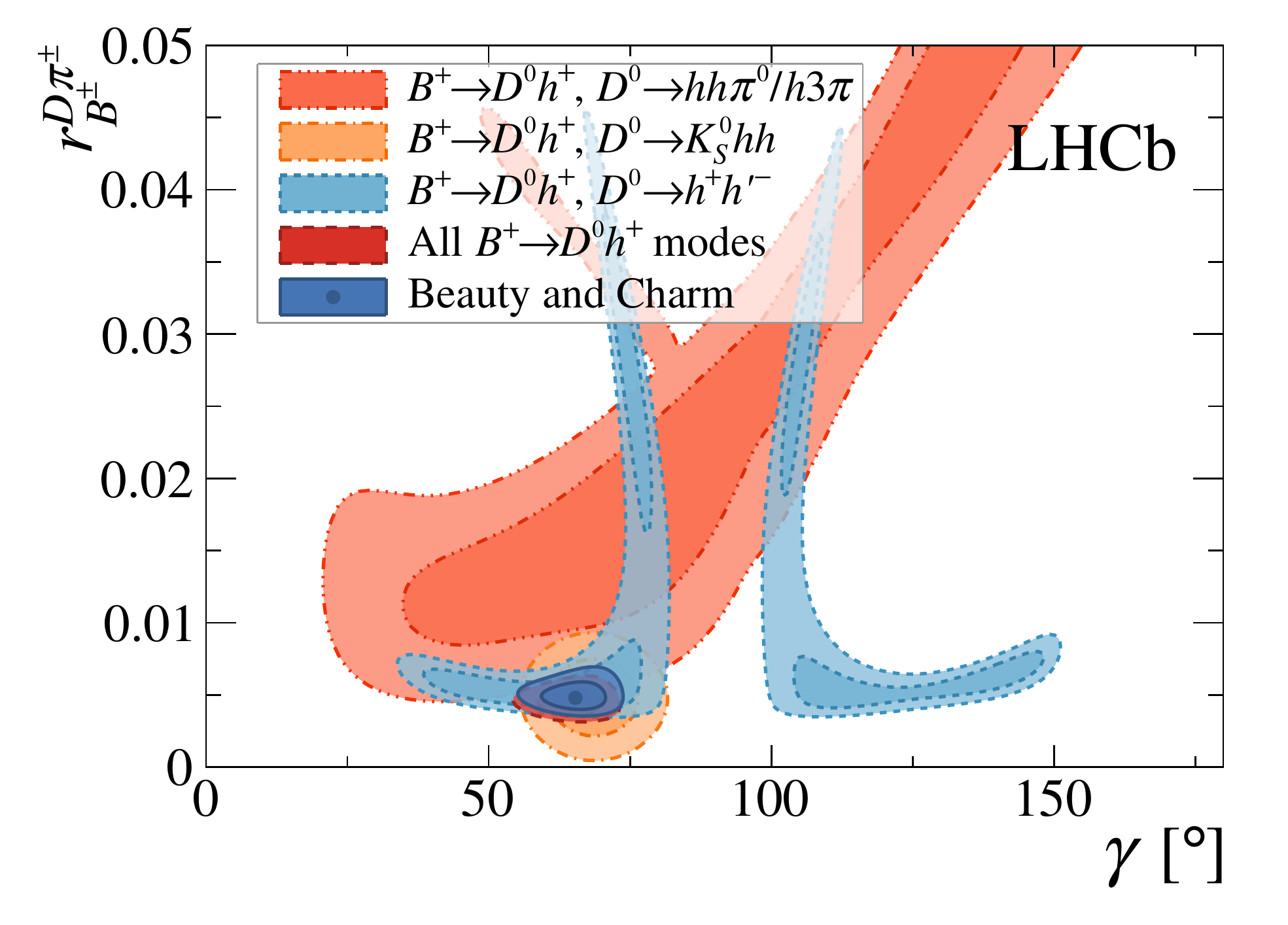}
  \includegraphics[width=0.48\textwidth]{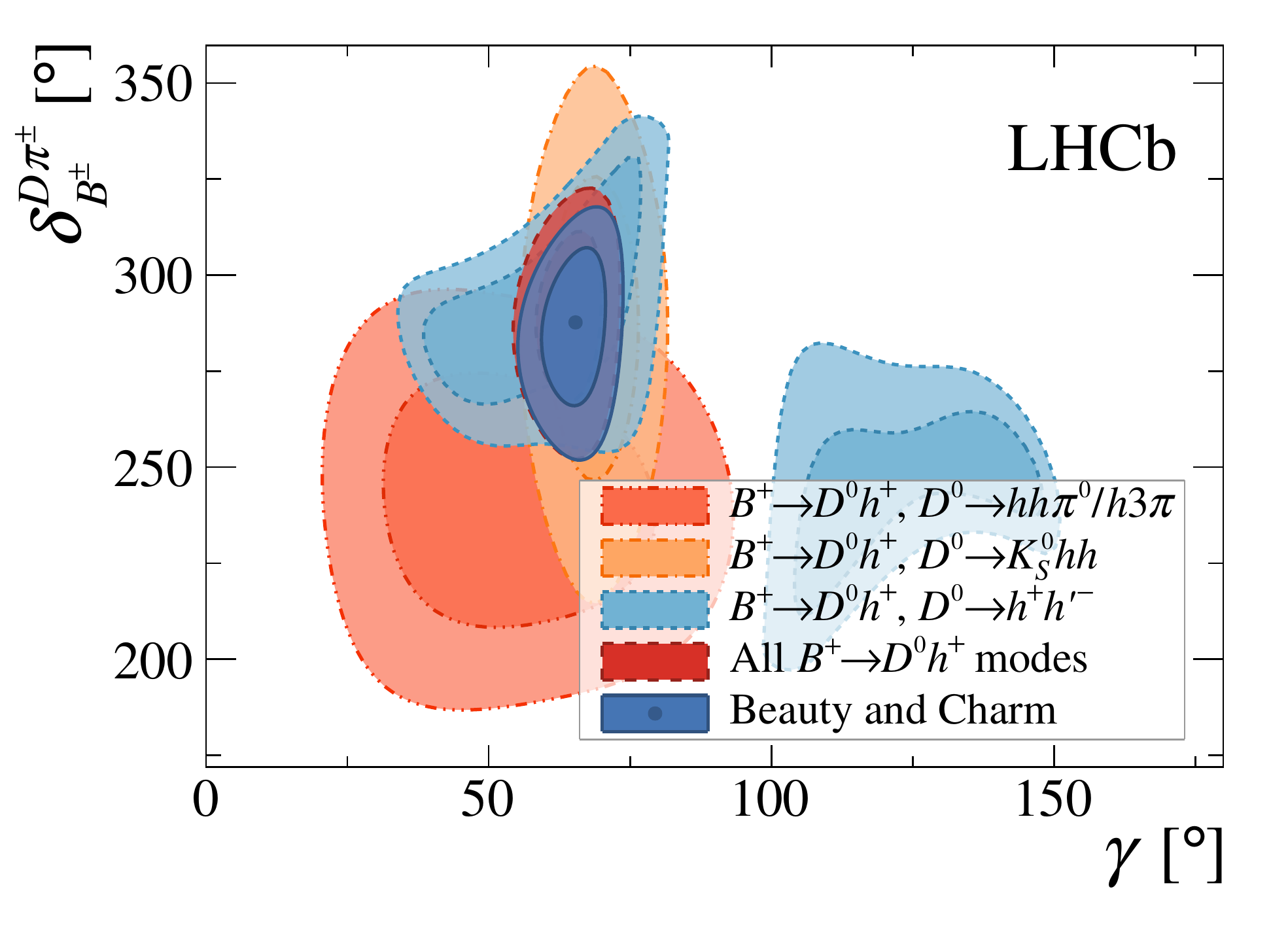}
  \includegraphics[width=0.48\textwidth]{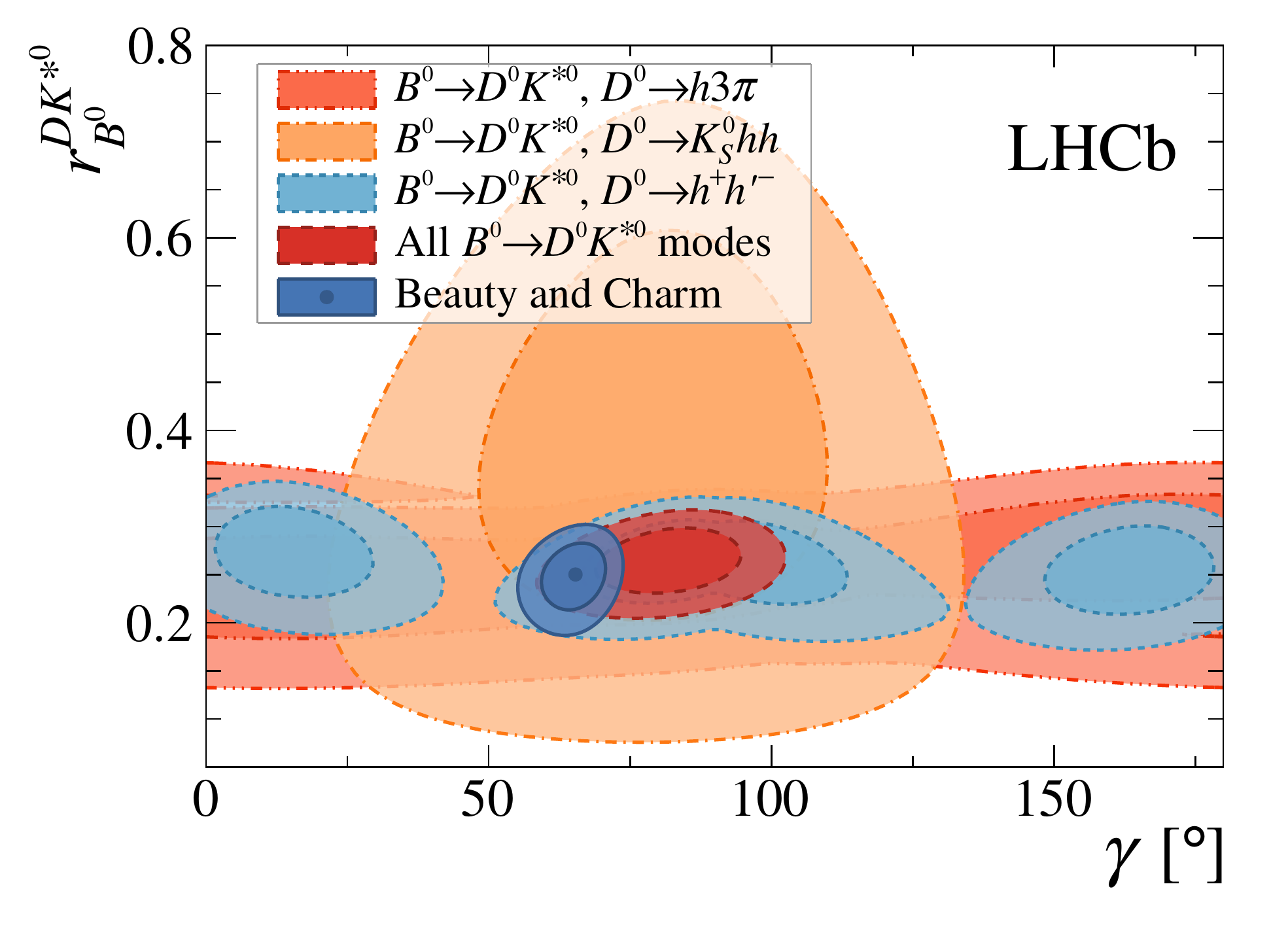}
  \includegraphics[width=0.48\textwidth]{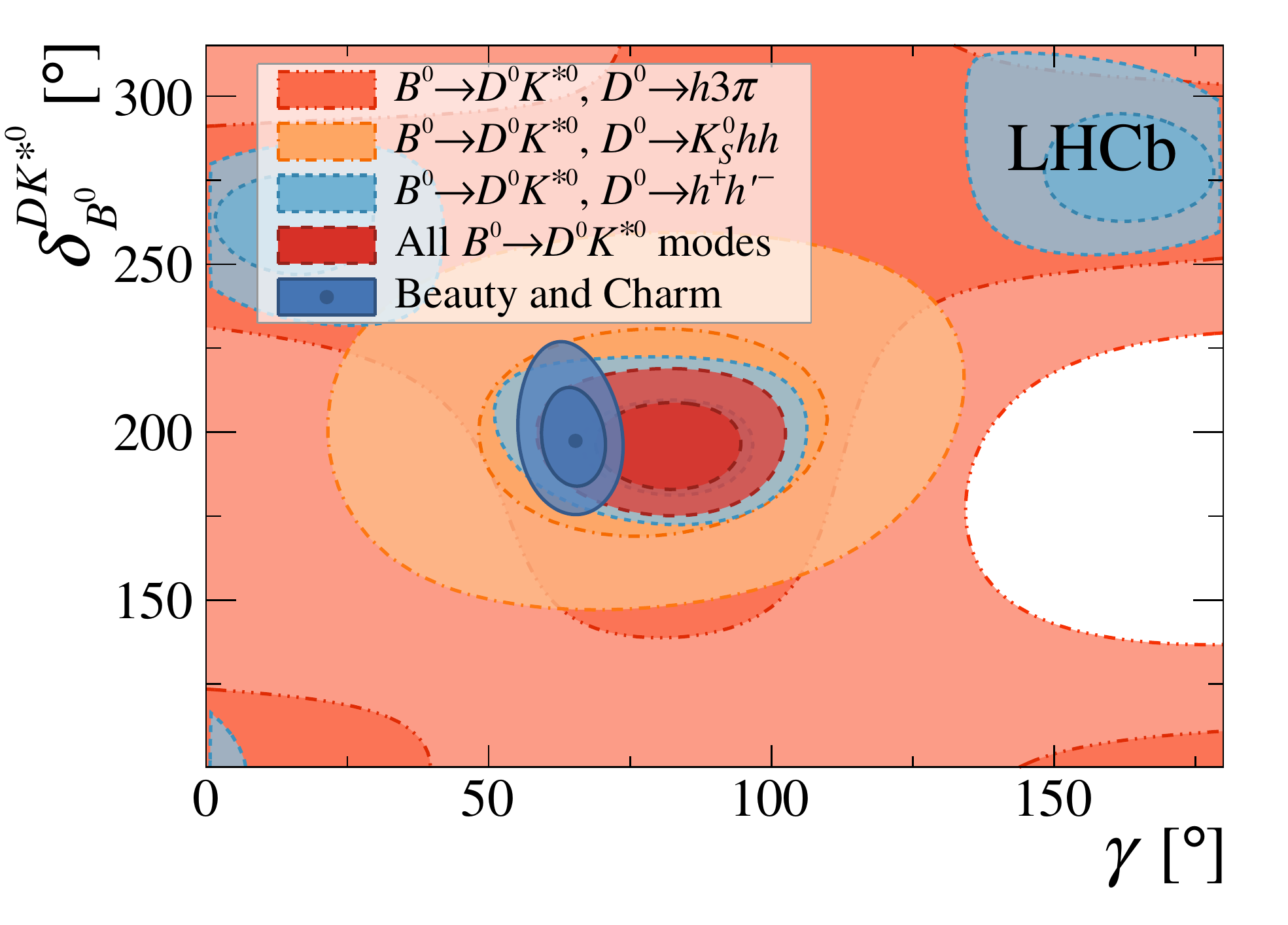}
  \caption{Profile likelihood contours for the beauty decay parameters versus \g, showing the breakdown of sensitivity amongst different sub-combinations of modes.
  The contours indicate the 68.3\% and 95.4\% confidence region.}
  \label{fig:res:breakdown_2d_dh}
\end{figure}

\begin{figure}[!tb]
  \centering
  \includegraphics[width=0.48\textwidth]{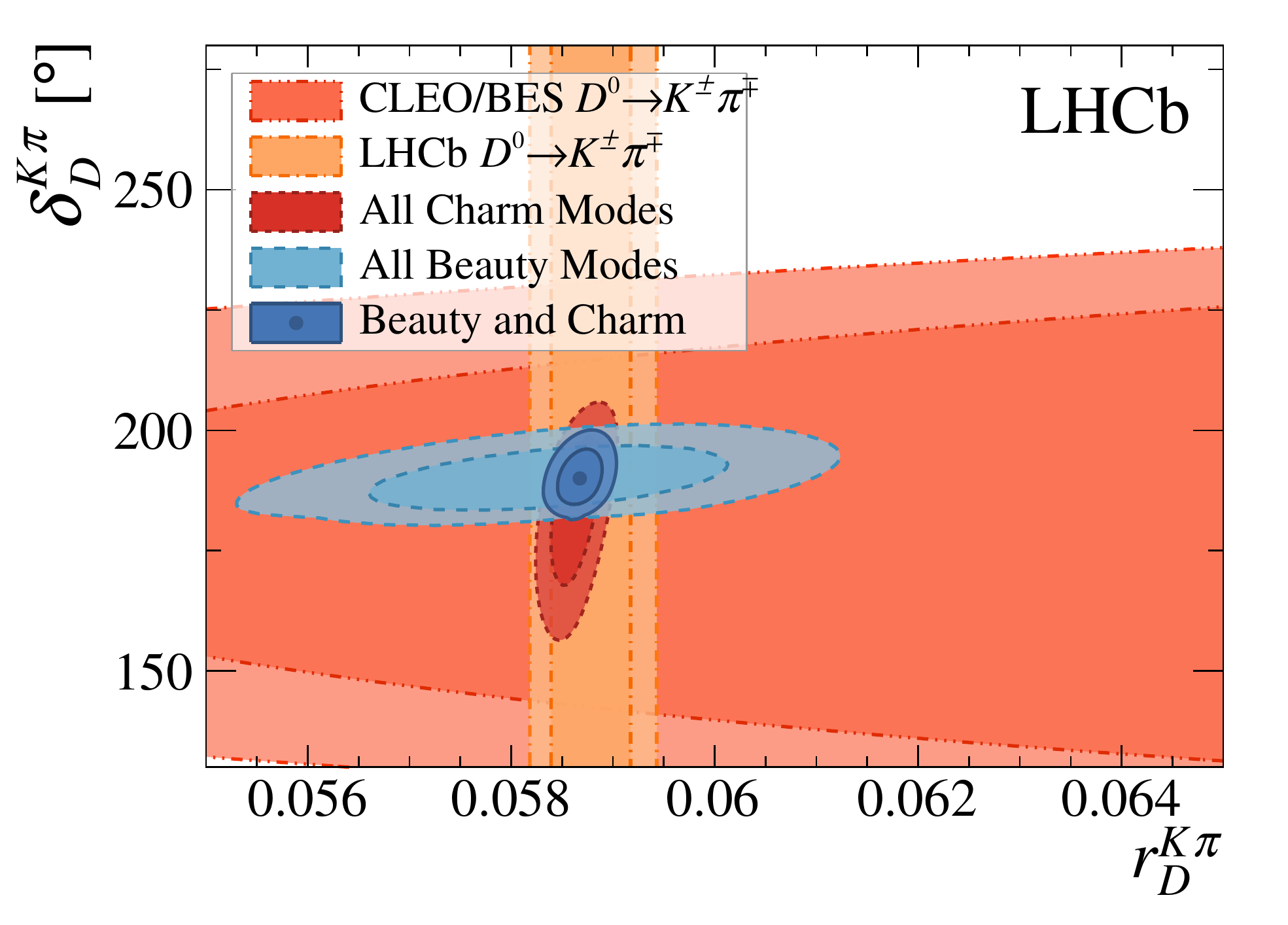}
  \includegraphics[width=0.48\textwidth]{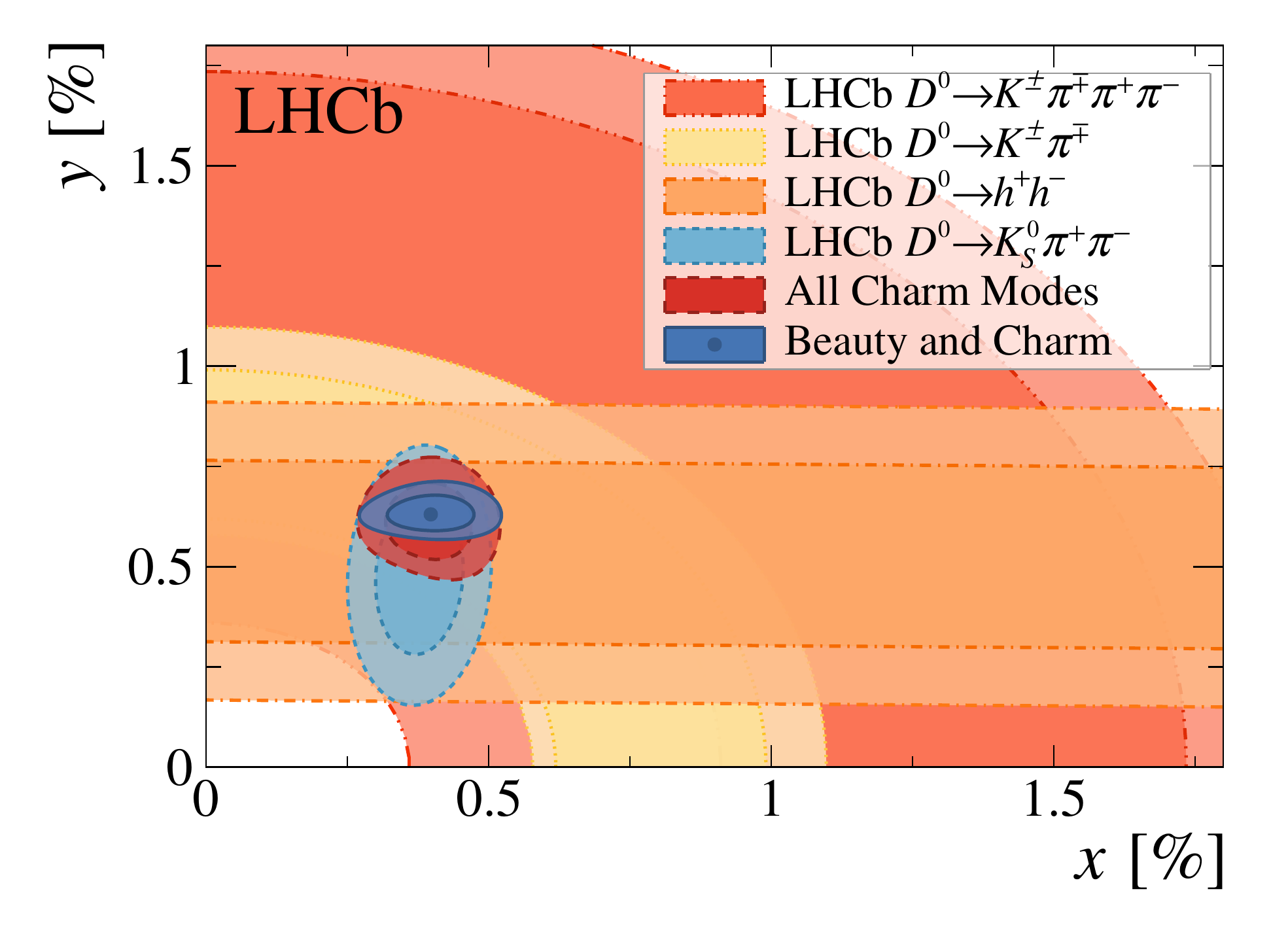}
  \includegraphics[width=0.48\textwidth]{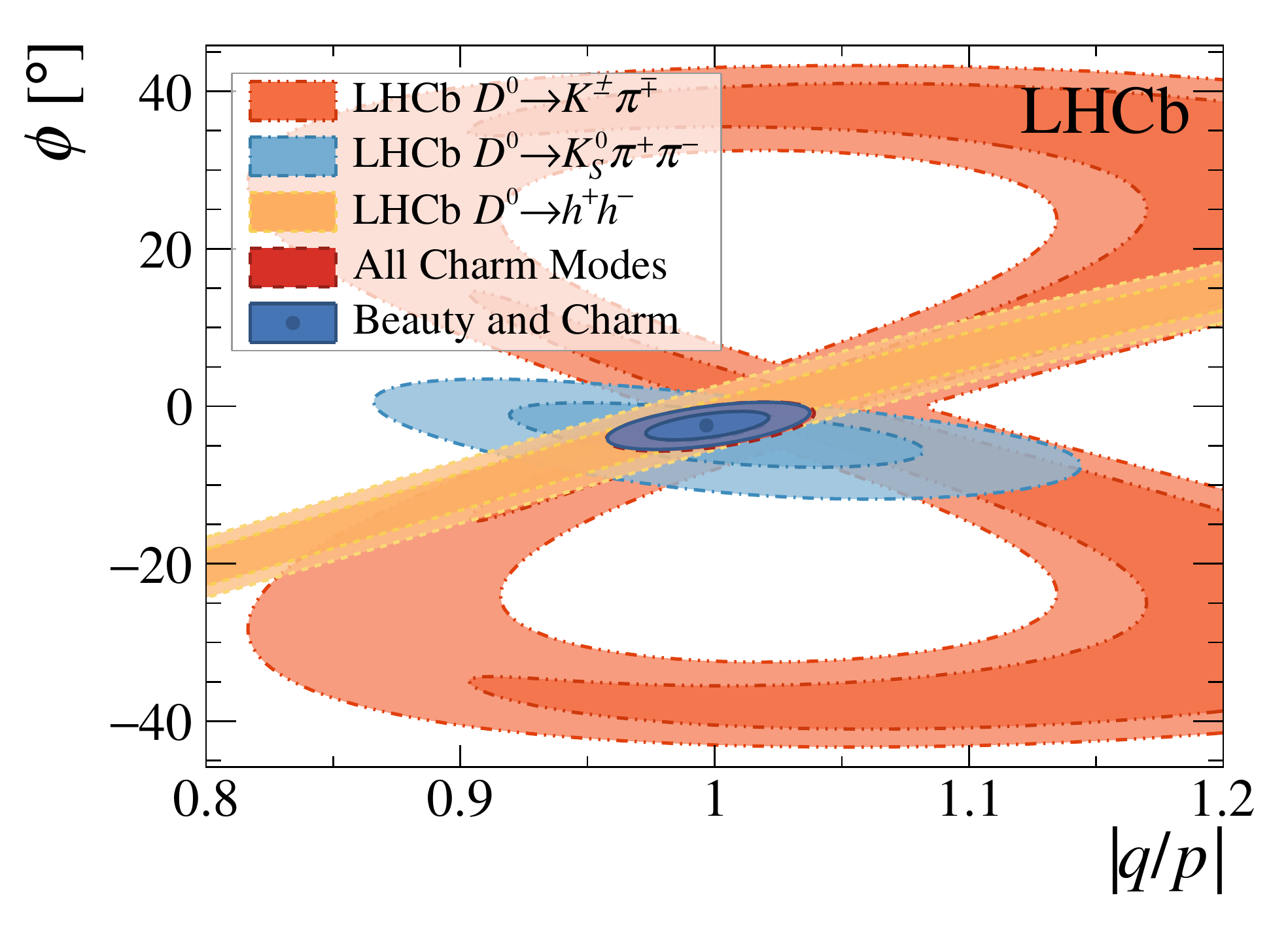}
  \caption{Profile likelihood contours for the charm decay parameters, showing the breakdown of sensitivity amongst different sub-combinations of modes.
  The contours indicate the 68.3\% and 95.4\% confidence region.}
  \label{fig:res:breakdown_2d_charm}
\end{figure}

Breakdowns of the contributing components in the combination are shown in Figs.~\ref{fig:res:breakdown_2d_dh} and~\ref{fig:res:breakdown_2d_charm}.
These highlight the complementary nature of the input measurements to constrain both \g and the charm mixing parameters.
In Fig.~\ref{fig:res:breakdown_2d_charm} (top left) the dark orange band shows external constraints from \mbox{CLEO-c}~\cite{Asner:2012xb} and BES-III~\cite{Ablikim:2014gvw}. These are required to constrain \ddkpi when obtaining the ``All Charm Modes" contours, but are not used in the full combination.
In the top right and bottom plots the orange bands show the constraints from \Dzhh modes, but these cannot provide bands in $(\xD,\yD)$ or $(\qopD,\phiD)$ without other constraints~\cite{Pajero:2021jev}. Consequently, when these orange bands are produced in the top right plot $(\qopD,\phiD,\rdkpi,\ddkpi)$ are fixed to their best fit values from Table~\ref{tab:results}, while in the bottom plot $(\xD,\yD,\rdkpi,\ddkpi)$ are fixed to their best fit values.
In the bottom figure the red contour is mostly hidden behind the blue; this is because no significant additional sensitivity to \CP violation in the charm system is provided by the inclusion of the beauty observables in the simultaneous fit.


The value of $\g = (65.4^{\,+3.8}_{\,-4.2})^\circ$ determined from this combination is compatible with, but lower than that of the previous LHCb combination $\g = (74^{\,+5.0}_{\,-5.8})^\circ$~\cite{LHCb-CONF-2018-002}. This change is driven by improved treatments of background sources in the major inputs described in Refs.~\cite{LHCb-PAPER-2020-019, LHCb-PAPER-2020-036}. An assessment of the compatibility between this and the previous combination, which considers the full parameter space and the correlation between the current set of inputs and the previous set of inputs, finds they are compatible at the level of $2.1\sigma$.
The new result is in excellent agreement with the global CKM fit results~\cite{CKMfitter2015,UTfit-UT}.

The charm mixing parameters, \xD and \yD, are determined simultaneously with \g in this combination for the first time. The precision on \xD is driven by the recent measurement described in Ref.~\cite{LHCb-PAPER-2021-009}. The result $\yD = (0.630^{\,+0.033}_{\,-0.030})\%$ is more precise than the world average, $\yD = (0.603^{\,+0.057}_{\,-0.056})\%$~\cite{HFLAV18}, by approximately a factor of two, driven entirely by the improved measurement of \ddkpi from the beauty system and the simultaneous averaging methodology employed in this article. The correlation between \ddkpi and \ddk is $-57\%$, highlighting \BuDK decays as the source of this improvement.

The beauty part of the combination is cross-checked with an independent framework using a Bayesian statistical treatment. A flat prior is used for \g and the relevant hadronic parameters and results in a value of $\g = (65.6^{\,+3.7}_{\,-3.8})^\circ$, in agreement with the default frequentist results.
Good agreement between the frequentist and Bayesian interpretations is also seen for the other hadronic parameters.
A second cross-check using an independent fitting framework with frequentist interpretation gives consistent results to better than $1\%$ precision.
Finally, the charm sector of the combination was validated by accurately reproducing the HFLAV results~\cite{HFLAV18}.

The relative impact of systematic uncertainties on the input observables is studied, and found to contribute approximately $1.2^\circ$ to the result for \g, demonstrating that the uncertainty of this combination is still dominated by the data sample size.

In previous combinations, the experimental input from \BdDpi decays was included with an external theoretical prediction of $\rbdmpi = 0.0182 \pm 0.0038$~\cite{LHCb-PAPER-2018-009}.
This prediction assumes SU(3) symmetry, and was the only input from theory.
This external input is no longer used, and the combination gives an experimental determination of $\rbdmpi = 0.030^{\,+0.014}_{\,-0.012}$.
This is in agreement with the theory-based prediction and provides confidence that the assumption of SU(3) symmetry is valid within the current precision. This change has a negligible impact on the determination of other parameters.

\section{Conclusion}

A simultaneous combination of LHCb measurements sensitive to the CKM angle \g and charm mixing parameters, along with auxiliary information from other experiments, is performed for the first time. This includes seven new and updated inputs from $B$-meson decays and eight inputs from $D$-meson decays. The result, 
\begin{equation*}
    \g = (65.4^{\,+3.8}_{\,-4.2})^\circ\,,
\end{equation*}
provides the most precise measurement from a single experiment. The charm mixing parameters are found to be
\begin{align*}
\xD &=(0.400^{\,+0.052}_{\,-0.053})\%\,,\\
\yD &=(0.630^{\,+0.033}_{\,-0.030})\%\,,
\end{align*}
which are the most precise determinations to date. In particular, the uncertainty on \yD is reduced by a factor of two by using the new procedure described in this paper.

\section*{Acknowledgements}
%
%
\noindent We express our gratitude to our colleagues in the CERN
accelerator departments for the excellent performance of the LHC. We
thank the technical and administrative staff at the LHCb
institutes.
We acknowledge support from CERN and from the national agencies:
CAPES, CNPq, FAPERJ and FINEP (Brazil); 
MOST and NSFC (China); 
CNRS/IN2P3 (France); 
BMBF, DFG and MPG (Germany); 
INFN (Italy); 
NWO (Netherlands); 
MNiSW and NCN (Poland); 
MEN/IFA (Romania); 
MSHE (Russia); 
MICINN (Spain); 
SNSF and SER (Switzerland); 
NASU (Ukraine); 
STFC (United Kingdom); 
DOE NP and NSF (USA).
We acknowledge the computing resources that are provided by CERN, IN2P3
(France), KIT and DESY (Germany), INFN (Italy), SURF (Netherlands),
PIC (Spain), GridPP (United Kingdom), RRCKI and Yandex
LLC (Russia), CSCS (Switzerland), IFIN-HH (Romania), CBPF (Brazil),
PL-GRID (Poland) and NERSC (USA).
We are indebted to the communities behind the multiple open-source
software packages on which we depend.
Individual groups or members have received support from
ARC and ARDC (Australia);
AvH Foundation (Germany);
EPLANET, Marie Sk\l{}odowska-Curie Actions and ERC (European Union);
A*MIDEX, ANR, Labex P2IO and OCEVU, and R\'{e}gion Auvergne-Rh\^{o}ne-Alpes (France);
Key Research Program of Frontier Sciences of CAS, CAS PIFI, CAS CCEPP, 
Fundamental Research Funds for the Central Universities, 
and Sci. \& Tech. Program of Guangzhou (China);
RFBR, RSF and Yandex LLC (Russia);
GVA, XuntaGal and GENCAT (Spain);
the Leverhulme Trust, the Royal Society
 and UKRI (United Kingdom).


\section*{Appendices}

\appendix

\section{Correlation matrix}
\label{sec:app:corr}

The global fit correlation matrix between each of the parameters presented in Table~\ref{tab:results} is provided in Tables~\ref{tab:corrs1},~\ref{tab:corrs2} and~\ref{tab:corrs3}. A subset of this matrix, including only the parameters of greatest interest, is given in Table~\ref{tab:corrs_small}. The correlation coefficients for $\beta$ and $\phi_s$ are not included as they are almost all smaller than $0.001$, an exception is $\rho(\gamma,\phi_s) = -0.009$.   

\begin{table}[!h]
\centering
\caption{Reduced correlation matrix for the parameters of greater interest. Values smaller than $0.001$ are replaced with a - symbol.}
\label{tab:corrs_small}
\renewcommand{\arraystretch}{1.1}
\resizebox{\textwidth}{!}{
\begin{tabular}{lrrrrrrrr}
\hline
{} &      \g &     $A_D$ &  \rDkpi &  \dDkpi &     \xD &     \yD &   \qopD &   \phiD \\
\hline
\g     &  $1.000$ &      - & $-0.003$ &  $0.003$ & $-0.002$ &  $0.003$ &      - &      - \\
$A_D$    &       &  $1.000$ & $-0.016$ & $-0.083$ & $-0.003$ & $-0.083$ & $-0.316$ &  $0.295$ \\
\rDkpi &  &  &  $1.000$ &  $0.295$ &  $0.282$ & $-0.095$ & $-0.070$ &  $0.015$ \\
\dDkpi &  &  &  &  $1.000$ &  $0.029$ &  $0.891$ & $-0.020$ &  $0.055$ \\
\xD    &  &  &  &  &  $1.000$ &  $0.013$ & $-0.129$ &  $0.083$ \\
\yD    &  &  & &  &  &  $1.000$ & $-0.018$ &  $0.040$ \\
\qopD  &      & &  & &  &  &  $1.000$ &  $0.554$ \\
\phiD  &       &  &  &  &  &  &  &  $1.000$ \\
\hline
\end{tabular}
}
\end{table}

\begin{table}[!tb]
\centering
\caption{Correlation matrix of the fit result, part 1 of 3. Values smaller than $0.001$ are replaced with a - symbol.}
\label{tab:corrs1}
\renewcommand{\arraystretch}{1.2}
\resizebox{\textwidth}{!}{
\begin{tabular}{lrrrrrrrrrrrrrr}
\hline
{} &      \g &    \rdk &    \ddk &   \rdpi &   \ddpi &  \rdstk &  \ddstk &  \rdstpi &  \ddstpi &  \rdkst &  \ddkst &  \rdkstz &  \ddkstz &   \ldsk \\
\hline
\g          &  $1.000$ &  $0.490$ &  $0.613$ & $-0.051$ &  $0.229$ &  $0.138$ &  $0.331$ &   $0.116$ &   $0.149$ & $-0.158$ & $-0.012$ &   $0.206$ &  $-0.111$ & $-0.003$\\
\rdk       & &  $1.000$ &  $0.442$ & $-0.048$ &  $0.113$ &  $0.058$ &  $0.174$ &   $0.070$ &   $0.100$ & $-0.079$ &  $0.026$ &   $0.098$ &  $-0.047$ & $-0.001$\\
\ddk      & & &  $1.000$ & $-0.055$ &  $0.236$ &  $0.062$ &  $0.231$ &   $0.102$ &   $0.156$ & $-0.094$ &  $0.054$ &   $0.120$ &  $-0.052$ & $-0.002$ \\
\rdpi      & & & &  $1.000$ &  $0.629$ & $-0.006$ & $-0.018$ &  $-0.008$ &  $-0.011$ &  $0.006$ & $-0.003$ &  $-0.009$ &   $0.004$ &      - \\
\ddpi     & & & & &  $1.000$ &  $0.025$ &  $0.085$ &   $0.034$ &   $0.052$ & $-0.035$ & $0.011$ &   $0.046$ &  $-0.022$ &      - \\
\rdstk    & & & & & &  $1.000$ &  $0.764$ &  $-0.211$ &  $-0.213$ & $-0.022$ & $-0.006$ &   $0.029$ &  $-0.016$ &      - \\
\ddstk   & & & & & & &  $1.000$ &  $-0.172$ &  $-0.164$ & $-0.052$ &  $0.001$ &   $0.067$ &  $-0.035$ &      - \\
\rdstpi   & & & & & & & &   $1.000$ &   $0.895$ & $-0.018$ &  $0.004$ &   $0.023$ &  $-0.011$ &      - \\
\ddstpi  & & & & & & & & &   $1.000$ & $-0.023$ &  $0.010$ &   $0.029$ &  $-0.013$ &      - \\
\rdkst    & & & & & & & & & &  $1.000$ & $-0.323$ &  $-0.033$ &   $0.017$ &      - \\
\ddkst   & & & & & & & & & & &  $1.000$ &  $-0.004$ &   $0.006$ &      - \\
\rdkstz  & & & & & & & & & & & &   $1.000$ &  $-0.139$ &      - \\
\ddkstz & & & & & & & & & & & & &   $1.000$ &      - \\
\ldsk     & & & & & & & & & & & & & &  $1.000$\\
\hline
\end{tabular}
}
\end{table}

\begin{table}[!tb]
\centering
\caption{Correlation matrix of the fit result, part 2 of 3. Values smaller than $0.001$ are replaced with a - symbol.}
\label{tab:corrs2}
\setlength{\tabcolsep}{5pt}
\renewcommand{\arraystretch}{1.2}
\resizebox{\textwidth}{!}{
\begin{tabular}{lrrrrrrrrrrrrr}
\hline
{} & \ddsk &  \ldskpipi &  \ddskpipi &  \ldmpi &  \ddmpi &  \rdkpipi &  \rdpipipi &  \rDkpi &  \dDkpi &     \xD &     \yD &   \qopD &   \phiD \\
\hline
\g          & $-0.019$ &    $-0.027$ &    $-0.144$ & $-0.065$ & $-0.168$ &    $0.024$ &    $-0.181$ & $-0.003$ &  $0.003$ & $-0.002$ &  $0.003$ &      - &      - \\
\rdk       & $-0.009$ &    $-0.013$ &    $-0.071$ & $-0.032$ & $-0.082$ &   $-0.006$ &     $0.038$ & $-0.079$ & $-0.236$ & $0.011$ & $-0.206$ &  $0.003$ & $-0.010$ \\
\ddk      & $-0.012$ &    $-0.017$ &    $-0.088$ & $-0.040$ & $-0.103$ &   $-0.028$ &     $0.196$ & $-0.161$ & $-0.573$ &  $0.020$ & $-0.510$ &  $0.007$ & $-0.029$ \\
\rdpi      & - &     $0.001$ &     $0.007$ &  $0.003$ &  $0.009$ &        - &    $-0.003$ &  $0.084$ &  $0.022$ & $-0.005$ & $-0.004$ & $-0.005$ & $-0.004$ \\
\ddpi     & $-0.004$ &    $-0.006$ &    $-0.033$ & $-0.015$ & $-0.038$ &   $-0.007$ &     $0.048$ &  $0.048$ & $-0.169$ &      - & $-0.180$ & $-0.004$ & $-0.016$ \\
\rdstk    & $-0.003$ &    $-0.004$ &    $-0.020$ & $-0.009$ & $-0.023$ &    $0.006$ &    $-0.046$ &  $0.013$ &  $0.040$ & $-0.002$ &  $0.035$ &      - &  $0.002$ \\
\ddstk   & $-0.006$ &    $-0.009$ &    $-0.048$ & $-0.022$ & $-0.055$ &    $0.004$ &    $-0.034$ & $-0.012$ & $-0.048$ &      - & $-0.044$ &      - & $-0.003$ \\
\rdstpi   & $-0.002$ &    $-0.003$ &    $-0.017$ & $-0.008$ & $-0.019$ &   $-0.001$ &     $0.008$ & $-0.024$ & $-0.053$ &  $0.002$ & $-0.045$ &  $0.001$ & $-0.002$ \\
\ddstpi  & $-0.003$ &    $-0.004$ &    $-0.021$ & $-0.010$ & $-0.025$ &   $-0.005$ &     $0.033$ & $-0.039$ & $-0.112$ &  $0.004$ & $-0.097$ &  $0.002$ & $-0.005$ \\
\rdkst    & $0.003$ &     $0.004$ &     $0.023$ &  $0.010$ &  $0.026$ &   $-0.004$ &     $0.032$ & $-0.003$ & -0.006 & $-0.002$ & $-0.005$ &      - &      - \\
\ddkst   &  - &         - &     $0.002$ &      - &  $0.002$ &   $-0.008$ &     $0.057$ & $-0.027$ & $-0.102$ &  $0.011$ & $-0.091$ &      - & $-0.004$ \\
\rdkstz  & $-0.004$ &    $-0.006$ &    $-0.030$ & $-0.013$ & $-0.034$ &    $0.006$ &    $-0.043$ &  $0.002$ &  $0.011$ & $-0.002$ &  $0.010$ &      - &      - \\
\ddkstz & $0.002$ &     $0.003$ &     $0.016$ &  $0.007$ &  $0.019$ &   $-0.005$ &     $0.035$ & $-0.007$ & $-0.028$ &  $0.002$ & $-0.025$ &      - & $-0.001$ \\
\ldsk     & $-0.034$ &         - &         - &      - &      - &        - &         - &      - &      - &      - &      - &      - &      - \\
\hline
\end{tabular}
}
\end{table}

\begin{table}[!tb]
\centering
\caption{Correlation matrix of the fit result, part 3 of 3. Values smaller than $0.001$ are replaced with a - symbol.}
\label{tab:corrs3}
\setlength{\tabcolsep}{4pt}
\renewcommand{\arraystretch}{1.2}
\resizebox{\textwidth}{!}{
\begin{tabular}{lrrrrrrrrrrrrr}
\hline
{} & \ddsk &  \ldskpipi &  \ddskpipi &  \ldmpi &  \ddmpi &  \rdkpipi &  \rdpipipi &  \rDkpi &  \dDkpi &     \xD &     \yD &   \qopD &   \phiD \\
\hline
\ddsk       & $1.000$ &         - &     $0.003$ &  $0.001$ &  $0.003$ &        - &     $0.003$ &      - &      - &      - &      - &      - &      - \\
\ldskpipi  & & $1.000$ &     $0.006$ &  $0.002$ &  $0.005$ &        - &     $0.005$ &      - &      - &      - &      - &      - &      - \\
\ddskpipi & & & $1.000$ &  $0.009$ &  $0.024$ &   -0.003 &     $0.026$ &      - &      - &      - &      - &      - &      - \\
\ldmpi      & & & & $1.000$ &  $0.485$ &   $-0.002$ &     $0.012$ &      - &      - &      - &      - &      - &      - \\
\ddmpi     & & & & & $1.000$ &   $-0.004$ &     $0.030$ &      - &      - &      - &      - &      - &      - \\
\rdkpipi    & & & & & & $1.000$ &    $-0.044$ &  $0.020$ &  $0.074$ & $-0.003$ &  $0.066$ &      - &  $0.004$ \\
\rdpipipi   & & & & & & & $1.000$ & $-0.148$ & $-0.531$ &  $0.025$ & $-0.473$ &  0.005 & $-0.026$ \\
\rDkpi      & & & & & & & & $1.000$ &  $0.295$ &  $0.282$ & $-0.095$ & -0.070 &  $0.015$ \\
\dDkpi     & & & & & & & & & $1.000$ &  $0.029$ &  $0.892$ & $-0.020$ &  $0.055$ \\
\xD          & & & & & & & & & & $1.000$ &  $0.013$ & $-0.129$ &  $0.083$ \\
\yD          & & & & & & & & & & & $1.000$ & $-0.018$ &  $0.040$ \\
\qopD      & & & & & & & & & & & & $1.000$ &  $0.554$ \\
\phiD       & & & & & & & & & & & & & $1.000$ \\
\hline
\end{tabular}
}
\end{table}
\clearpage

\section{Contribution of each input measurement to the global \texorpdfstring{\boldmath\chisq}{chi2}}
\label{sec:app:chisq}

The contribution of each input measurement to the global \chisq is shown in Table~\ref{tab:chi2_contrib}.

\begin{table}[!b]
	\caption{Contributions to the total \chisq and the number of observables of each input measurement.}
	\label{tab:chi2_contrib}
	\centering
	\begin{tabular}{l|lrc}
	\hline
		 & Measurement & \chisq & No.~of obs.\\
		\hline
		\multirow{13}{*}{\rotatebox{90}{Beauty sector}} & $B^{\pm}\rightarrow Dh^{\pm}, D\rightarrow h^{\pm}h^{\prime\mp}$               & $ 2.71$ & 8  \\
																										& $B^{\pm}\rightarrow Dh^{\pm}, D\rightarrow h^{\pm}\pi^{\mp}\pi^{+}\pi^{-}$     & $ 7.36$ & 8  \\
																										& $B^{\pm}\rightarrow Dh^{\pm}, D\rightarrow h^{\pm}h^{\prime\mp}\pi^{0}$        & $ 7.14$ & 11 \\
																										& $B^{\pm}\rightarrow Dh^{\pm}, D\rightarrow K_{S}^{0}h^{+}h^{-}$                & $ 4.67$ & 6  \\
																										& $B^{\pm}\rightarrow Dh^{\pm}, D\rightarrow K_{S}^{0}K^{\pm}\pi^{\mp}$          & $ 7.57$ & 7  \\
																										& $B^{\pm}\rightarrow D^{*}h^{\pm}, D\rightarrow h^{\pm}h^{\prime\mp}$           & $ 7.31$ & 16 \\
																										& $B^{\pm}\rightarrow DK^{*\pm}, D\rightarrow h^{\pm}h^{\prime\mp}(\pi^+\pi^-)$  & $ 3.71$ & 12 \\
																										& $B^{0}\rightarrow DK^{*0}, D\rightarrow h^{\pm}h^{\prime\mp}(\pi^+\pi^-)$      & $ 9.45$ & 12 \\
																										& $B^{0}\rightarrow DK^{*0}, D\rightarrow K_{S}^{0}h^{+}h^{-}$                   & $ 3.26$ & 4  \\
																										& $B^{\pm}\rightarrow Dh^{\pm}\pi^+\pi^-, D\rightarrow h^{\pm}h^{\prime\mp}$     & $ 1.34$ & 11 \\
																										& $B^{0}_{s}\rightarrow D^{\mp}_{s}K^{\pm}$                                      & $ 5.71$ & 5  \\
																										& $B^{0}_{s}\rightarrow D^{\mp}_{s}K^{\pm}\pi^+\pi^-$                            & $ 2.88$ & 5  \\
																										& $B^{0}\rightarrow D^{\mp}\pi^{\pm}$                                            & $ 0.00$ & 2  \\
		\hline
		\multirow{9}{*}{\rotatebox{90}{Charm sector}}	  & $D\rightarrow K_{S}^{0}\pi^{+}\pi^{-}\;2011$                                   & $ 5.38$  & 2 \\
																										& $D\rightarrow K_{S}^{0}\pi^{+}\pi^{-}\;\mathrm{Run\;1}$                        & $ 0.77$  & 4 \\
																										& $D\rightarrow K_{S}^{0}\pi^{+}\pi^{-}\;\mathrm{Run\;2}$                        & $ 1.37$  & 4 \\
																										& $D\rightarrow K^{\pm}\pi^{\mp}\;\mathrm{Run\;1}$                               & $ 1.29$  & 6 \\
																										& $D\rightarrow h^{+}h^{-} \; \Delta A_{CP}$                                     & $ 0.00$  & 2 \\
																										& $D\rightarrow K^{\pm}\pi^{\mp}\pi^+\pi^-$                                      & $ 3.59$  & 1 \\
																										& $D\rightarrow h^{+}h^{-} \; y_{CP}$                                            & $ 0.40$  & 1 \\
																										& $D\rightarrow h^{+}h^{-} \; \Delta Y$                                          & $ 0.15$  & 1 \\
																										& $D\rightarrow K^{\pm}\pi^{\mp}\;\mathrm{Run\;2}$                               & $ 2.23$  & 6 \\
		\hline
		\multirow{9}{*}{\rotatebox{90}{External constraints}}	  & 	$D\rightarrow K^{\pm}\pi^{\mp}\pi^{0},\;D\rightarrow K^{\pm}\pi^{\mp}\pi^{+}\pi^{-}$  & $ 0.79$ & 6 \\
																										& 	$D\rightarrow \pi^{+}\pi^{-}\pi^{+}\pi^{-}$                                           & $ 0.03$ & 1 \\
																										& 	$D\rightarrow h^{+}h^{-}\pi^{0}$                                                      & $ 0.01$ & 2 \\
																										& 	$D\rightarrow K_{S}^{0}K^{\pm}\pi^{\mp}\;\mathrm{WS}$                                 & $ 0.60$ & 1 \\
																										& 	$D\rightarrow K_{S}^{0}K^{\pm}\pi^{\mp}$                                              & $ 3.79$ & 3 \\
																										& 	$B^{\pm}\rightarrow DK^{*\pm}$                                                        & $ 0.02$ & 1 \\
																										& 	$B^{0}\rightarrow DK^{*0}$                                                            & $ 0.01$ & 1 \\
																										& 	$\phi_s$                                                                              & $ 0.00$ & 1 \\
																										& 	$\beta$                                                                               & $ 0.00$ & 1 \\
		\hline
		& \textbf{Total} & 83.53 & 151 \\
		\hline
	\end{tabular}
\end{table}

\clearpage

\section{Pull distribution of each input observable}
\label{sec:app:pull}

\begin{figure}[!tb]
	\centering
	\includegraphics[width=0.32\textwidth]{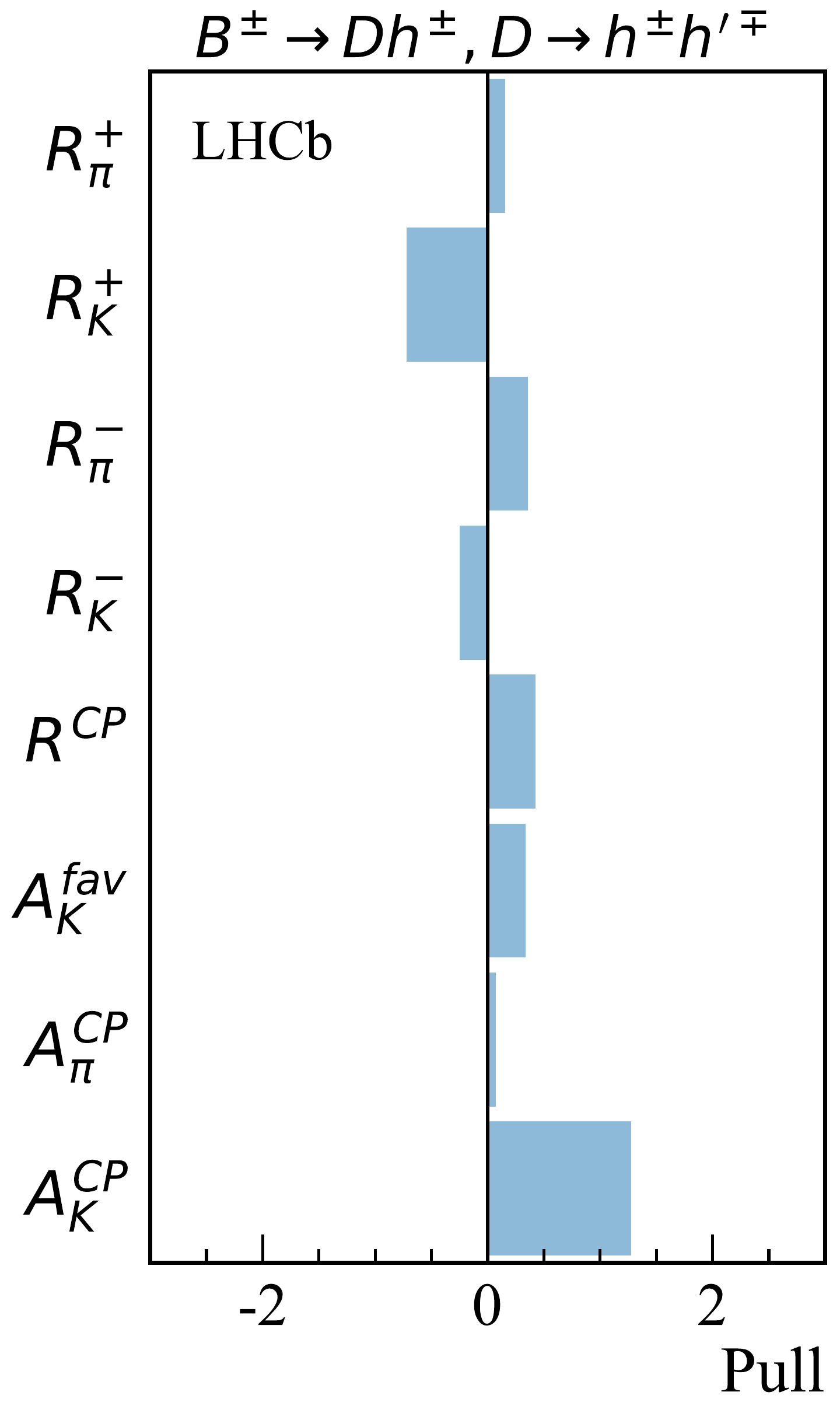}
	\includegraphics[width=0.32\textwidth]{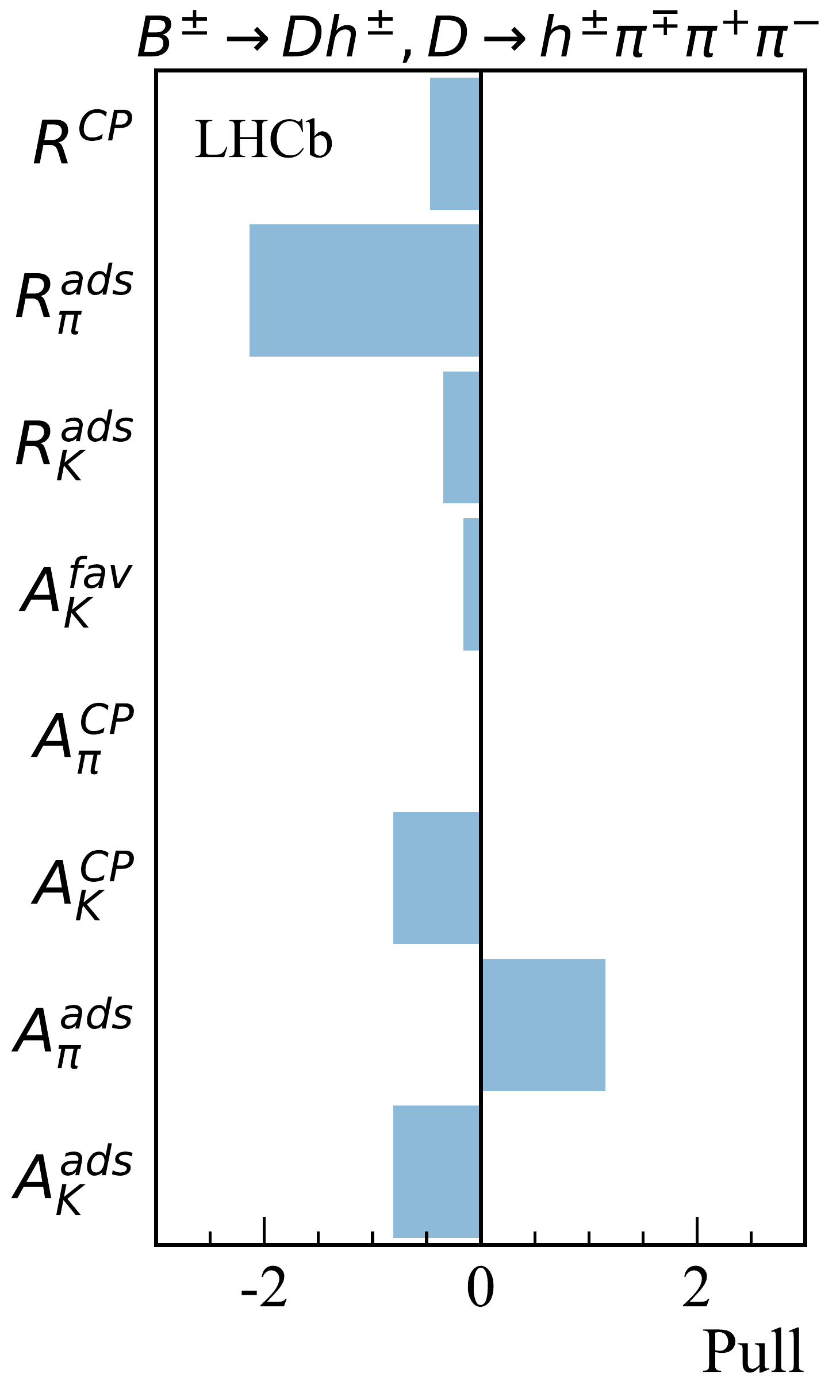}
	\includegraphics[width=0.32\textwidth]{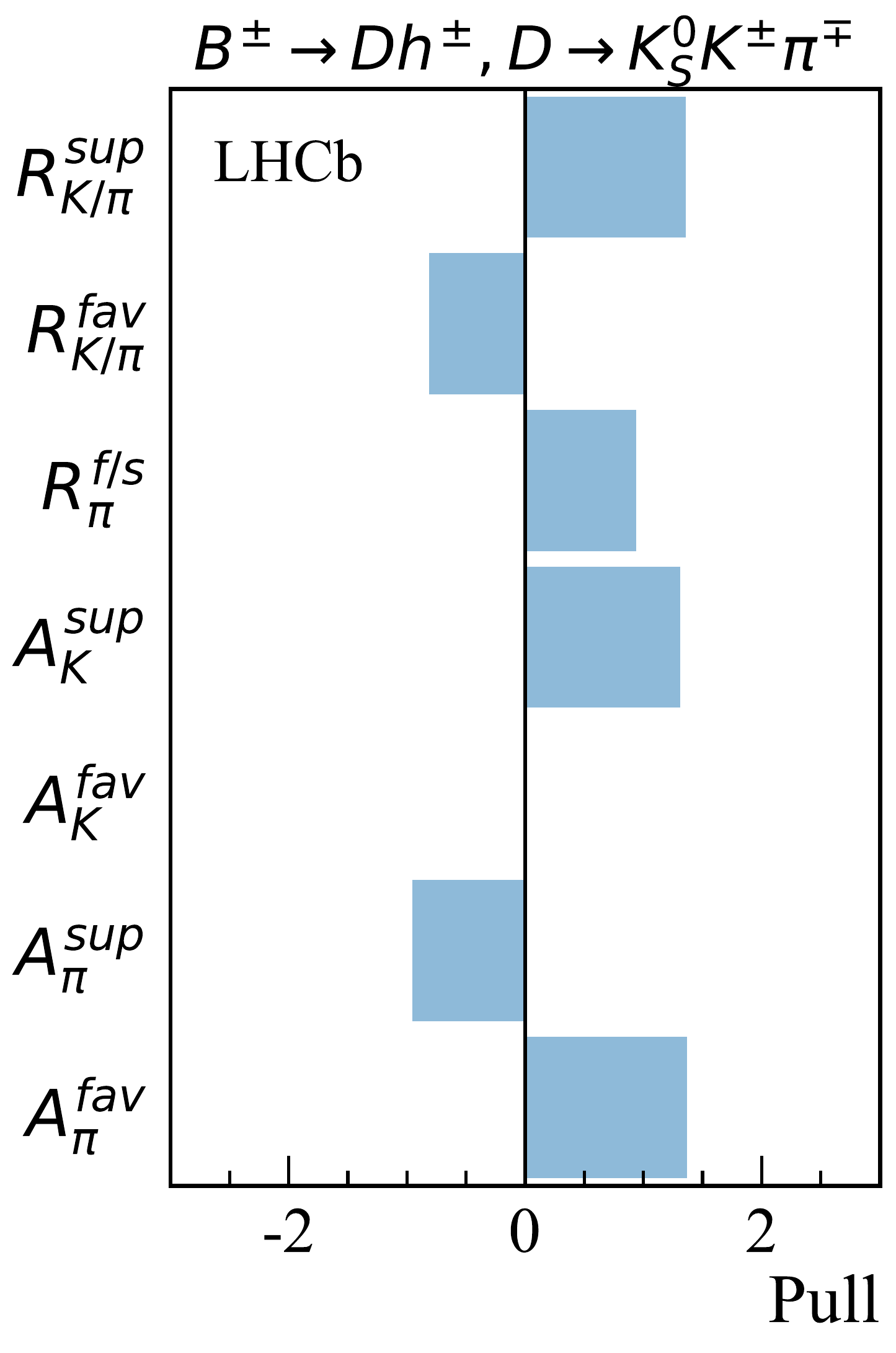}
	\includegraphics[width=0.32\textwidth]{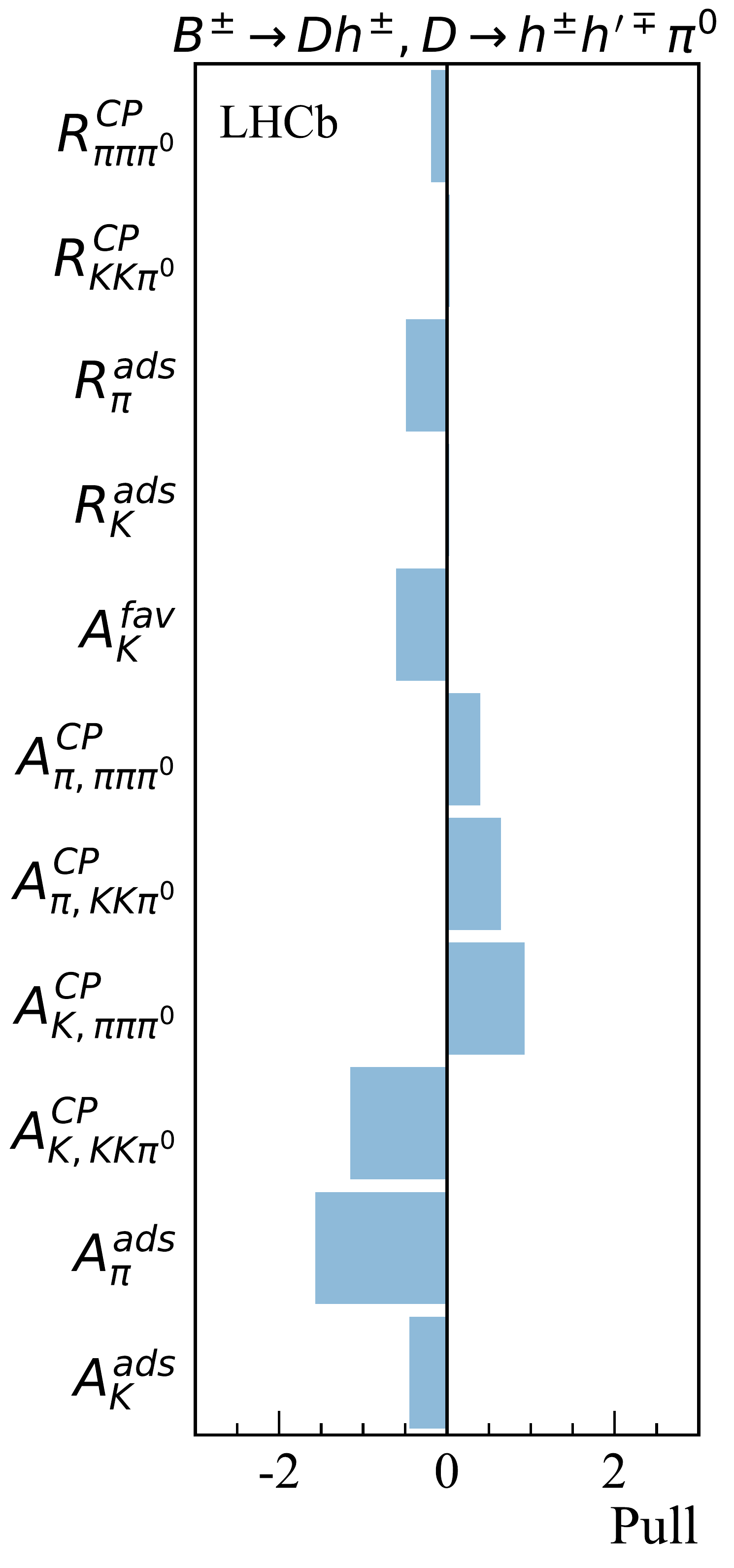}
	\includegraphics[width=0.32\textwidth]{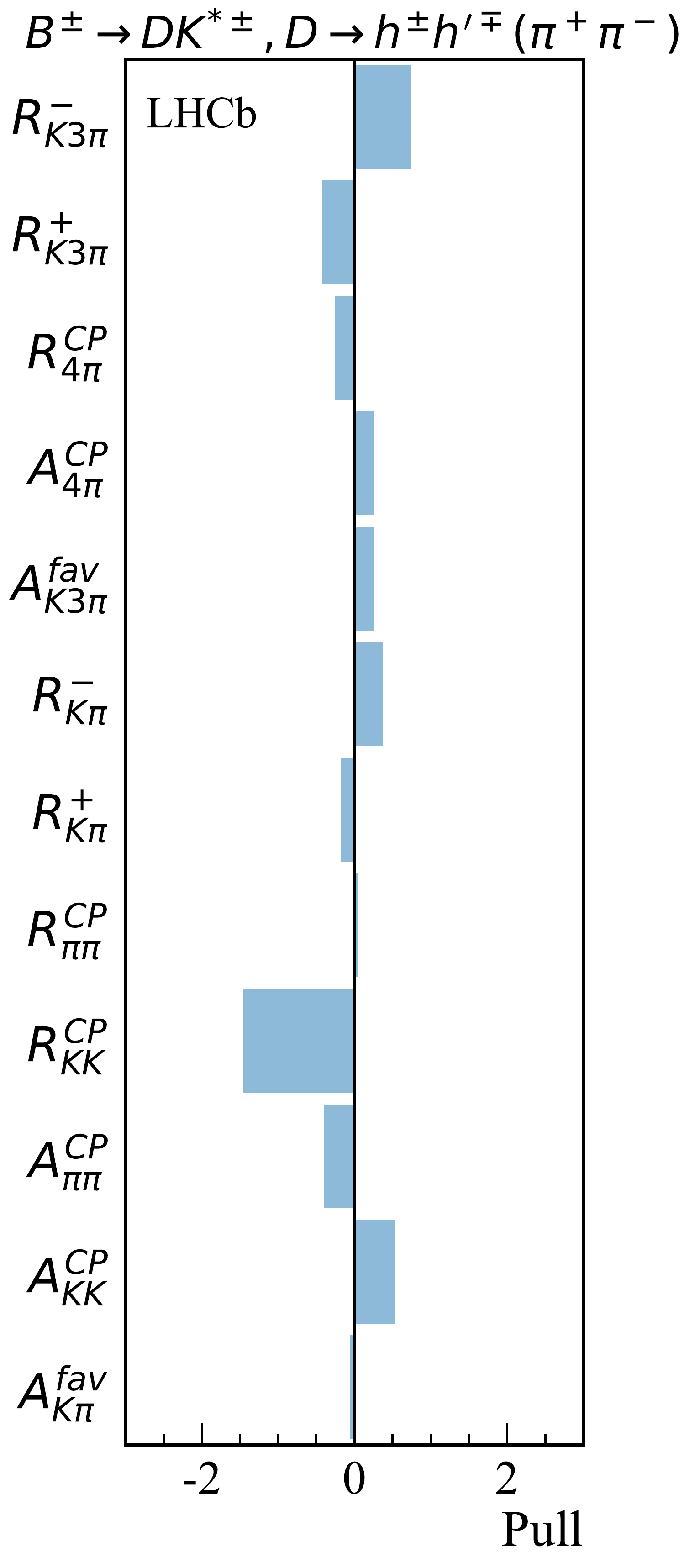}
	\includegraphics[width=0.32\textwidth]{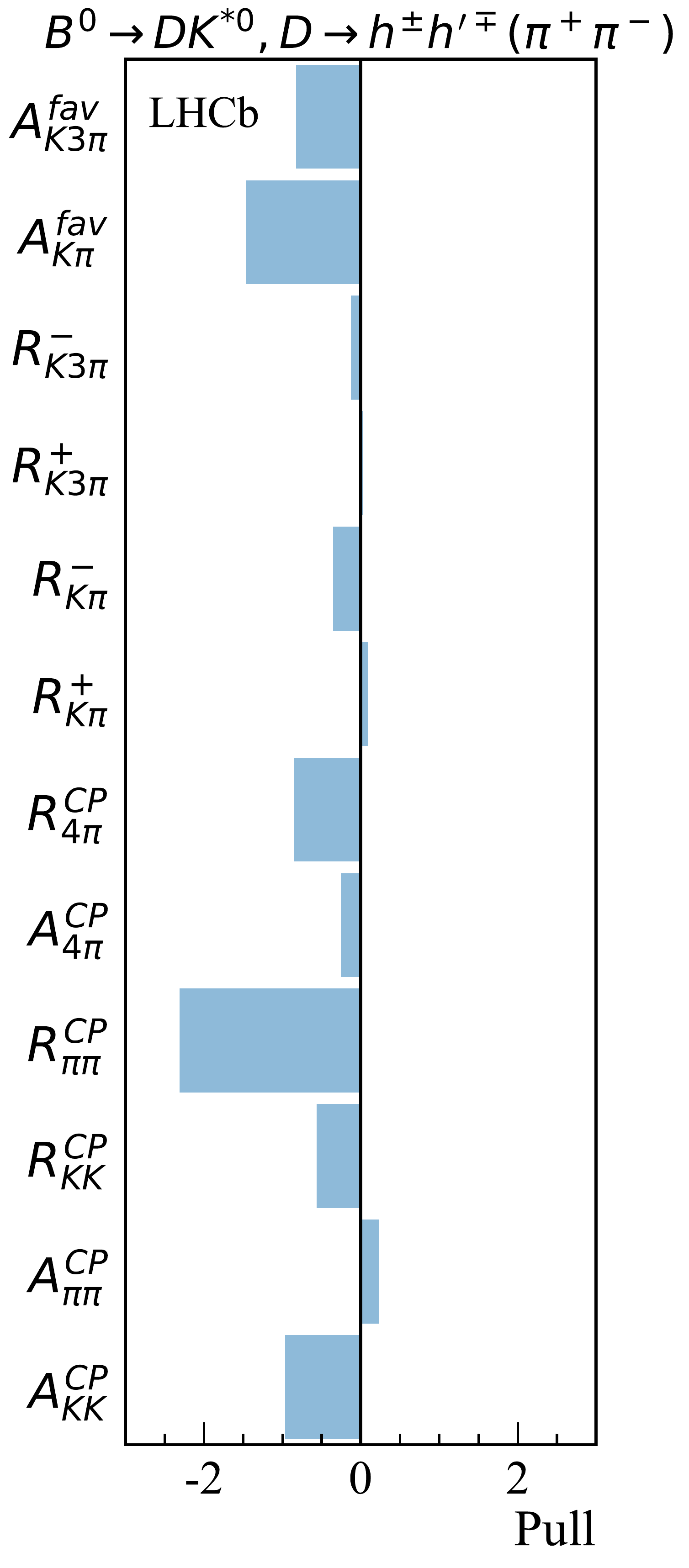}
	\caption{Pulls of the input observables, part 1 of 3.}
	\label{fig:pulls1}
\end{figure}

\begin{figure}[!tb]
	\centering
	\begin{minipage}[b]{0.32\textwidth}
		\includegraphics[width=\textwidth]{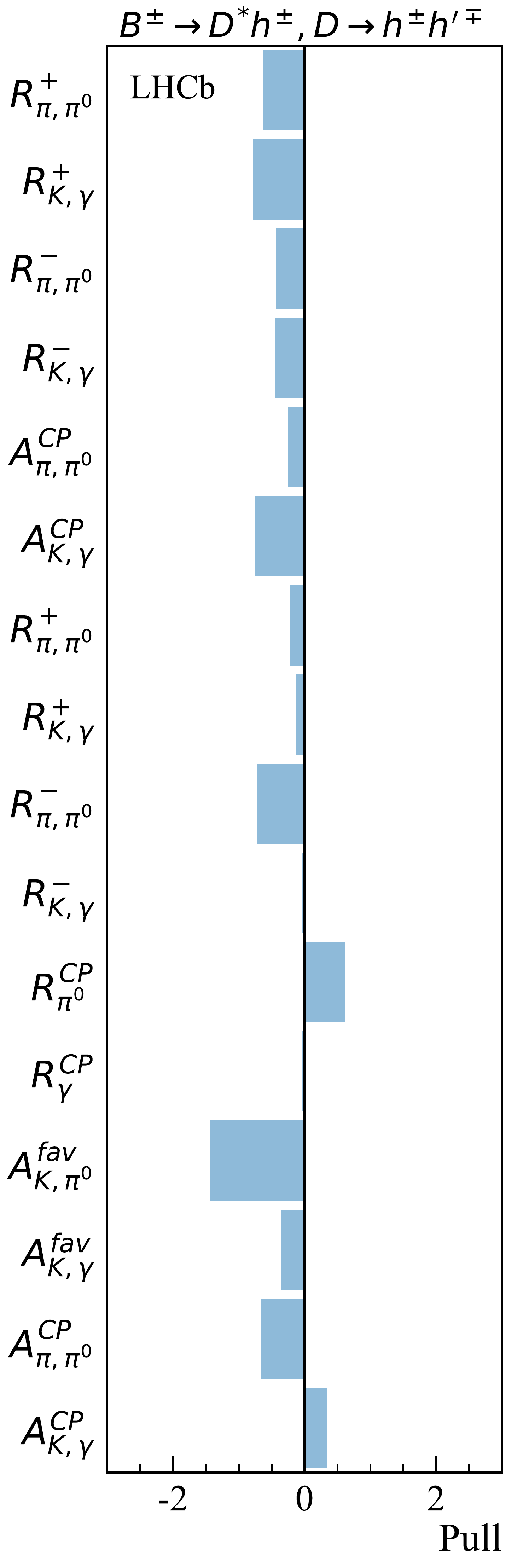}
	\end{minipage}
	\begin{minipage}[b]{0.32\textwidth}
		\includegraphics[width=\textwidth]{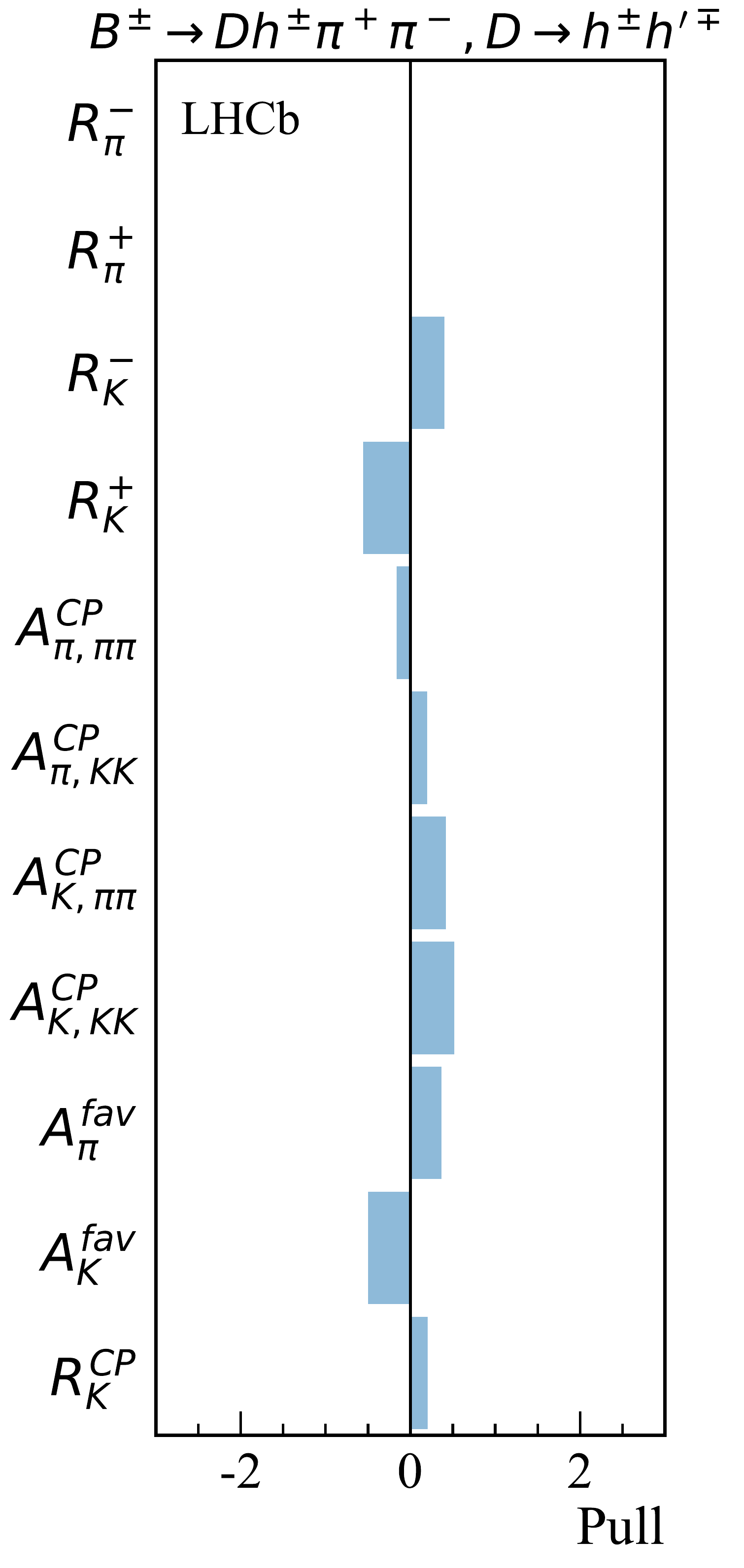}
	\end{minipage}
	\begin{minipage}[b]{0.32\textwidth}
		\includegraphics[width=\textwidth]{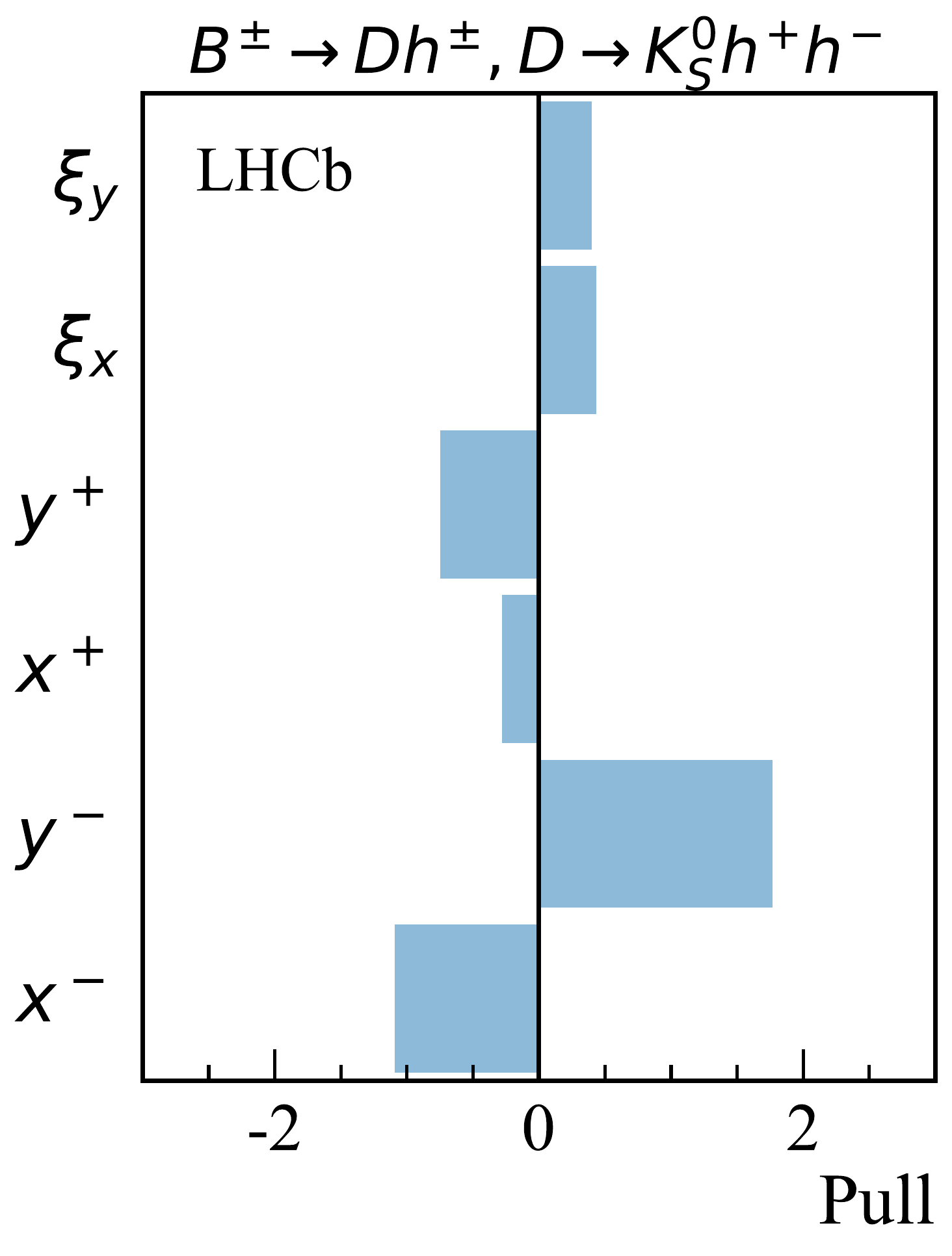} \\
		\includegraphics[width=\textwidth]{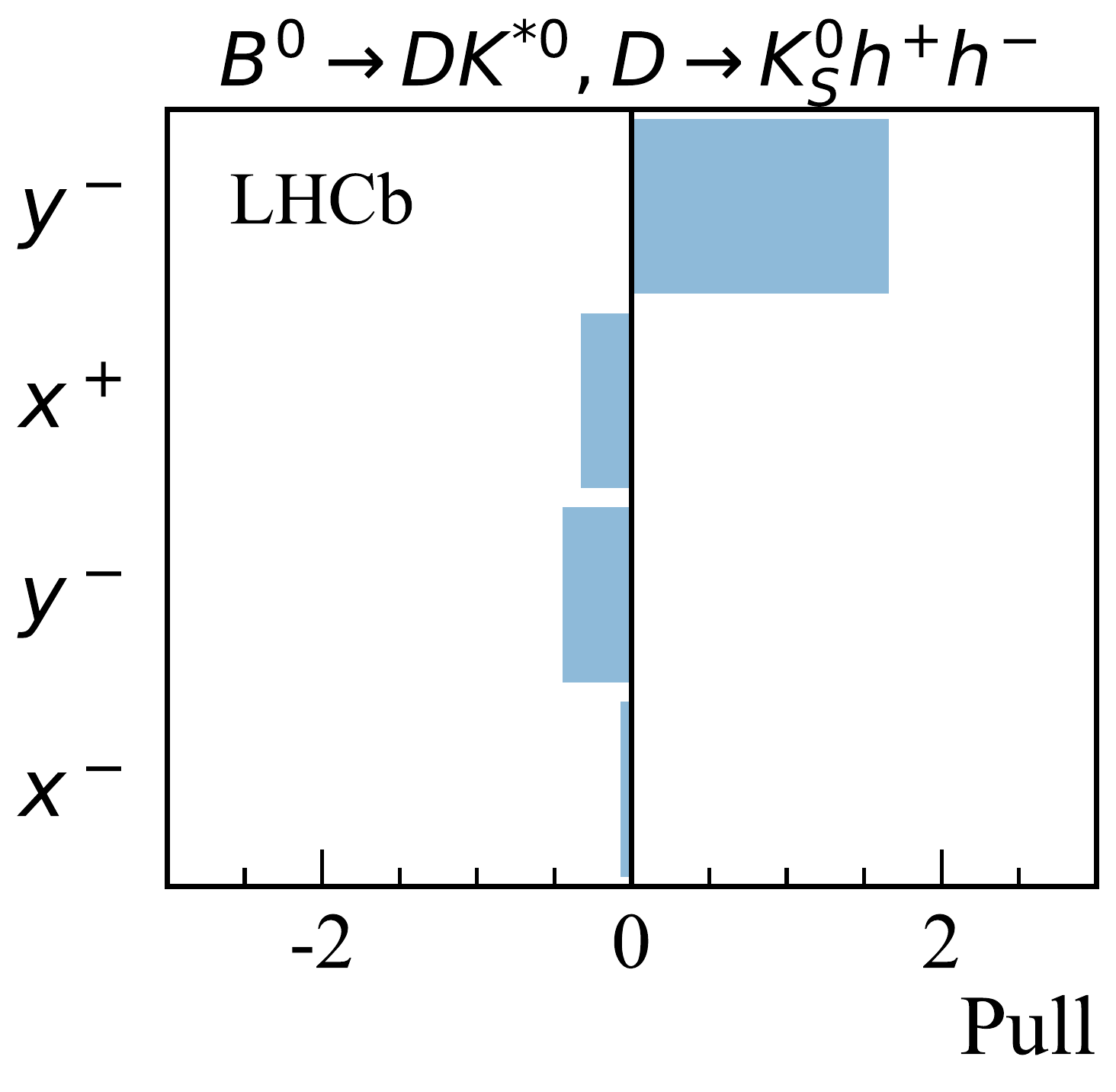}
	\end{minipage}
	\includegraphics[width=0.32\textwidth]{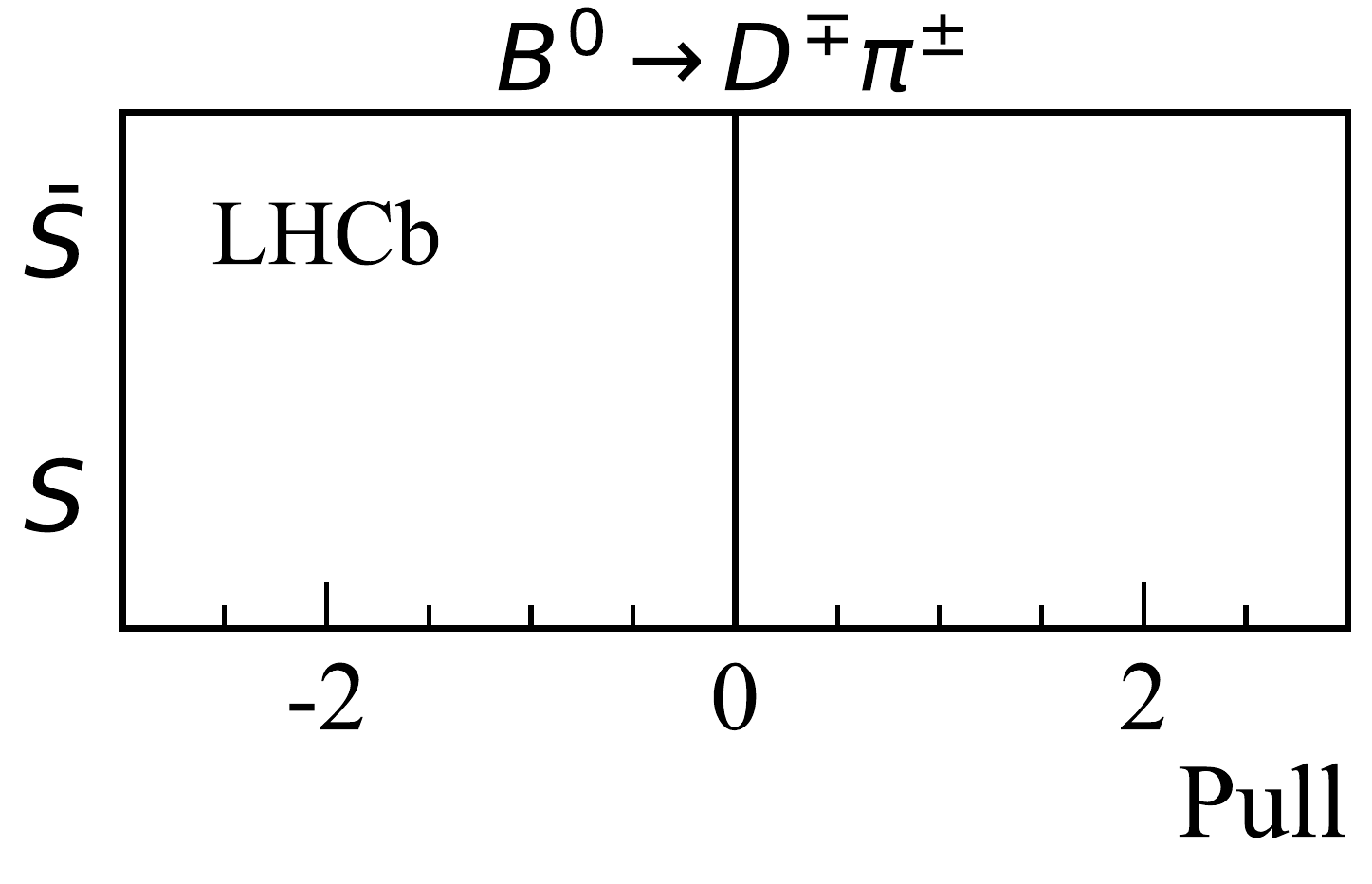}
	\includegraphics[width=0.32\textwidth]{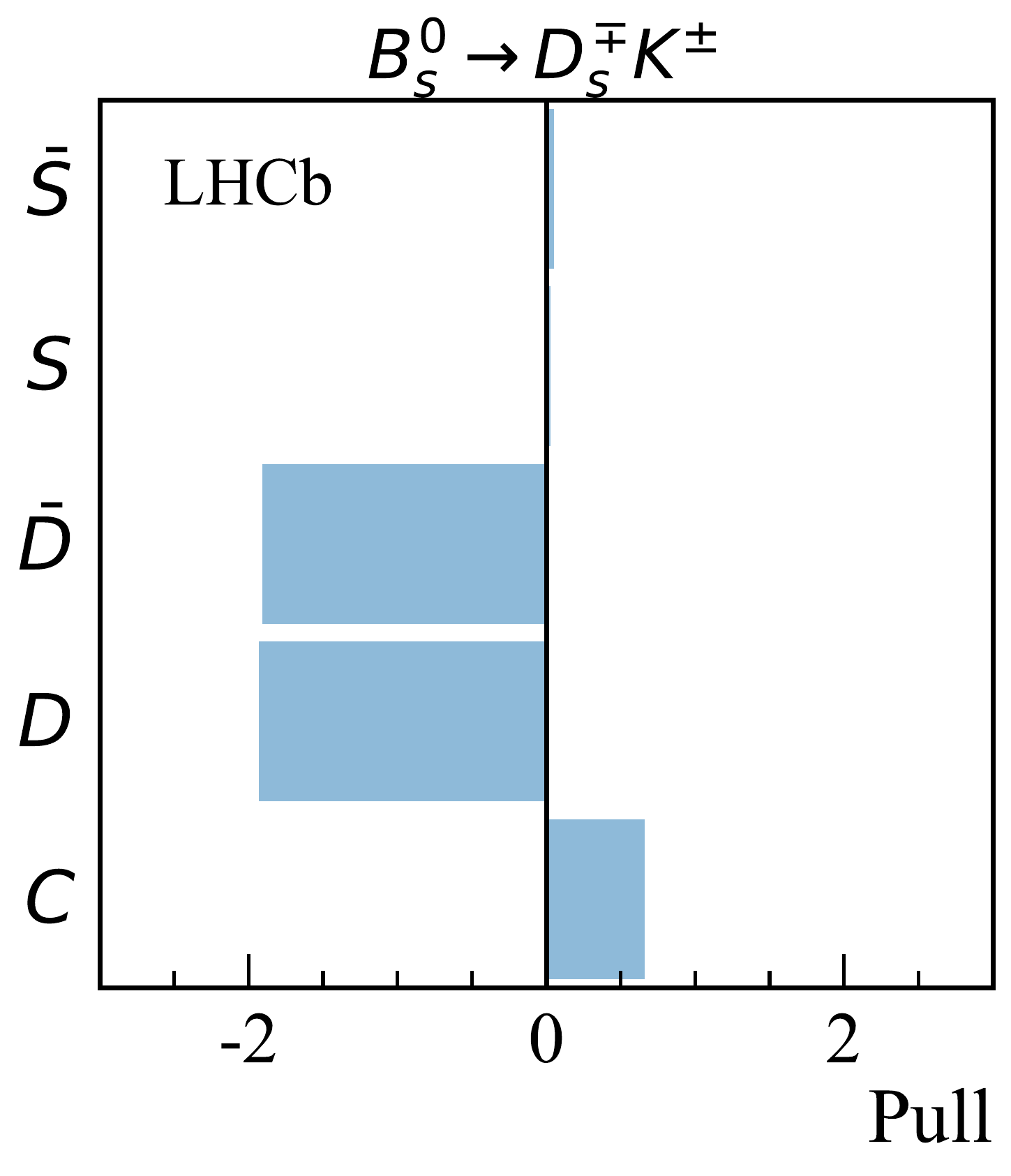}
	\includegraphics[width=0.32\textwidth]{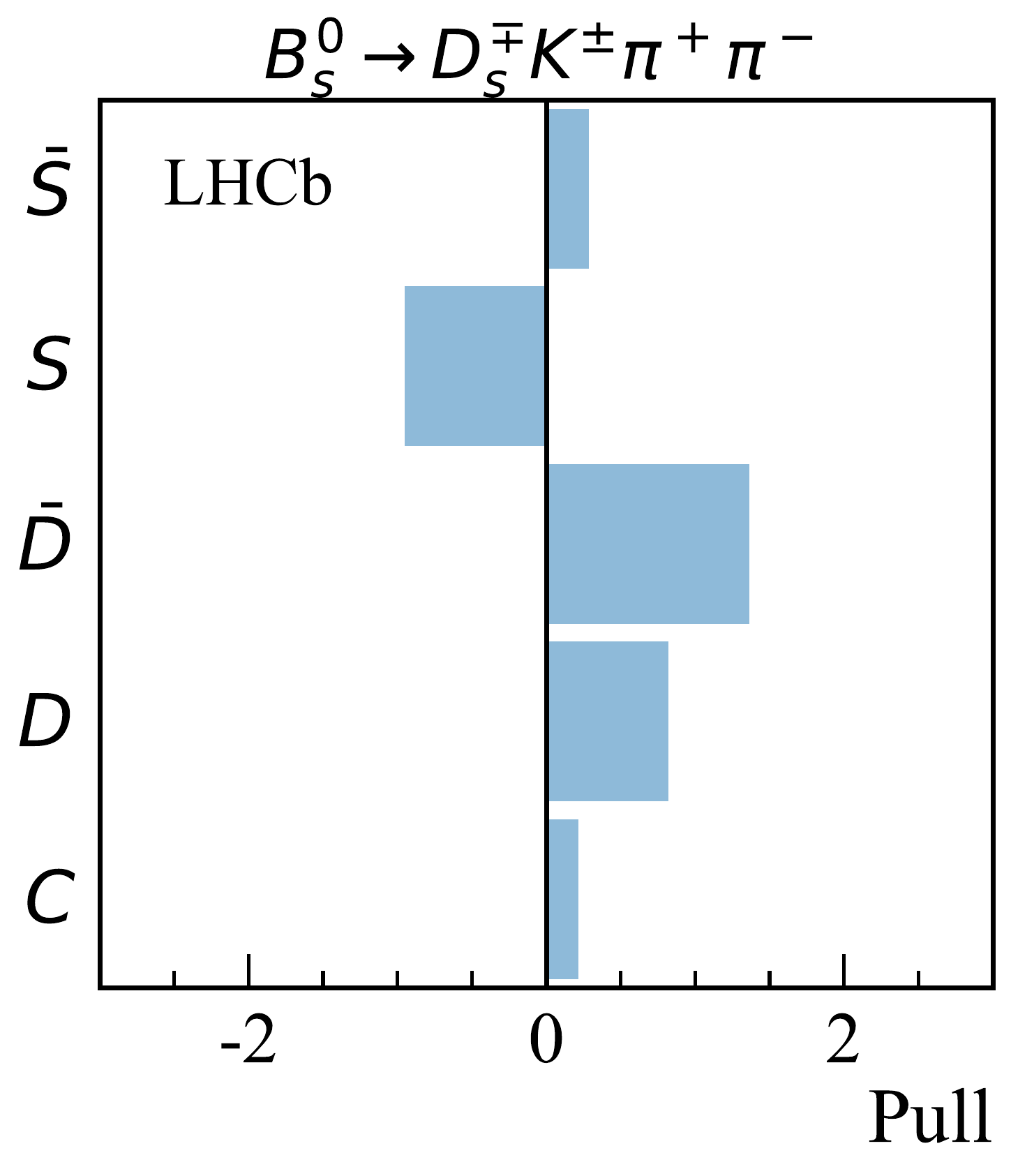}
	\caption{Pulls of the input observables, part 2 of 3.}
	\label{fig:pulls2}
\end{figure}

\begin{figure}[!tb]
	\centering
	\includegraphics[width=0.32\textwidth]{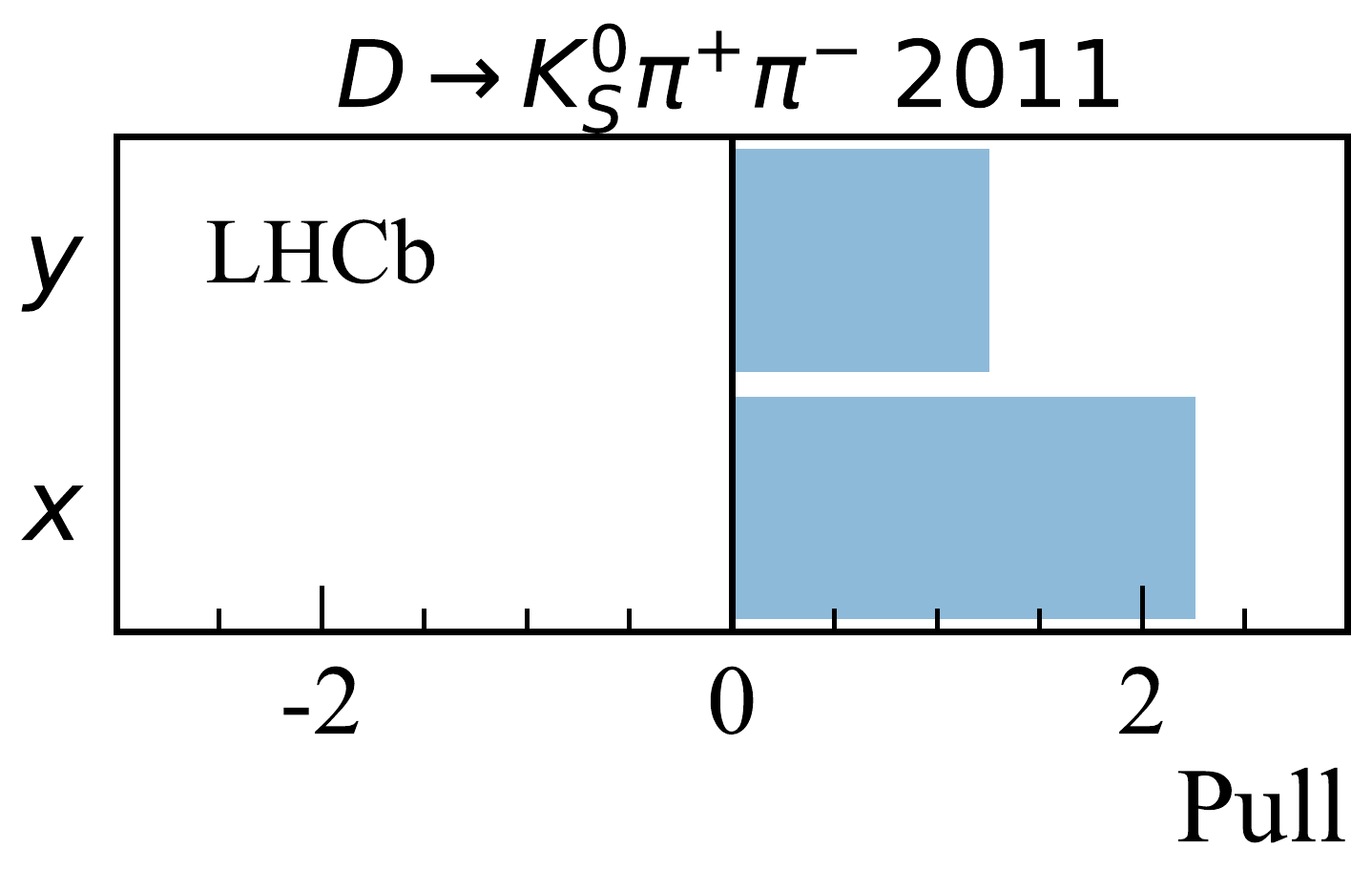}
	\includegraphics[width=0.32\textwidth]{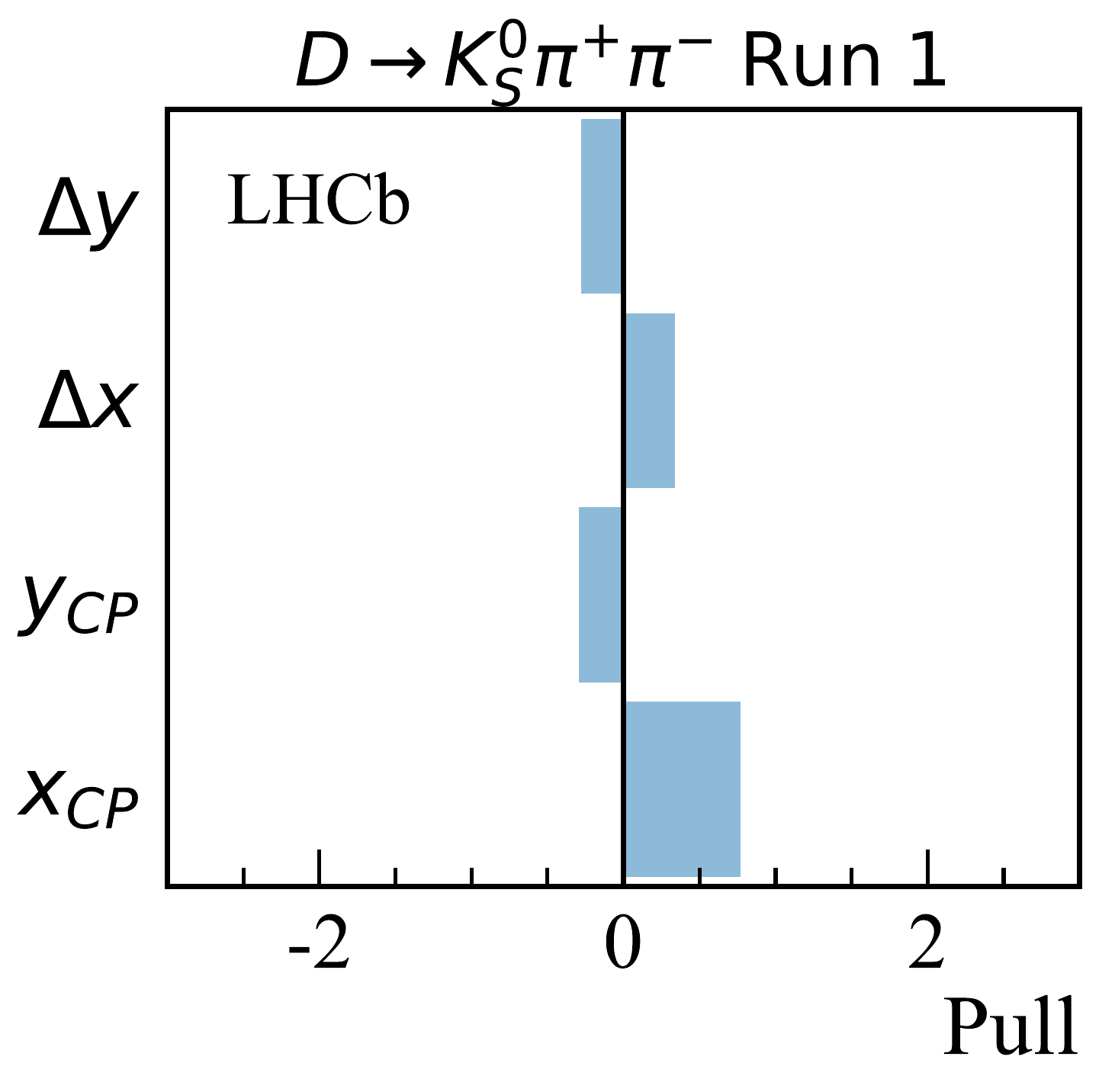}
	\includegraphics[width=0.32\textwidth]{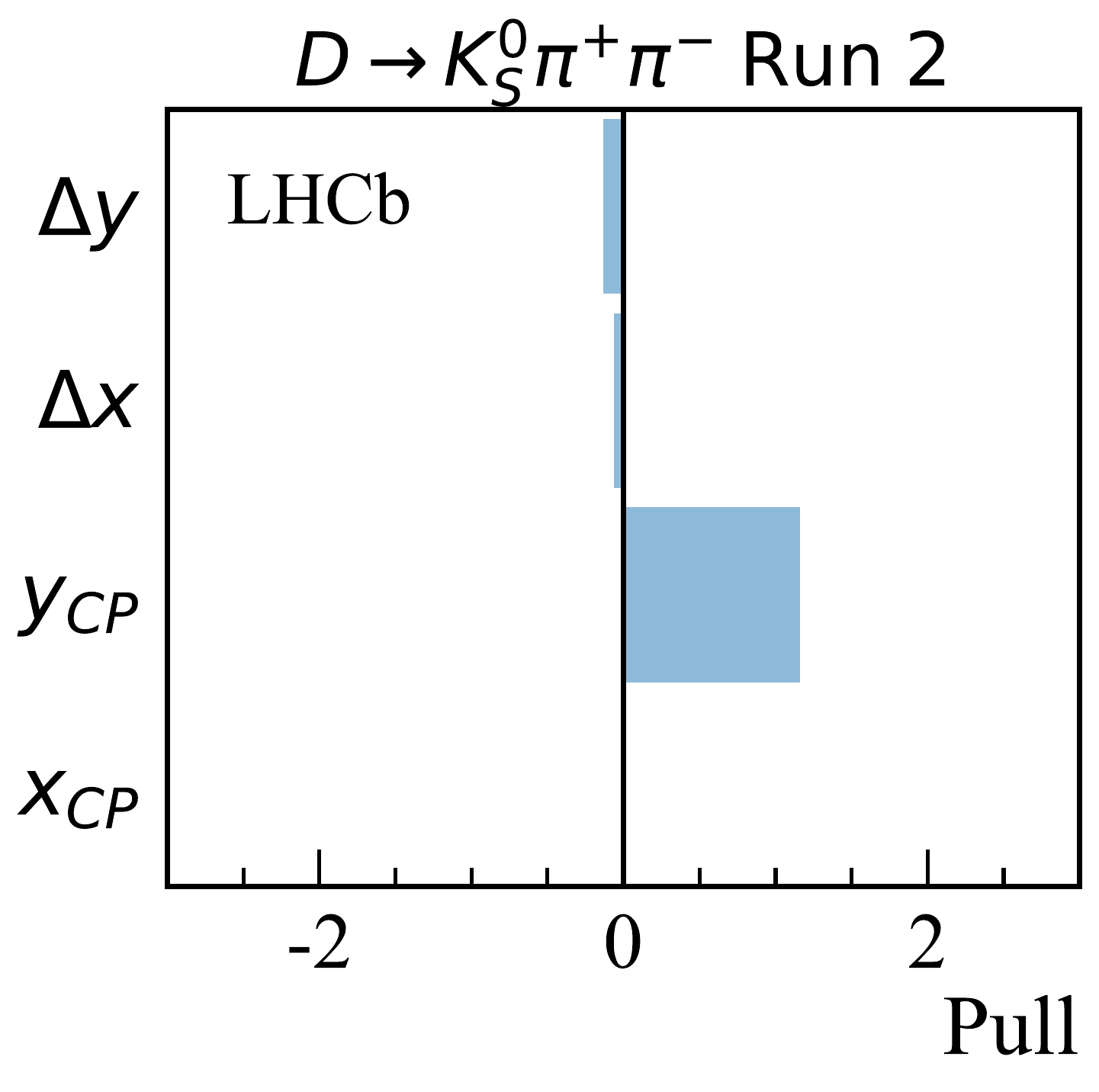}
	\begin{minipage}[b]{0.32\textwidth}
		\includegraphics[width=\textwidth]{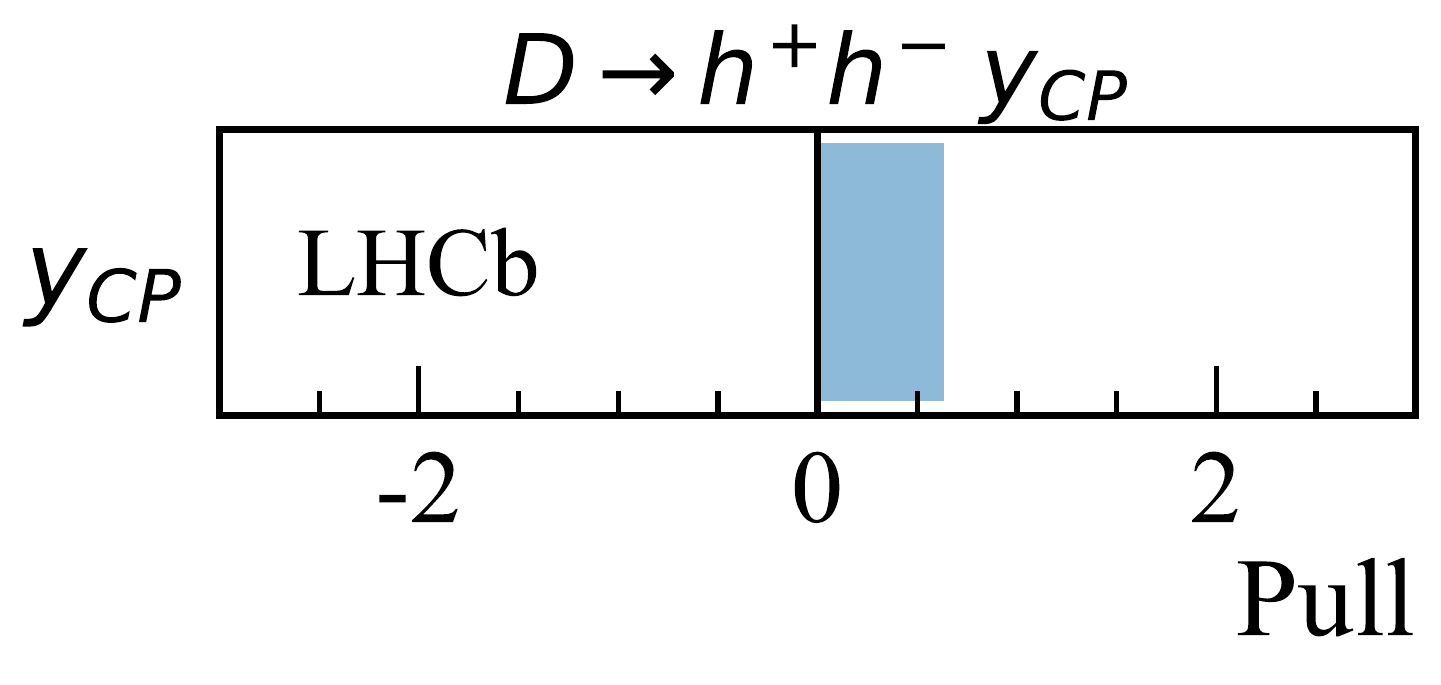} \\
		\includegraphics[width=\textwidth]{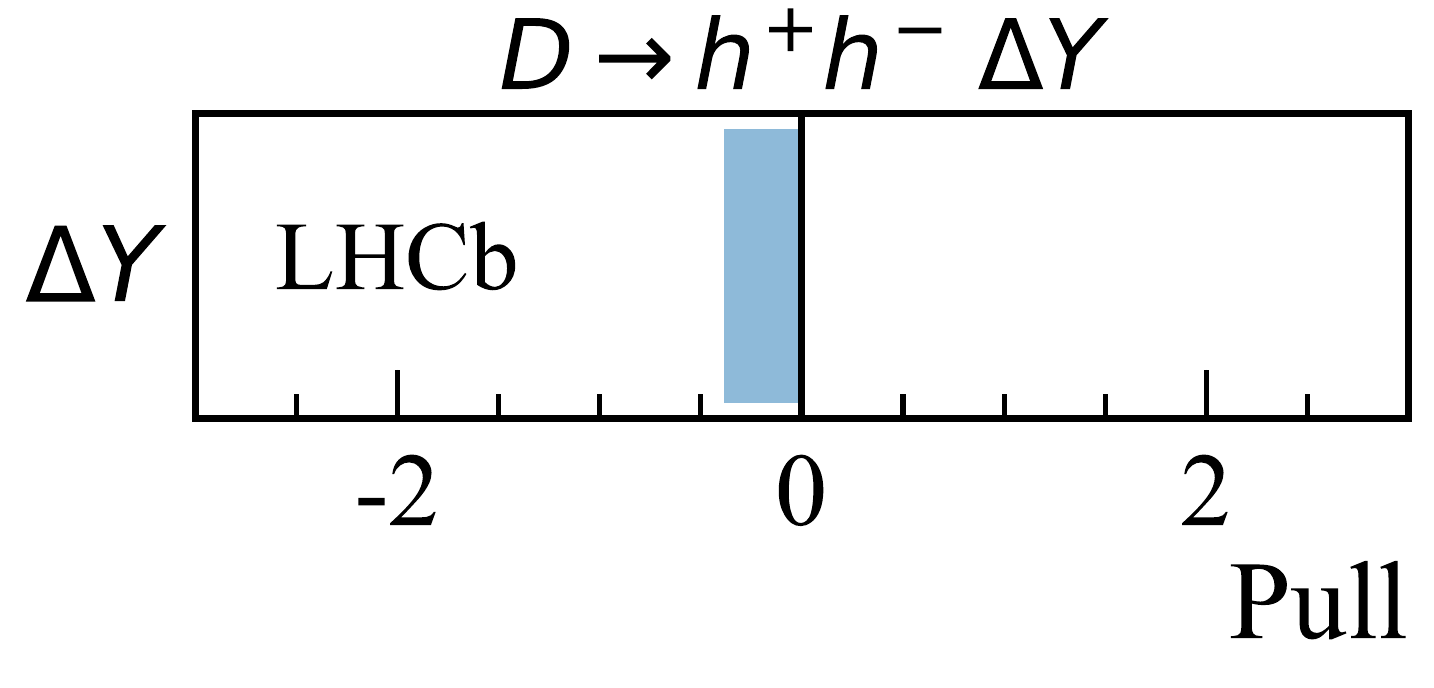} \\
		\includegraphics[width=\textwidth]{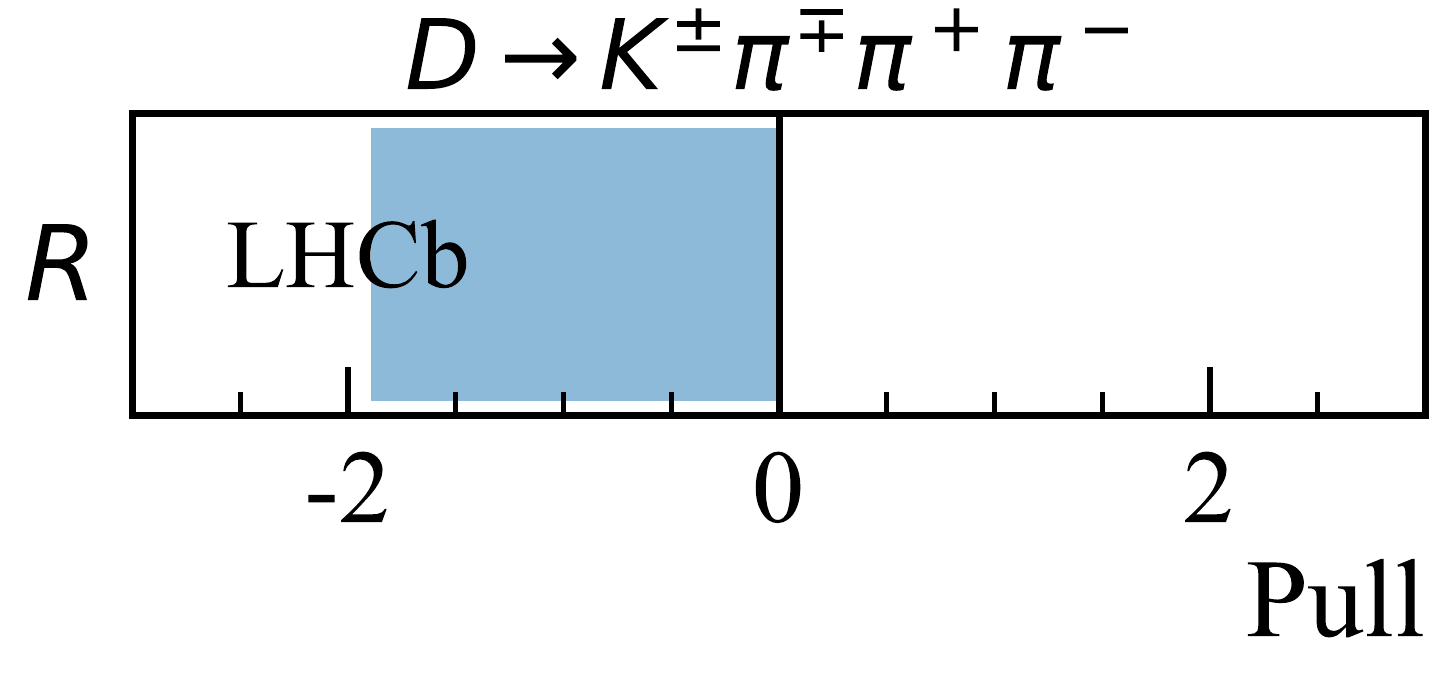}
	\end{minipage}
	\begin{minipage}[b]{0.32\textwidth}
		\includegraphics[width=\textwidth]{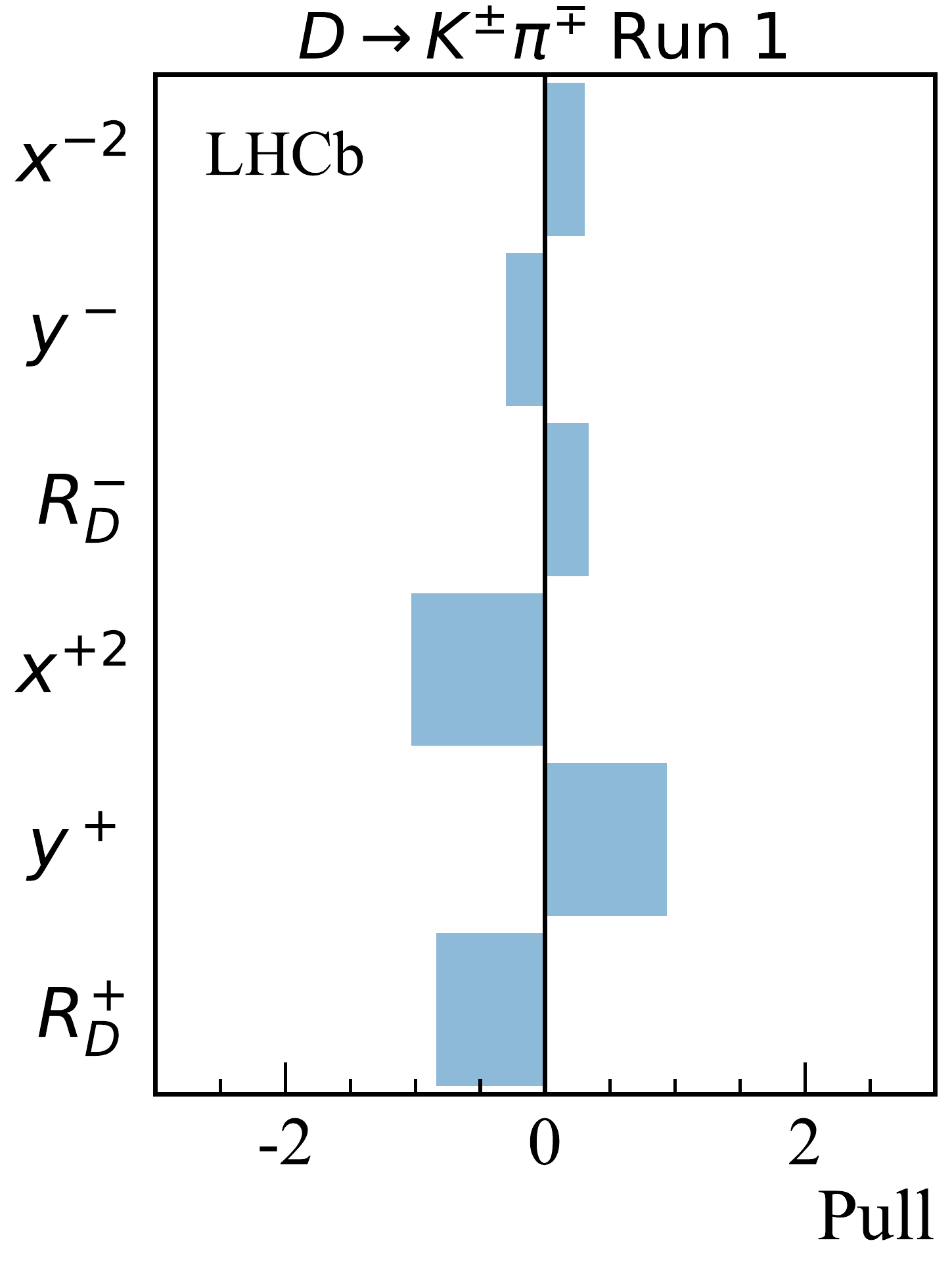}
	\end{minipage}
	\begin{minipage}[b]{0.32\textwidth}
		\includegraphics[width=\textwidth]{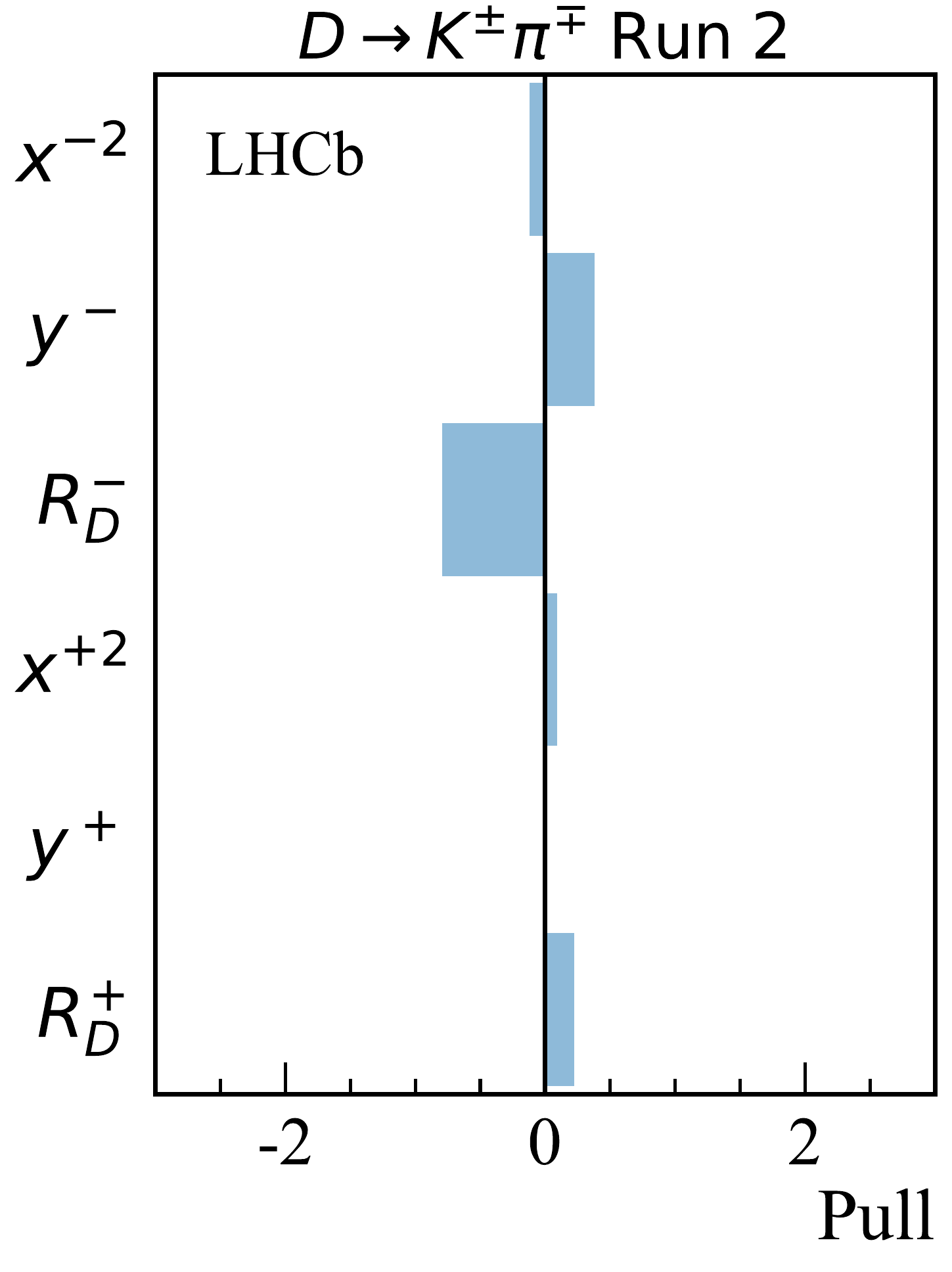}
	\end{minipage}
	\includegraphics[width=0.32\textwidth]{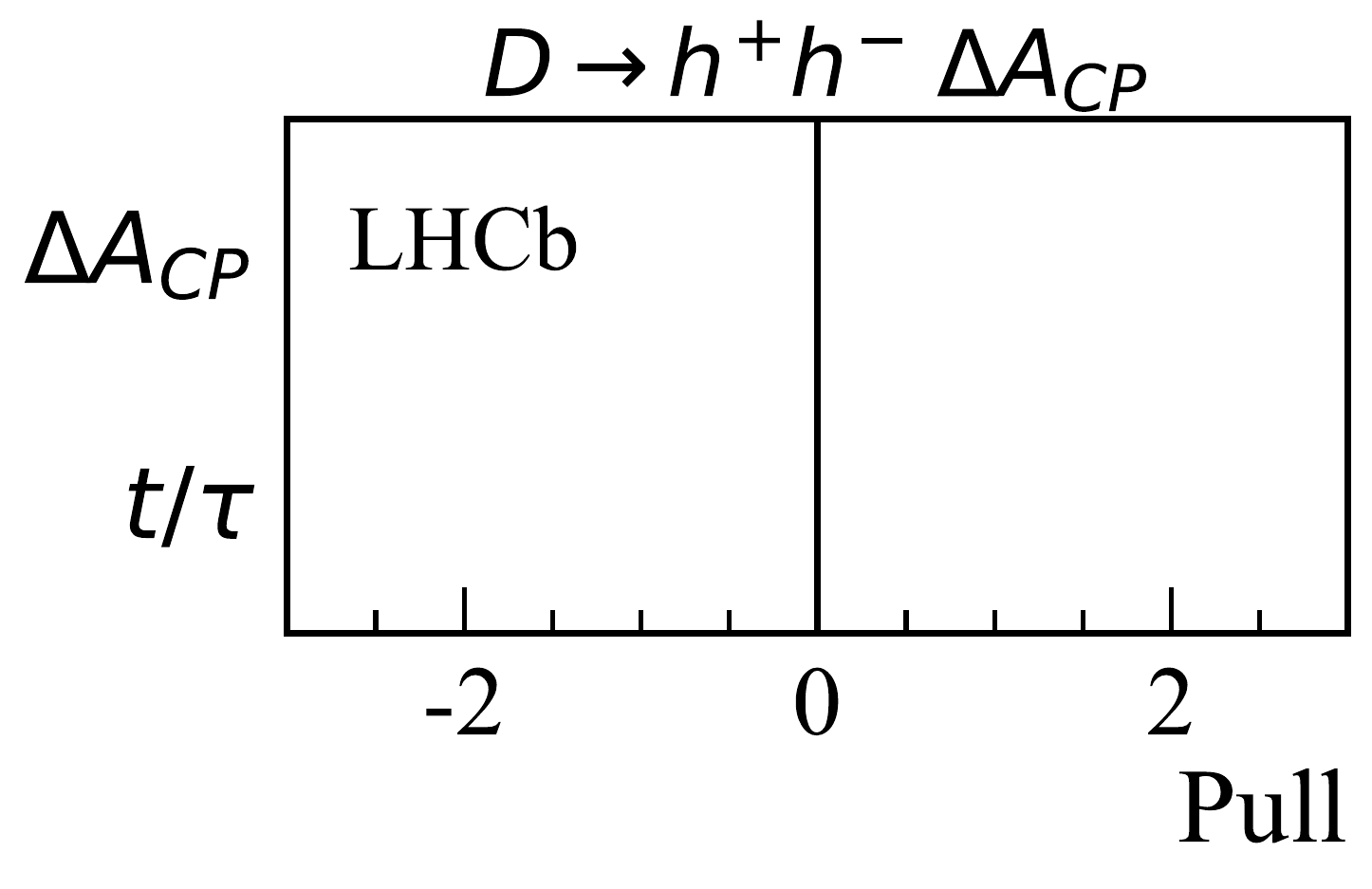}
	\caption{Pulls of the input observables, part 3 of 3.}
	\label{fig:pulls3}
\end{figure}

The pull of each input observable with respect to the global best fit point is shown in Figs.~\ref{fig:pulls1}--\ref{fig:pulls3}. The pull is defined as $(A_{\rm exp}-A_{\rm fit})/\sigma(A_{\rm exp})$, where $A_{\rm exp}$ and $A_{\rm fit}$ are the experimental input value and value at the best-fit point, respectively, and $\sigma(A_{\rm exp})$ is the experimental uncertainty.

\clearpage

\section{Additional figures}

\begin{figure}[!htb]
  \centering
  \includegraphics[width=0.6\textwidth]{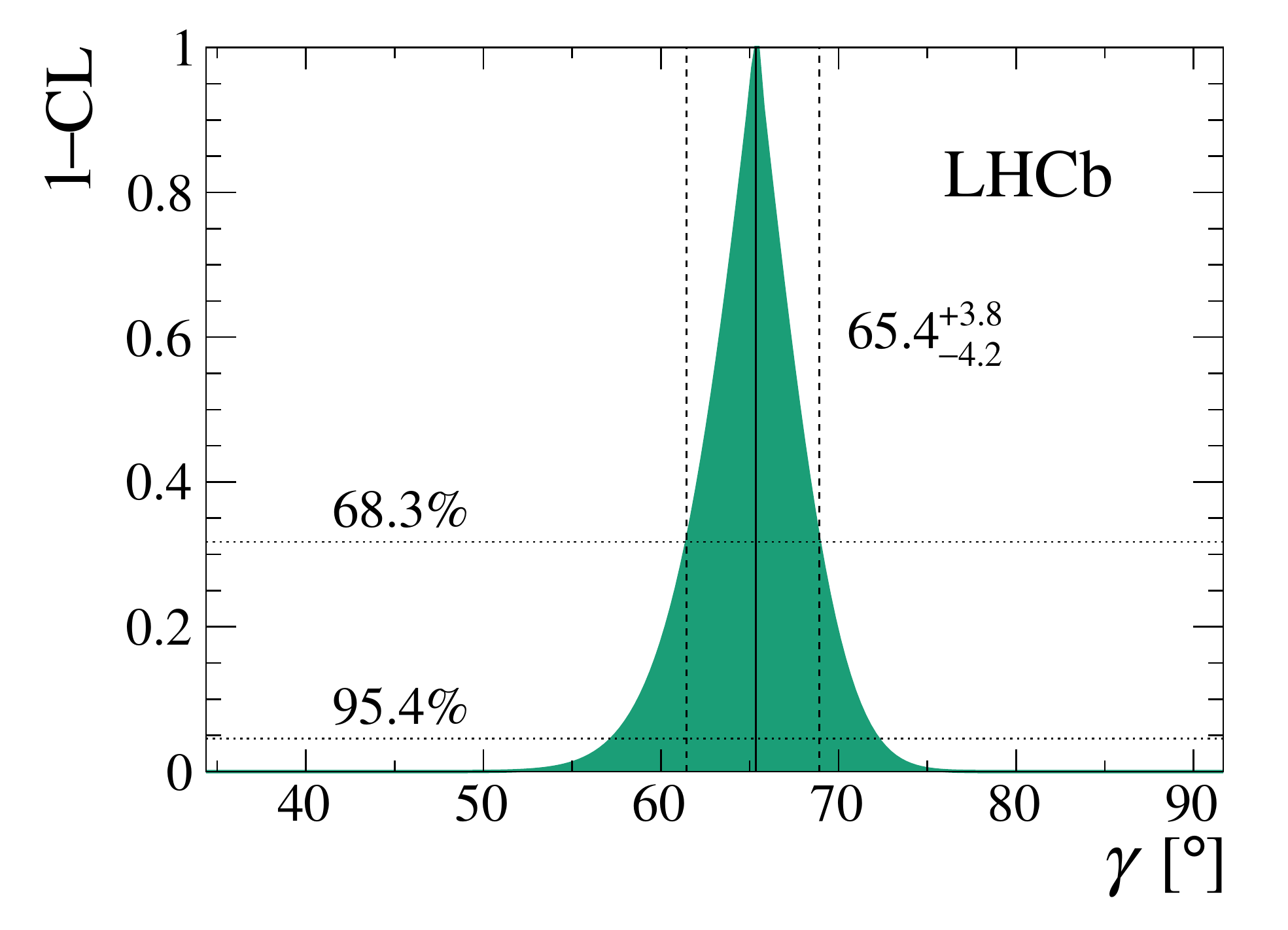}
  \caption{One dimensional \omcl profile for \g from all inputs used in the combination.}
  \label{fig:gamma_only}
\end{figure}

\begin{figure}[!htb]
  \centering
  \includegraphics[width=0.6\textwidth]{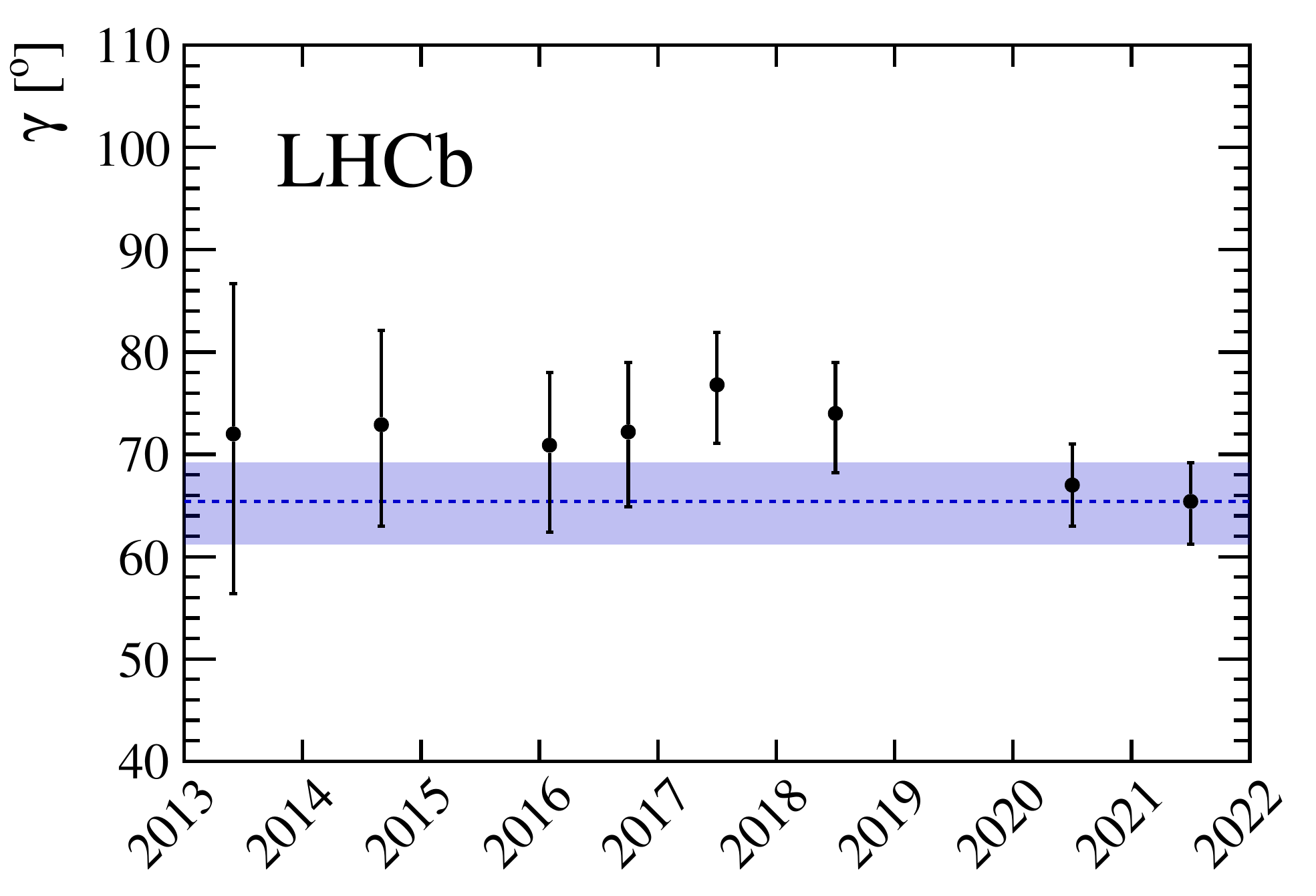}
  \caption{Evolution of the LHCb combination result for \g, with the central values and $1\sigma$ uncertainties in black. This result is the 2021 data point, the value and uncertainty are highlighted by the dashed blue line and band, respectively. }
  \label{fig:gammaevolution}
\end{figure}

Figure~\ref{fig:gamma_only} shows the $p$-value distribution as a function of \g for the global fit.
A summary of LHCb \g combination results as a function of time is given in Fig.~\ref{fig:gammaevolution}.


\clearpage

\FloatBarrier

\addcontentsline{toc}{section}{References}
\bibliographystyle{LHCb}
\bibliography{main,standard,LHCb-PAPER,LHCb-CONF,LHCb-DP,LHCb-TDR}

\newpage
\centerline
{\large\bf LHCb collaboration}
\begin
{flushleft}
\small
R.~Aaij$^{32}$,
A.S.W.~Abdelmotteleb$^{56}$,
C.~Abell{\'a}n~Beteta$^{50}$,
F.~Abudin{\'e}n$^{56}$,
T.~Ackernley$^{60}$,
B.~Adeva$^{46}$,
M.~Adinolfi$^{54}$,
H.~Afsharnia$^{9}$,
C.~Agapopoulou$^{13}$,
C.A.~Aidala$^{87}$
% ...

{\footnotesize \it

% ...
$^{9}$Universit{\'e} Clermont Auvergne, CNRS/IN2P3, LPC, Clermont-Ferrand, France\\
% ...
$^{13}$LPNHE, Sorbonne Universit{\'e}, Paris Diderot Sorbonne Paris Cit{\'e}, CNRS/IN2P3, Paris, France\\
% ...
$^{32}$Nikhef National Institute for Subatomic Physics, Amsterdam, Netherlands\\
% ...
$^{46}$Instituto Galego de F{\'\i}sica de Altas Enerx{\'\i}as (IGFAE), Universidade de Santiago de Compostela, Santiago de Compostela, Spain\\
% ...
$^{50}$Physik-Institut, Universit{\"a}t Z{\"u}rich, Z{\"u}rich, Switzerland\\
% ...
$^{54}$H.H. Wills Physics Laboratory, University of Bristol, Bristol, United Kingdom\\
% ...
$^{56}$Department of Physics, University of Warwick, Coventry, United Kingdom\\
% ...
$^{60}$Oliver Lodge Laboratory, University of Liverpool, Liverpool, United Kingdom\\
% ...
$^{68}$Syracuse University, Syracuse, NY, United States\\
% ...
$^{87}$University of Michigan, Ann Arbor, United States, associated to $^{68}$\\
\bigskip
% ...
}
\end{flushleft}

\end{document}